\documentclass[usegraphicx,useAMS,usenatbib]{mn2e}
\usepackage[T1]{fontenc}
\usepackage{verbatim}
\usepackage[normalem]{ulem}
\usepackage{amsfonts,amsmath,amssymb}
\usepackage{times}
\usepackage[italic]{mathastext}
\usepackage{enumitem}
\usepackage[colorlinks,linkcolor=blue,citecolor=blue,urlcolor=blue,breaklinks]{hyperref}
\usepackage{color}
\usepackage{upgreek}
\usepackage{flushend}
\usepackage{lineno}
\usepackage{hyperref}
\usepackage{siunitx}
\usepackage{mathrsfs}
\usepackage[normalem]{ulem}

\usepackage{eso-pic}

\AddToShipoutPictureBG*{%
  \AtPageUpperLeft{%
    \hspace{0.75\paperwidth}%
    \raisebox{-4.5\baselineskip}{%
      \makebox[0pt][l]{\textnormal{DES 2016-0209}}
}}}%

\AddToShipoutPictureBG*{%
  \AtPageUpperLeft{%
    \hspace{0.75\paperwidth}%
    \raisebox{-5.5\baselineskip}{%
      \makebox[0pt][l]{\textnormal{FERMILAB-PUB-18-131-PPD}}
}}}%

\newcommand{\deltasig}{$\Delta \Sigma$}

\newcommand{\photoz}{\mbox{photo-$z$}}
\newcommand{\photozs}{\mbox{photo-$z$s}}

\newcommand{\redmapper}{redMaPPer}
\newcommand{\redmagic}{redMaGiC}

\newcommand{\bpz}{\textsc{bpz}}

\newcommand{\imshape}{{\textsc{im3shape}}}
\newcommand{\snr}{S/N}
\newcommand{\metacal}{\textsc{metacalibration}}

\newcommand{\msun}{\mathrm{M}_{\odot}}

\newcommand\avg[1]{\langle #1 \rangle}

\newcommand{\fcl}{f_\mathrm{cl}}

\newcommand{\Mobs}{M_{\rm obs}}

\newcommand{\Var}{{\rm Var}}

\newcommand{\Rmis}{R_{\rm mis}}
\newcommand{\fmis}{f_{\rm mis}}

\newcommand{\lkhd}{{\cal L}}
\newcommand{\calB}{{\cal B}}
\newcommand{\calM}{{\cal M}}

\usepackage{todonotes}

\title[DES Y1 WL Mass Calibration of redMaPPer Clusters]{Dark Energy Survey Year 1 Results:\\ Weak Lensing Mass Calibration of redMaPPer Galaxy Clusters}

\vspace{-30pt}

\author[DES Collaboration]{
\parbox{\textwidth}{
\Large
T.~McClintock$^{1}\star$,
T.~N.~Varga$^{2,3}\dagger$,
D.~Gruen$^{4,5}\ddagger$,
E.~Rozo$^{1}$,
E.~S.~Rykoff$^{4,5}$,
T.~Shin$^{6}$,
P.~Melchior$^{7}$,
J.~DeRose$^{8,4}$,
S.~Seitz$^{2,3}$,
J.~P.~Dietrich$^{9,10}$
% ...
\begin{center} (DES Collaboration) \end{center}
}
\vspace{0.4cm}
\\
\parbox{\textwidth}{
(affiliations are listed at the end of the paper)\\
$\star$ corresponding author: \href{mailto:tmcclintock@email.arizona.edu}{tmcclintock@email.arizona.edu}\\
$\dagger$ corresponding author: \href{mailto:t.varga@physik.lmu.de}{t.varga@physik.lmu.de}\\
$\ddagger$ Einstein Fellow
}
}

\vspace{-30pt}

\begin{document}
\date{\today}
\pagerange{\pageref{firstpage}--\pageref{lastpage}}
\pubyear{2017}
\maketitle
\label{firstpage}

\begin{abstract}
We constrain the mass--richness scaling relation of \redmapper\ galaxy clusters identified in the Dark Energy Survey Year 1 data using weak gravitational lensing. We split clusters into $4\times3$ bins of richness $\lambda$ and redshift $z$ for $\lambda\geq20$ and $0.2 \leq z \leq 0.65$ and measure the mean masses of these bins using their stacked weak lensing signal. 
By modeling the scaling relation as $\langle M_{\rm 200m}|\lambda,z\rangle = M_0 (\lambda/40)^F ((1+z)/1.35)^G$, we constrain the normalization of the scaling relation at the 5.0 per cent level, finding $M_0 = [3.081 \pm 0.075 ({\rm stat}) \pm 0.133 ({\rm sys})] \cdot 10^{14}\ {\rm M}_\odot$ at $\lambda=40$ and $z=0.35$. The recovered richness scaling index is $F=1.356 \pm 0.051\ ({\rm stat})\pm 0.008\ ({\rm sys})$ and the redshift scaling index $G=-0.30\pm 0.30\ ({\rm stat})\pm 0.06\ ({\rm sys})$. These are the tightest measurements of the normalization and richness scaling index made to date from a weak lensing experiment. We use a semi-analytic covariance matrix to characterize the statistical errors in the recovered weak lensing profiles. Our analysis accounts for the following sources of systematic error: shear and photometric redshift errors, cluster miscentering, cluster member dilution of the source sample, systematic uncertainties in the modeling of the halo--mass correlation function, halo triaxiality, and projection effects. We discuss prospects for reducing our systematic error budget, which dominates the uncertainty on $M_0$.  Our result is in excellent agreement with, but has significantly smaller uncertainties than, previous measurements in the literature, and augurs well for the power of the DES cluster survey as a tool for precision cosmology and upcoming galaxy surveys such as LSST, Euclid and WFIRST.
\end{abstract}



\section{Introduction} 
\label{sec:introduction}

Galaxy clusters have the potential to be the most powerful cosmological probe \citep{CosmicVisions16}. Current constraints are dominated by uncertainties in the calibration of cluster masses \citep[e.g.,][]{mantzetal15,planck_clusters_15,rozoetal10}. Weak lensing allows us to determine the mass of galaxy clusters: gravitational lensing of background galaxies by foreground clusters induces a tangential alignment of the background galaxies around the foreground cluster. This alignment is a clear observational signature predicted from clean, well-understood physics.  Moreover, the resulting signal is explicitly sensitive to all of the cluster mass, not just its baryonic component, and is insensitive to the dynamical state of the cluster. For all these reasons, weak lensing is the most robust method currently available for calibrating cluster masses.  It is therefore not surprising that the community has invested in a broad range of weak lensing experiments specifically designed to calibrate the masses of galaxy clusters \citep{WtGI,vonderLinden14, applegateetal14,Hoekstra2015,okabesmith16,Mantz2015,rmsva,Simet2017,Murata2018,Dietrich2017,Miyatake2018_ACTPolHSC,Medezinski2018_PlanckHSC}.

The Dark Energy Survey (DES) is a 5,000 square degree photometric survey of the southern sky. It uses the 4-meter Blanco Telescope and the Dark Energy Camera \citep{Flaugher2015} located at the Cerro Tololo Inter-American Observatory. As its name suggests, the primary goal of the DES is to probe the physical nature of dark energy, in addition to constraining the properties and distribution of dark matter. Owing to its large area, depth, and image quality, at its conclusion DES will support optical identification of $\sim100,000$ galaxy clusters and groups up to redshift $z\approx1$. We use galaxy clusters identified using the \redmapper\ algorithm \citep{Rykoff2014_RM1}, which assigns each cluster a photometric redshift and optical richness $\lambda$ of red galaxies. To fully utilize these clusters, one must understand mass-observable relations (MORs), such as that between cluster mass and optical richness. Weak lensing can establish this relation -- with high statistical uncertainty for individual clusters, but low systematic uncertainty in the mean mass scale derived from the joint signal of large samples.

In this work, we use stacked weak lensing to measure the mean galaxy cluster mass of \redmapper\ galaxy clusters identified in DES Year 1 (Y1) data. We use these data to calibrate the mass--richness--redshift relation of these clusters. In \citet{rmsva} we provided a first calibration of this relation using DES Science Verification (SV) data.  There, we were able to achieve a 9.2 per cent statistical and 5.1 per cent systematic uncertainty. Here, we update that result using the first year of regular DES observations, incorporating a variety of improvements to the analysis pipeline.  Our results provide the tightest, most accurate calibration of the richness--mass relation of galaxy clusters to date, at 2.4 per cent statistical and 4.3 per cent systematic uncertainty.

The structure of this paper is as follows. In \autoref{sec:desy1}, we introduce the DES Y1 data used in this work. In \autoref{sec:stacked_lensing_measurements} we describe our methodology for obtaining ensemble cluster density profiles from stacked  weak lensing shear measurements, with a focus on updates relative to \citet{rmsva}. A comprehensive set of tests and corrections  for systematic effects is presented in \autoref{sec:systematics}. The model of the lensing data and the inferred stacked cluster masses are given in \autoref{sec:modeling}. The main result, the mass--richness--redshift relation of \redmapper\ clusters in DES, is presented in \autoref{sec:mass_richness_relation}. We compare our results to other published works in the literature in \autoref{sec:comparisons}, discuss systematic improvements made in this work compared to \citet{rmsva} in \autoref{sec:future_improvements}, and conclude in \autoref{sec:summary}. In Appendix~\ref{app:rm} we present the DES Y1 \redmapper\ catalog used in this work for public use. Supplementary information on the analysis is given in additional appendices.

Unless otherwise stated, we assume a flat $\Lambda$CDM cosmology with $\Omega_{\rm m}=0.3$ and $H_0=70$ km s$^{-1}$ Mpc$^{-1}$, with distances defined in physical coordinates, rather than comoving. Finally, unless otherwise noted all cluster masses refer to $M_{200\rm{m}}$. That is, cluster mass is defined as the mass enclosed within a sphere whose average density is 200 times higher than the mean cosmic matter density $\bar \rho_m$ at the cluster's redshift, matching the mass definition used in the cosmological analyses that make use of our calibration.


\section{The DES year 1 data}
\label{sec:desy1}

\begin{figure}
 	\includegraphics[width=\linewidth]{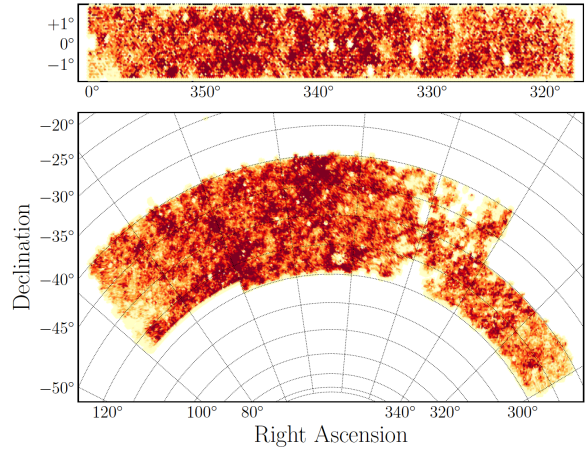}
	\caption{Surface density of source galaxies in the \metacal\ catalog within the DES Y1 footprint in the ``S82'' field (\emph{top}) and the ``SPT'' field (\emph{bottom}).}
	\label{fig:footprint}
\end{figure}

DES started its main survey operations in 2013, with the Year One (Y1) observational season running from August 31, 2013 to February 9, 2014 \citep{Y1gold}. During this period 1839 deg$^2$ of the southern sky were observed in three to four tilings in each of the four DES bands $g,r,i,z$, as well as $\sim$1800 deg$^2$ in the $Y$-band.  The resulting imaging is shallower than the SV data release but covers a significantly larger area. In this study we utilize approximately 1500 deg$^2$ of the main survey, split into two large non-contiguous areas. This is a reduction from the 1800 deg$^2$ area due to a series of veto masks. These masks include masks for bright stars and the Large Magellanic Cloud, among others.  The two non-contiguous areas are the ``SPT'' area (1321 deg$^2$), which overlaps the footprint of the South Pole Telescope Sunyaev-Zel'dovich Survey \citep{Carlstrom11.1}, and the ``S82'' area (116 deg$^2$), which overlaps the Stripe-82 deep field of the Sloan Digital Sky Survey \citep[SDSS;][]{S82}. The DES Y1 footprint is shown in \autoref{fig:footprint}.

In the following we briefly describe the main data products used in this analysis, and refer the reader to the corresponding papers for more details. The input photometric catalog, as well as the photometric redshift and weak lensing shape catalogs used in this study have already been employed in the cosmological analysis combining galaxy clustering and weak lensing by the DES collaboration \citep{desy1kp}.

\subsection{Photometric Catalog}
\label{sec:photocat}

Input photometry for the \redmapper{} cluster finder (\autoref{sec:redmapper}) and photometric redshifts (\autoref{sec:photo-z}) was derived from the DES Y1A1 Gold catalog~\citep{Y1gold}. Y1A1 Gold is the science-quality internal photometric catalog of DES created to enable cosmological analyses. This data set includes a catalog of objects as well as maps of survey depth and foreground masks, and star-galaxy classification.  In this work we make use of the multi-epoch, multi-object fitting (MOF) composite model ({\tt CM}) galaxy photometry.  The MOF photometry simultaneously fits a psf-convolved galaxy model to all available epochs and bands for each object, while subtracting and masking neighbors.  The typical $10\sigma$ limiting magnitude inside $2''$ diameter apertures for galaxies in Y1A1 Gold using MOF {\tt CM} photometry is $g \approx 23.7$, $r \approx 23.5$, $i \approx 22.9$, and $z \approx 22.2$. Due to its low depth and calibration uncertainty, we do not use $Y$ band photometry for shape measurement or photometric redshift estimation.

The galaxy catalog used for the \redmapper{} cluster finder is constructed as follows.  Bad objects that are determined to be catalog artifacts, including having unphysical colors, astrometric discrepancies, and PSF model failures are rejected~\citep[Section 7.4][]{Y1gold}.  Galaxies are then selected via the more complete {\tt MODEST\_CLASS} classifier \citep[Section 8.1][]{Y1gold}.  Only galaxies that are brighter in $z$ band than the local $10\,\sigma$ limiting magnitude are used by \redmapper.  The average survey limiting magnitude is deep enough to image a $0.2\ L^*$ galaxy at $z\approx0.7$.  Finally, we remove galaxies in regions that are contaminated by bright stars, bright nearby galaxies, globular clusters, and the Large Magellanic Cloud.

\subsection{Cluster catalog}
\label{sec:redmapper}

We use a volume limited sample of galaxy clusters detected in the DES Y1 photometric data using the \redmapper\ cluster finding algorithm v6.4.17 \citep{Rykoff2014_RM1,Rykoff2016}. This \redmapper\ version is fundamentally the same as the v6.3 algorithm described in Rykoff et al. (2016), with minor updates.

Two versions of the \redmapper\ cluster catalog are generated: a ``flux limited'' version, which includes high redshift clusters for which the richness requires extrapolation along the cluster luminosity function, and one that is locally volume-limited.  By ``locally volume-limited'' we mean that at each point in the sky, a galaxy cluster is included in the sample if and only if all cluster galaxies brighter than the luminosity threshold used to define cluster richness in \redmapper\ lies above $10\,\sigma$ in $z$, $5\,\sigma$ in $i$ and $r$, and $3\,\sigma$ in $g$ according to the survey MOF depth maps~\citep{Y1gold}.  That is, no extrapolation in luminosity is required when estimating cluster richness. At the threshold the galaxy sample is $>90-95$ per cent complete. It is this volume-limited cluster sample that is used in follow-up work deriving cosmological constraints from the abundance of galaxy clusters. Consequently, we focus exclusively on this volume-limited sample in this work. It contains more than 76,000 clusters down to $\lambda>5$, of which more than 6,500  are above $\lambda=20$. The format of the catalogs are described in \autoref{app:rm}.

\begin{figure}
  \includegraphics[width=\linewidth]{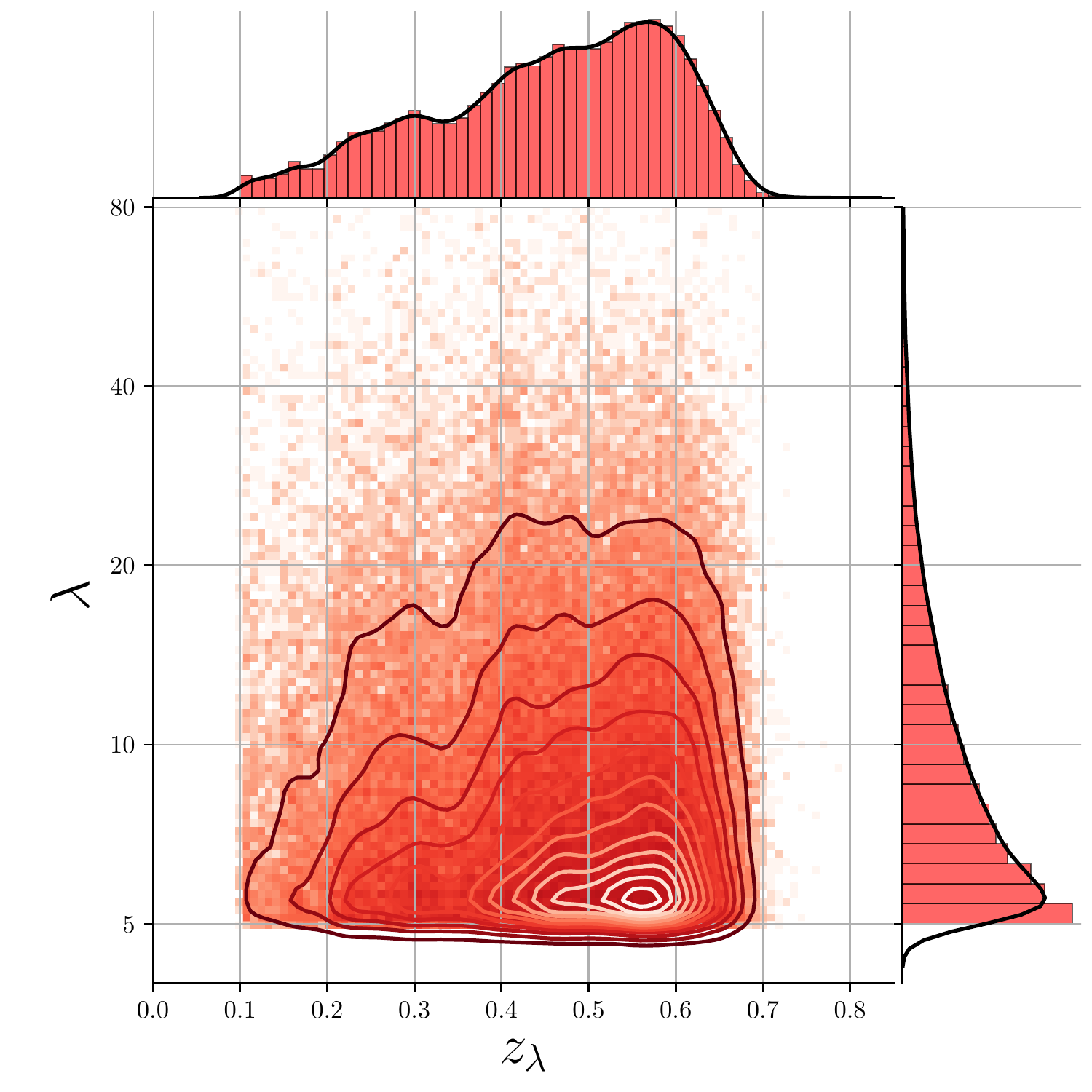}
  \caption{Redshift--richness distribution of \redmapper\, clusters in the volume limited DES Y1 cluster catalog, overlaid with density contours to highlight the densest regions. At the top and on the right are histograms of the projected quantities, $z_\lambda$ and $\lambda$, respectively, with smooth kernel density estimates overlaid.}
  \label{fig:rm_z_lambda}
\end{figure}

\redmapper\ identifies galaxy clusters as overdensities of red-sequence galaxies. Starting from an initial set of spectroscopic seed galaxies, the algorithm iteratively fits a model for the local red-sequence, and finds cluster candidates while assigning a membership probability to each potential member. Clusters are centered on bright galaxies selected using an iteratively self-trained matched-filter method. The method allows for the inherent ambiguity of selecting a central galaxy by assigning a probability to each galaxy of being the central galaxy of the cluster. The final membership probabilities of all galaxies in the field are assigned based on spatial, color, and magnitude filters. 

The distribution of cluster richness and redshift of the DES volume-limited cluster sample is shown in \autoref{fig:rm_z_lambda}. The richness estimate $\lambda$ is the sum over the membership probabilities of all galaxies within a pre-defined, richness--dependent projected radius $R_\lambda$. The radius $R_\lambda$ is related to the cluster richness via $R_\lambda= 1.0(\lambda/100)^{0.2}\ h^{-1}{\rm Mpc}$.  This relation was found to minimize the scatter between richness and X-ray luminosity in \cite{Rykoff12}. A redshift estimate for each cluster is obtained by maximizing the probability that the observed color-distribution of likely members matches the self-calibrated red-sequence model of \redmapper.

\begin{figure}
  \includegraphics[width=\linewidth]{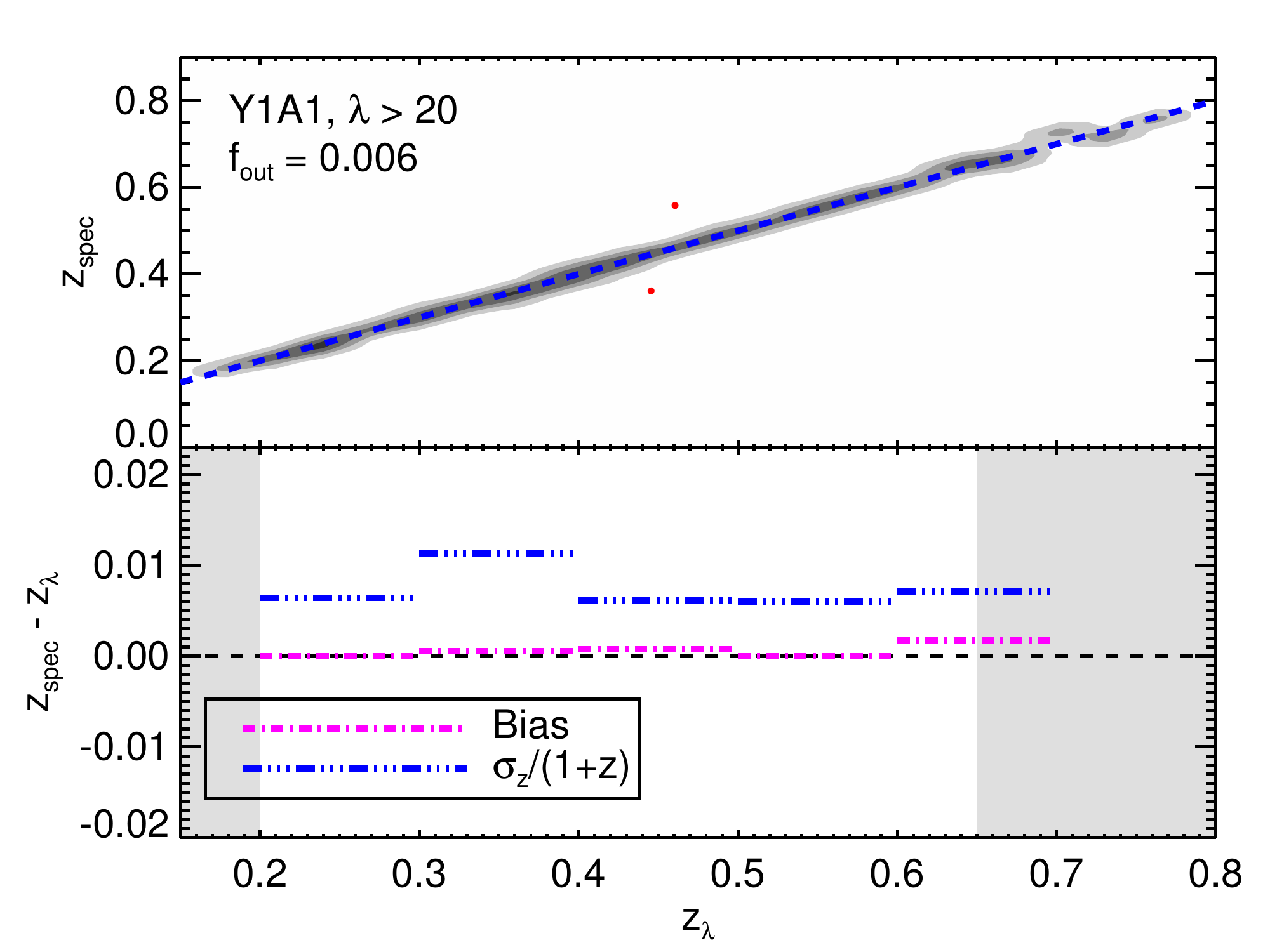}
  \caption{Photometric redshift performance of the DES Y1 \redmapper\ cluster catalog, as evaluated using available spectroscopy (333 clusters). {\it Upper panel}: Gray contours are $3\sigma$ confidence intervals, and the two red dots are the only $4\sigma$ outliers, caused by miscentering on a foreground/background galaxy. {\it Lower Panel}: \photoz\ bias and uncertainty. The comparatively large uncertainty from $0.3<z<0.4$ is due to a filter transition. }
  \label{fig:z_lambda}
\end{figure}

\autoref{fig:z_lambda} shows the photometric redshift performance of the DES Y1 volume-limited \redmapper\ cluster sample. The photometric redshift bias and scatter are calculated by comparing the photometric redshift of the clusters to the spectroscopic redshift of the central galaxy of the cluster, where available. Unfortunately, the small overlap with existing spectroscopic surveys means that our results are limited by small-number statistics: there are only 333 galaxy clusters with a spectroscopic central galaxy, and only 34 (six) with redshift $z\geq 0.6$ ($z\geq 0.65$).  Nevertheless, the photometric redshift performance is consistent with our expectations: our redshifts are very nearly unbiased, and have a remarkably tight scatter --- the median value of $\sigma_z/(1+z)$ is $\approx 0.006$. An upper limit for the photometric redshift bias of $0.003$ is consistent with our data.

Of particular importance to this work is the distribution of miscentered clusters -- both the frequency and severity of their miscentering. Based on the \redmapper\ centering probabilities, we would expect $\approx 80$ per cent of the clusters to be correctly centered, meaning the most likely \redmapper\ central galaxy is at the center of the potential well of the host halo. In practice, the fraction of correctly centered galaxy clusters is closer to $\approx 70$ per cent, as estimated from a detailed comparison of the \redmapper\ photometric centers to the X-ray centers of \redmapper\ clusters for which high-resolution X-ray data is available \citep{Zhang2018,vonderLinden2018}. The expected impact of this miscentering effect, and the detailed model for the miscentered distribution from \citet{Zhang2018,vonderLinden2018} is described in \autoref{sec:miscentering_correction}. 


\subsection{Shear catalogs}
\label{sec:shearcat}

Our work uses the DES Y1 weak lensing galaxy shape catalogs presented in \cite{Y1shape}. Two independent catalogs were created: \metacal\ \citep{SheldonMETA, HuffMETA} based on \textsc{ngmix} \citep{ngmix2015}, and \imshape\ \citep{Zuntz13}. Both pass a multitude of tests for systematics, making them suitable for cosmological analyses. While the Y1 data is shallower than the DES SV data, improvements in the shear estimation pipelines and overall data quality enabled us to reach a  number density of sources similar to that from DES SV data \citep{Jarvis2016}. 

In this study we will focus exclusively on the \metacal\ shear catalog because of its larger effective source density (6.28 arcmin$^{-2}$) compared to the \imshape\ catalog (3.71 arcmin$^{-2}$).  The difference mainly arises because \metacal\ utilizes images taken in $r,i,z$ bands, whereas \imshape\ relies exclusively on $r$-band data. In the \metacal\ shear catalog the fiducial shear estimates are obtained from a single Gaussian fit via the \textsc{ngmix} algorithm. As a supplementary data product \metacal\ provides $(g,r,i,z)$-band fluxes and the corresponding error estimates for objects using its internal model of the galaxies.  

Galaxy shape estimators, such as the \textsc{ngmix} model-fitting procedure used for \metacal, are subject to various sources of systematic errors. For a stacked shear analysis, the dominant problem is a multiplicative bias, i.e. an over- or underestimation of gravitational shear as inferred from the mean tangential ellipticity of lensed galaxies. This bias needs to be characterized and corrected. Traditionally, this is done using simulated galaxy images -- with the critical limitation that simulations never fully resemble the observations.

The \metacal\ catalog, in contrast, uses the galaxy images themselves to de-bias shear estimates. Specifically, each galaxy image is deconvolved from the estimated point spread function (PSF), and a small positive and negative shear is applied to the deconvolved image in both the $\hat{e}_1$ and $\hat{e}_2$ directions.  The resulting images are then convolved once again with a representation of the PSF, and an ellipticity $\mathbf{e}$ is estimated for these new images \citep{Y1shape}. These new measurements can be used to directly estimate the response of the ellipticity measurement to a gravitational shear $\boldsymbol{\gamma}$ using finite difference derivatives:
\begin{equation} \label{eq:r}
	\mathsf{R}_\gamma=\frac{\partial\mathbf{e}}{\partial\mathbf{\bgamma}}\,.
\end{equation}
Selection effects can also be accounted for by examining the response of the selections to shear. The application of a weight when calculating the mean shear over an ensemble is effectively a type of smooth selection, and is accounted for in the same way.  We describe this effect with a selection response $\mathsf{R}_\mathrm{sel}$, which leads to the response-corrected mean shear estimate
\begin{equation}
	\label{eq:metacalibration}
        \langle {\bgamma} \rangle \approx \langle \mathsf{R} \rangle^{-1} \langle \mathsf{R} \cdot  \mathbf{\bgamma}_\mathrm{true} \rangle \approx \langle \mathsf{R} \rangle^{-1} \langle \mathbf{e} \rangle
\end{equation}
from biased measurements $\mathbf{e}$ with a joint response $\mathsf{R}\approx\mathsf{R}_\gamma+\mathsf{R}_\mathrm{sel}$ \citep{SheldonMETA}. Here the left hand side represents our estimate of the mean shear, while $\mathbf{\bgamma}_\mathrm{true}$ refers to the actual value.

$\mathsf{R}$ is a $2\times 2$ Jacobian matrix for the two ellipticity components $e_{1},e_{2}$ in a celestial coordinate system. For the \metacal\ mean shear measurements in this work, we calculate the response of mean \emph{tangential} shear on mean tangential ellipticity. $\mathsf{R}$ is close to isotropic on average, which is why other recent weak lensing analyses \citep{Troxel2017,Prat2017,Gruen2018_splits,Chang2017} have assumed it to be a scalar. For the larger tangential shears measured on small scales around clusters, however, we account for the fact that the response might not be quite isotropic by explicitly rotating it to the tangential frame.

The tangential ellipticity $e_\mathrm{T}$ is related to $e_1,e_2$ (and likewise $\gamma_T$ to $\gamma_1$ and $\gamma_2$) by 
\begin{equation} \label{eq:et}
	e_\mathrm{T}=-e_1\cos(2\phi)-e_2\sin(2\phi) \; ,
\end{equation}
where $\phi$ is the polar angle of the source in a coordinate system centered on the lens. For the shear response, the corresponding rotation is derived from \autoref{eq:r} and \autoref{eq:et} as
\begin{equation}
\begin{split}
\mathsf{R}_{\gamma,\mathrm{T}} =\   &\mathsf{R}_{\gamma,11}\cos^2(2\phi) + \mathsf{R}_{\gamma,22}\sin^2(2\phi) +\\&\left(\mathsf{R}_{\gamma,12} + \mathsf{R}_{\gamma,21}\right)\sin(2\phi)\cos(2\phi)\,.
\end{split}
\end{equation}
For the \metacal\ selection response, no such rotation can be performed as the term itself is only meaningful for ensembles of galaxies. In this case, we exploit that the orientation of source galaxies should be random relative to the clusters, which suggest a symmetrized version of the response in the tangential frame:
\begin{equation}
\label{eq:meta-gamma}
	\langle\mathsf{R}_{\mathrm{sel}}^{(\mathrm{T})} \rangle 
	\approx \frac{1}{2}\mathrm{Tr}\langle\mathsf{R}_{\mathrm{sel}}\rangle
   \; \; \mathrm{where} \; \; \langle\mathsf{R}_\mathrm{sel}\rangle_{i, j}  \approx \frac{\langle e_i  \rangle^{S+}- \langle e_i\rangle^{S-} }{ \Delta\gamma_j}\,.
\end{equation}
In the above equation $\langle e_i \rangle^{S\pm}$ denotes the mean \emph{un-sheared} ellipticity of galaxies when selected based on quantities measured on their artificially \emph{sheared} images. Four such sheared images are created by applying positive ($+$) and negative ($-$) shears of magnitude $\Delta\gamma_j = 0.01$ along the $j\in\left\{\hat{e}_1; \hat{e}_2\right\}$ directions separately. The response in the tangential and cross directions are consistent, however both depend on the cluster-centric distance and richness of the lensing clusters. These dependencies will be investigated in a future work.
Errors introduced from this approximation are sub-dominant due to the already small bias associated with source galaxy selection. 
A detailed discussion of additional possible systematics in our specific analysis is presented in \autoref{sec:shear_systematics}.

\subsubsection*{Blinding procedure}
\label{sec:catalog_blinding}

As a precaution against unintentional confirmation bias in the scientific analyses, both weak lensing shape catalogs produced for DES Y1 had an unknown \emph{blinding} factor in the magnitude of $\mathbf{e}$ \citep{Y1shape} applied to them. This unknown factor was constrained between 0.9 and 1.1. While we made initial blinded measurements for this work, the factor was revealed as part of unblinding the cosmology results of \citet{desy1kp}.

In accordance with the practices of other DES Y1 cosmology analyses, we have further adopted a secondary layer of blinding.  Specifically, we blindly transform the chains from our MCMCs to hide our in-progress results, and to prevent comparison between our cluster masses and those estimated using mass--observable relations from the literature. Chains of the parameters in the modeled lensing profiles and the mass--richness relation were unaltered after unblinding.

\subsection{Photometric redshift catalog}
\label{sec:photo-z}

In interpreting the weak gravitational lensing signal of galaxy clusters as physical mass profiles we need to employ information about the geometry of the source-lens systems by considering the relevant angular-diameter distances. To calculate these distances we rely on estimates of the overall redshift distribution of source galaxies, and also on information about the individual $P(z)$ of source galaxies.

We use the DES Y1 photometric redshifts estimated and validated by \cite{Y1pz} using the template-based \bpz\  algorithm \citep{Benitez2000, coe06}. It was found by \cite{Y1pz} that these photo-z estimates were modestly biased, introducing an overall multiplicative systematic correction in the recovered weak lensing profiles.  We determine this correction and its systematic uncertainty in \autoref{sec:photometric_redshift_systematics}.

In order to be able to correct selection effects due to the change of photo-$z$ with shear while utilizing the highest signal-to-noise flux measurements for determining the source redshift distribution, we use two separate BPZ catalogs: one generated from \metacal-measured photometry (for selecting and weighting sources), and one from MOF (see \autoref{sec:photocat}) photometry (for determining the resulting source redshift distributions). Details of this are described in the following section.


\section{Stacked lensing measurements}
\label{sec:stacked_lensing_measurements}


\subsection{Mass density profiles}
\label{sec:lensing_methodology}


Gravitational lensing induces distortions in the images of background ``source'' galaxies. In the limit of weak gravitational lensing, these are characterized by the ``reduced shear''
\begin{equation}
	\label{eq:reduced_shear_definition}
	\boldsymbol{g}\equiv \frac{\bgamma}{1- \kappa}\,,
\end{equation}
where  $\bgamma$ is the shear and $\kappa$ is the convergence. In the presence of non-negligible convergence, the ellipticity estimator $\mathbf{e}$ introduced in \autoref{sec:shearcat}  relates to the reduced shear as $\langle\mathbf{g}\rangle\approx \langle \mathsf{R} \rangle^{-1} \langle\mathbf{e} \rangle$.

The gravity of a localized mass distribution, such as a galaxy cluster, induces positive shear along the tangential direction with respect to the center of the overdensity. This net tangential shear results in the stretching and preferential alignment of the images of background galaxies along the tangential direction. The magnitude of the azimuthally averaged tangential shear $\bgamma_{\rm T}$ at projected radius $R$ can be predicted from the line-of-sight projected surface mass density $\Sigma$ of the lens mass distribution by the relation
\begin{equation}
	\label{eq:dsdef}
	\gamma_{\rm T} = \frac{\overline{\Sigma}(<R) - \overline{\Sigma}(R)}{\Sigma_{{\rm crit}}} \equiv \frac{\Delta\Sigma(R)}{\Sigma_{{\rm crit}}}.
\end{equation}
Here $\overline{\Sigma}(<R)$ represents the average surface mass density within projected radius $R$, and $\overline{\Sigma}(R)$ represents the (azimuthal) average of the surface mass density at $R$. For the case of reduced shear this equation holds only in linear order, therefore we account for the effect of $\kappa$ in our model described in \autoref{sec:reduced_shear}.

The geometry of the source--lens system modulates the amplitude of the induced shear signal, and is characterized by the critical surface mass density
\begin{equation}
	\label{eq:sigma_crit}
	\Sigma_{\rm crit}(z_{\rm s}, z_{\rm l}) =\frac{c^2}{4\pi G} \frac{D_{\rm s}}{D_{\rm l} D_{\rm ls}}
\end{equation}
in \autoref{eq:dsdef}. Here $D_{\rm s}$, $D_{\rm l}$ and $D_{\rm ls}$ are the angular diameter distances to the source, to the lens, and between the lens and the source.
Estimating the $\Delta\Sigma$ signal thus relies on robustly estimating the redshifts of the galaxy clusters and the source galaxies.
The lens redshifts are the photometric redshift estimates from the \redmapper\ algorithm. The statistical uncertainty on these estimates is found to be $\Delta z_{\rm l} \approx 0.01$ \citep{Rykoff2016}, which is negligible compared to other sources of error in the lensing measurement, allowing us to treat these redshifts as point estimates.

Source redshifts are also estimated from photometry, and are described by a probability distribution $p_\mathrm{phot}(z_{\rm s})$ for each source galaxy. We can therefore only estimate an {\em effective} critical surface density
\begin{equation}
	\label{eq:sigma_critinv_eff}
	\langle\Sigma_{\rm crit}^{-1}\rangle_{i, j} = \int {\rm d}z_{\rm s} \; p_\mathrm{phot}(z_{{\rm s}, i})\; \Sigma_{\rm crit}^{-1}(z_{{\rm l}, j}, z_{{\rm s}, i})\,,
\end{equation}
where $i$ and $j$ index the source and the lens in a lens-source pair. Note that here we choose to express the \emph{inverse} critical surface density, which is the predicted amplitude of the lensing signal in \autoref{eq:dsdef}. We consistently define it as zero if $z_s \leq z_l$. For reasons of data compression, we will in fact \emph{not} use the full integral over $p_{\rm phot}(z)$ later, but rather replace \autoref{eq:sigma_critinv_eff} by $\Sigma_{\rm crit}^{-1}$ evaluated at a random sample of the $p_{\rm phot}(z)$.


\subsubsection{The lensing estimator} 
\label{sec:ideal_estimator}

Due to the low signal-to-noise of individual source-lens pairs we measure the stacked (mean) signal of many source galaxies around a selection of clusters. 

\cite{Sheldon04.1} show that the minimum variance estimator for the weak lensing signal is
\begin{equation}
	\label{eq:ds_ideal}
	\widetilde{\Delta\Sigma} = \frac{\sum\limits^{{\rm lens}}_{j}  \sum\limits^{{\rm src}}_{i} \left. w_{i,j}\;  e_{{\rm T};\,i,j} \middle/ \langle \Sigma_{{\rm crit}}^{-1} \rangle_{i, j} \right.}{\sum\limits_{j,i} w_{i,j}} \,,
\end{equation}
where the summation goes over all source--lens pairs in some radius bin and $e_{{\rm T};\,i,j}$ is the tangential component of the ellipticity of source $i$ relative to lens $j$. The optimal weights, proportional to the inverse variance of $e_{{\rm T};\,i,j} / \left\langle \Sigma_{{\rm crit}}^{-1}\right\rangle$, are
\begin{equation}
	\label{eq:w_ideal}
	w_{i,j} =  \left. \langle \Sigma_{{\rm crit}}^{-1} \rangle_{i, j}^{2} \middle/ \sigma^2_{\gamma, i} \right.\,, 
\end{equation}
where $\sigma^2_{\gamma,i }$ is the estimate on the variance of the measured shear estimate of galaxy $i$ relating to both the intrinsic variance of shapes and also to the uncertainty originating from shear estimation. 


\subsubsection{Practical lensing estimator}
\label{sec:practical_estimator}

This estimator can be equivalently understood as a mean tangential ellipticity, weighted by the expected shear signal amplitude of each galaxy $\langle \Sigma_{{\rm crit}}^{-1}\rangle$. It is normalized by the expected signal per unit $\Delta\Sigma$, i.e.~the $\langle \Sigma_{{\rm crit}}^{-1}\rangle$-weighted mean of the $\langle \Sigma_{{\rm crit}}^{-1}\rangle$. With this in mind, and including shear and selection response (see \autoref{sec:shearcat}), we define the estimator we use in practice as
\begin{equation}
	\label{eq:updated_deltasigma_estimate}
	\widetilde{\Delta\Sigma} \; \equiv\; \frac{\sum\limits_{j,i} \omega_{i,j}\;  e_{\rm T;\,i,j} }{\sum\limits_{j,i} \omega_{i,j}\, \Sigma_{{\rm crit;}i,j}^{'-1}\, \mathsf{R}^{\rm T}_{\gamma,i}  + \left(\sum\limits_{j,i} \omega_{i,j}\, \Sigma_{{\rm crit;}i,j}^{'-1}\right)\langle \mathsf{R}^{\rm T}_\mathrm{sel} \rangle } \; .
\end{equation}
In the above, $\langle \mathsf{R}^{\rm T}_\mathrm{sel} \rangle $ is calculated via \autoref{eq:meta-gamma} separately for source galaxies selected in each radial bin and each richness -- redshift bin, where the corresponding selections were defined by the photometric redshift estimates derived from the \emph{sheared} \metacal\ photometries. The small number of source galaxies at small radii introduces some noise to the estimated response, however due to the intrinsic environmental dependence of $\mathsf{R}^{\rm T}_\mathrm{sel}  $, this cannot be readily substituted or approximated with other, less noisy quantities. By considering the expectation value
\begin{equation}
	\langle e_{\rm T;\,i,j}\rangle=\Delta\Sigma \, \Sigma_{{\rm crit;}i,j}^{-1}\, \mathsf{R}^{\rm T}_{i} \; ,
\end{equation}
it is easy to see that the definition of \autoref{eq:updated_deltasigma_estimate} yields an unbiased estimate of $\Delta\Sigma$.

\autoref{eq:updated_deltasigma_estimate} includes two simplifications to make calculations less computationally demanding. First, for the normalization, we replace the expectation value of $\Sigma_{\rm crit}^{-1}$ by a Monte Carlo estimate
\begin{equation}
	\label{eq:sigma_crit_mc_estimate}
	\Sigma_{{\rm crit;}i,j}^{'-1}=\Sigma_{\rm crit}(z_{{\rm l}_j},z_{{\rm s}_i}^{\rm MC}) \,,
\end{equation}
where $z_{s_i}^{\rm MC}$ is a random sample from the $p_\mathrm{phot}(z_s)$ distribution estimated with BPZ using MOF photometry. Second, the weights are chosen as
\begin{equation}
	\label{eq:updated_weights}
	\omega_{i,j} \equiv \Sigma_{{\rm crit}}^{-1}\left(z_{{\rm l}_j}, \langle z_{{\rm s}_i}^{{\rm MCAL}}\rangle\right)\; \mathrm{if}\; \langle z_{{\rm s}_i}^{{\rm MCAL}}\rangle > z_{{\rm l}_j} + \Delta z\,,
\end{equation}
with $\langle z_{\rm s}^{{\rm MCAL}}\rangle$ being the mean redshift of the source galaxy estimated from \metacal\ photometry. Given the width of our photometrically estimated $p(z)$, this is close to the optimal weight. We use a padding of $\Delta z=0.1$ for source selection. We found that including the source weights provided by \metacal\ does not introduce a significant improvement in the signal-to-noise of the measurement.

The use of two different photometric estimators is necessary because when calculating the selection response, the internal photometry of the \metacal, with measurements on sheared images, must be used for all selection and weighting of sources. \cite{Y1pz} find this photometric redshift estimate to have a greater scatter than the default MOF photometry. We therefore opt to use the \metacal\ photo-$z$ estimates only for {\em selecting and weighting} source-lens pairs. When normalizing the shear signal to find $\Delta\Sigma$, we utilize the MOF-based photo-$z$ estimates.


\subsubsection{Data vector binned in redshift and richness}
\label{sec:data_vector}

In estimating the lensing signal through \autoref{eq:updated_deltasigma_estimate} we utilize a modified version of the publicly available \textsc{xshear} code\footnote{\url{https://github.com/esheldon/xshear}} and the custom built \textsc{xpipe} python package.\footnote{\url{https://github.com/vargatn/xpipe}} The core implementation of the measurement code is identical to the one used by \cite{rmsva}.

\begin{figure*}
	\includegraphics[width=\linewidth]{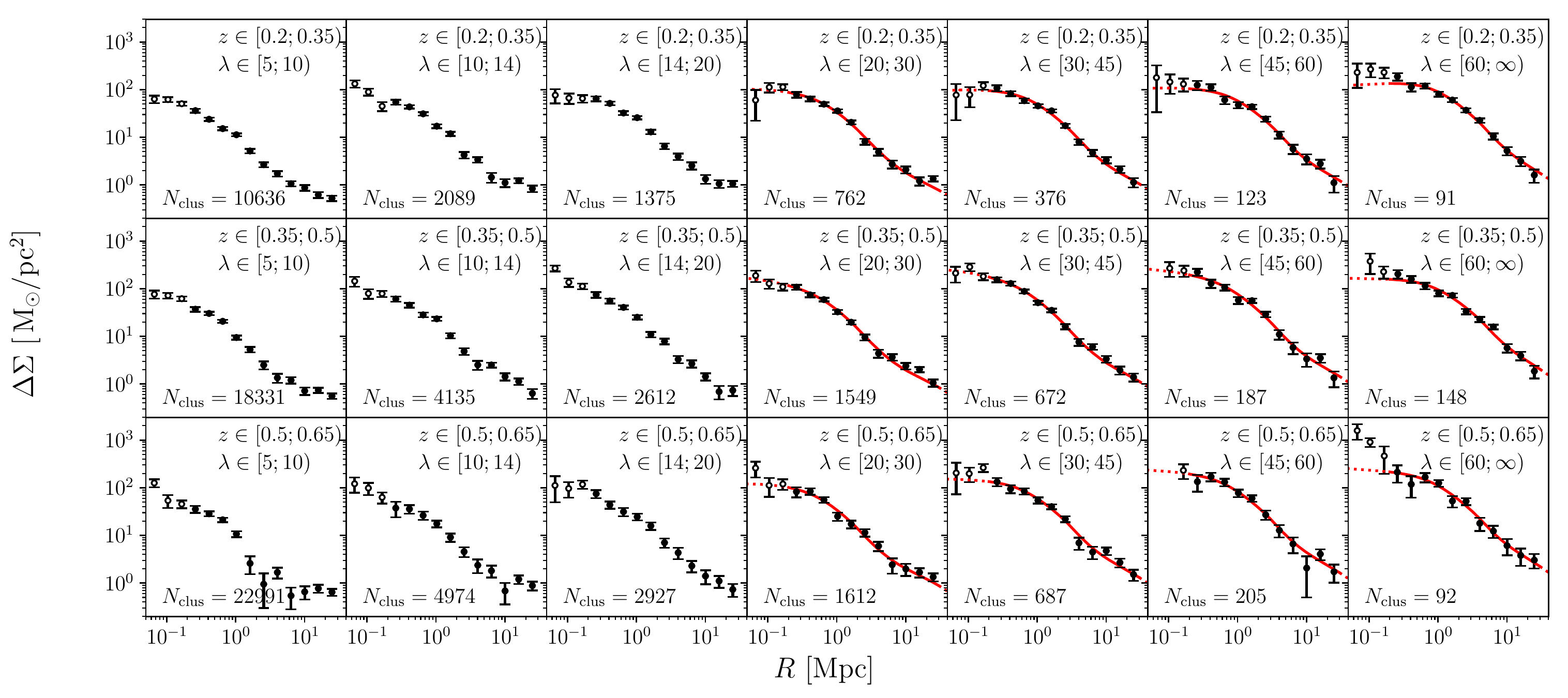}
	\caption{Mean $\Delta\Sigma$ for cluster subsets split in redshift $z_{\rm l}$ ({\rm increasing from top to bottom}) and $\lambda$ ({\rm increasing from left to right}), as labeled.  The error bars shown are the diagonal entries of our semi-analytic covariance matrix estimate (see \autoref{sec:covariance_matrices}) for bins with $\lambda > 20$ and the jackknife estimated covariance matrix for bins with $\lambda < 20$. The best-fit model ({\rm red curve}) is shown for bins with $\lambda > 20$, and includes dilution from cluster member galaxies (\autoref{sec:boost_factor_model}) and miscentering (\autoref{sec:miscentering_correction}); see \autoref{sec:modeling} for details. Semi-analytic covariances were not computed for stacks with $\lambda < 20$ due to the significant computational cost. Below 200 kpc we consider data points unreliable and therefore exclude them from our analysis; these are indicated by open symbols and dashed lines. The profiles and jackknife errors are calculated after the subtraction of the random-point shear signal (see \autoref{sec:randoms}).}
	\label{fig:all_profiles}
\end{figure*}

We group the clusters into three bins in redshift: $z\in[0.2;0.4)$, $[0.4;0.5)$, and $[0.5;0.65)$, as well as seven bins in richness: $\lambda\in[5;10)$, $[10; 14)$, $[14; 20)$, $[20; 30)$, $[30; 45)$, $[45; 60)$, and $[60; \infty)$.  The redshift limit $z=0.65$ of our highest redshift corresponds roughly to the highest redshift for which the \redmapper\ cluster catalog remains volume limited across the full DES Y1 survey footprint. The $\Delta\Sigma$ profiles were measured in 15 logarithmically spaced radial bins ranging from $0.03\;$Mpc to $30\;$Mpc. For our later results we will only utilize the radial range above 200~kpc. Scales below this cut are included only in our figures and for reference purposes, and are excluded from the analysis to avoid systematic effects such as obscuration, significant membership contamination, and blending. This radial binning scheme yields similar \snr\ across all bins. The measured shear profiles are shown in \autoref{fig:all_profiles}. 

We find a mild radial dependence in the typical value for \metacal\ shear response $\langle \mathsf{R}_{\gamma, \mathrm{T}} \rangle$, the asymptotic values are 0.6, 0.58 and 0.55 as a function of increasing cluster redshift. For the selection response we find an asymptotic value of $\langle \mathsf{R}_{\rm sel} \rangle\approx $ 0.013, 0.014, and 0.015.  

\subsection{Covariance matrices}
\label{sec:covariance_matrices}

The $\Delta\Sigma$ profiles estimated in the previous section deviate from the true signal due to statistical uncertainties and systematic biases. We construct a description for the covariance of our data vector below and calibrate the influence of systematic effects in \autoref{sec:systematics}.

Statistical uncertainties originate from the large intrinsic scatter in the shapes of source galaxies, the uncertainty in estimating their photometric redshifts, and due to the intrinsic variations in the properties and environments of galaxy clusters. Furthermore, our typical maximum radii: 2, 1.5 and 1.3 degrees for the different redshift ranges respectively are much larger than the 0.22 degree median separation between clusters in the catalog. This means that source galaxies are paired with multiple clusters, possibly generating covariance between different radial ranges and/or across different cluster bins in richness and redshift.

To quantify the correlation and uncertainty involved in the measurement we construct a semi-analytic model for the data covariance matrix following the framework developed by \cite{Gruen2015}. Our use of a semi-analytic covariance (SAC) matrix is motivated by explicit covariance estimators exhibiting non-negligible uncertainty and possible biases, for instance from jackknife regions that are not completely independent. Both of these problems lead to a biased estimate of the precision matrix (i.e. the inverse covariance matrix), which in turn will bias the posteriors of likelihood inference \citep{Friedrich2016}.

Instead, we \emph{predict} several key contributions of the observed covariances, namely those due to correlated and uncorrelated large scale structure, stochasticity in cluster centering, the intrinsic scatter in cluster concentrations at fixed mass, cluster ellipticity, and the scatter in the richness--mass relation of galaxy clusters.
Only the shape noise contribution is estimated directly from the data, as detailed below.

While we rely on the SAC matrix estimates in the remainder of our analysis, we compare the SAC matrices to those derived using a standard jackknife method.  We use jackknife (JK) resampling with $K=100$ simply-connected spatial regions $\mathcal{R}_k$ selected via a \textit{k-means} algorithm on the sphere.\footnote{\url{https://github.com/esheldon/kmeans_radec}} The jackknife covariance is defined following \cite{Efron82.1}:
\begin{equation}
\label{eq:jackknife_cov}
C_{\widetilde{\Delta\Sigma}} = \frac{K - 1}{K} \sum\limits_{k}^{K}\left(\widetilde{\Delta\Sigma}_{(k)} - \widetilde{\Delta\Sigma}_{(\cdot)}\right)^T \cdot \left(\widetilde{\Delta\Sigma}_{(k)} - \widetilde{\Delta\Sigma}_{(\cdot)}\right)\,,
\end{equation}
where $\widetilde{\Delta\Sigma}_{(\cdot)} = \frac{1}{K}\sum_k \widetilde{\Delta\Sigma}_{(k)}$ and $\widetilde{\Delta\Sigma}_{(k)}$ denotes the lensing signal estimated via \autoref{eq:updated_deltasigma_estimate} using all lenses except those in region $\mathcal{R}_k$. Using this method, we calculate the covariance between all radial bins in a single richness and redshift bin, as well as the covariance between adjacent richness and redshift bins.

\begin{figure}
	\includegraphics[width=\linewidth]{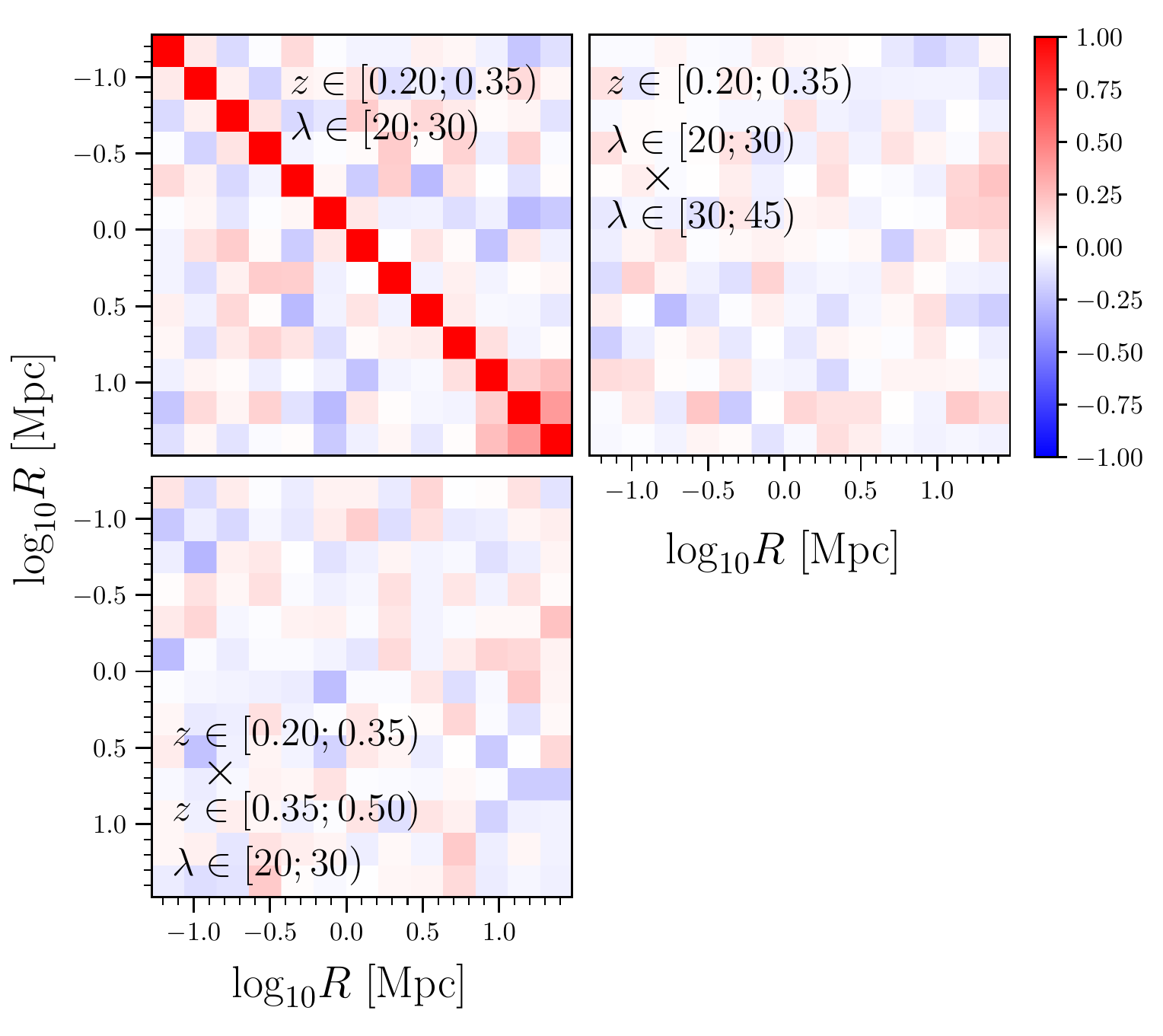}
	\caption{Jackknife estimated correlation matrix of $\widetilde{\Delta\Sigma}$ of a single richness-redshift selection with $\lambda \in [20;\, 30)$ and $z \in [0.2;\, 0.35)$ \emph{(upper left panel)}. The off-diagonal blocks display the correlation matrix between the reference profile and the neighboring richness bin $\lambda \in [30;\, 45)$ \emph{(upper right panel)}, and the neighboring redshift bin $z \in [0.35;\, 0.5)$ \emph{(lower left panel)}}
    \label{fig:corr_subset}
\end{figure}

\autoref{fig:corr_subset} shows an example of the structure of the jackknife estimated correlation matrix between neighboring bins in richness and redshift. We find no significant correlation between richness/redshift bin and therefore treat each bin independently, even though some systematic parameters may be shared between bins.

\subsubsection{Shape noise}
\label{sec:shape_noise}

The large intrinsic variations of the shapes of galaxies (\emph{shape noise}) in the source catalog constitute a dominant source of uncertainty in lensing measurements. We estimate the covariance originating from both the random intrinsic alignments and also the stochastic positions of source galaxies. In order to do so, we make use of the measurement setup outlined in \autoref{sec:data_vector}, but each source is randomly \emph{rotated} to create a new source catalog. We generated 1000 such independent \emph{rotated} source catalogs, and performed the lensing measurement with each. The resulting data vectors are consistent with zero, as the random rotation washes away the imprint of the weak lensing signal.  However, their scatter is indicative of the covariance due to shape noise.

We estimate the shape noise covariance matrix for each of the 1000 realizations using the spatial jackknife scheme outlined above in \autoref{sec:covariance_matrices}. The final shape noise covariance matrix estimate is obtained by averaging all 1000 of these jackknife covariance matrices. We expect this method to be less noisy compared to estimating the covariance matrix from the 1000 independent measurements of the rotated $\Delta\Sigma$ vector only.


\subsubsection{Uncorrelated LSS}
\label{sec:uncorrelated_lss}

Line of sight structures which are not physically connected to the cluster leave an impact on the lensing signal. We cannot remove them from the signal, but we can estimate their expected contribution to the covariance. For an individual cluster, the covariance of the $\Delta\Sigma$ profile at radii $\theta_i,\theta_j$ due to uncorrelated large scale structure (uLSS) can be written as \citep[e.g.][]{Schneider1998,Hoekstra03,Umetsu2011}
\begin{equation}
	\label{eq:ulss_SAC_contribution}
    C_{ij}^{\rm uLSS} = \langle\Sigma_{\rm crit}^{-1}\rangle^{-2} \int \frac{\ell{\rm d}\ell}{2\pi}\ P_\kappa(\ell) J_2 (\ell\theta_i) J_2(\ell\theta_j)\,,
\end{equation}
where $P_\kappa$ is the power spectrum of the convergence, and $J_2$ is the Bessel function of the first kind of order 2.

Naively, one would expect that the variance of a cluster stack due to uncorrelated large scale structure to scale simply as $1/N_{\rm clusters}$.  In practice, however, the positions of galaxy clusters are correlated, and the area around them overlaps on large scales. Consequently, we expect the variance due to uncorrelated structure to decrease somewhat more slowly than $1/N_{\rm clusters}$.

We estimate this source of noise by measuring random realizations of the signal due to shear fields induced by log-normal density fields with the appropriate power spectra and skewness. We calculate the latter with the perturbation theory model of \cite{Friedrich2017b} for the Buzzard cosmology \citep{DeRose2018,Wechsler2018}, using the log-normal parameter $\kappa_0$ at a 10' aperture radius. As our cluster sample spans a range in redshift, a different shear field is calculated for each of the three redshift bins. This is done such that the shear fields are calculated at the lens-weighted mean source galaxy redshifts found during the initial measurement in \autoref{sec:data_vector}. We then pass these shear fields through the measurement pipeline using a spatial mask reflecting the actual source number density variations across the footprint, and estimate the covariance matrix for each realizations using 100 spatial jackknife regions for each bin in richness and redshift. 

This above procedure was repeated 300 times, and the final covariance matrix due to uncorrelated LSS is taken to be the mean of the 300 jackknife covariance estimates.


\subsubsection{Correlated LSS and halo ellipticity}
\label{sec:correlated_lss}

Following \citet{Gruen2015}, we model correlated large scale structure using a halo model approach.  We assume correlated halos can be adequately described using only two parameters: the mass $M$ of the correlated halo, and the projected distance $R_{\rm h}$ from the cluster.  The mass distribution of the halos is assumed to follow the halo mass function, while their spatial distribution is modeled as a Poisson realization of the density field defined by the appropriate halo--cluster correlation function.  That is, the excess density of halos of mass $M$ a distance $R$ from the halo is
\begin{equation}
	\rho_{\rm h}(M,R_{\rm h}) = b(M_{\rm cl})b(M)w_{\rm mm}(R_{\rm h}) \frac{{\rm d}n}{{\rm d}M}\,,
\end{equation}
where $w_{\rm mm}$ is the projected linear correlation function at the redshift of the cluster, $b(M)$ is the halo bias, and ${\rm d}n/{\rm d}M$ is the halo mass function, or the number of halos per unit volume per unit mass \citep{Tinker2008}.

Given a model for the halo profile $\Sigma(R|M)$, the contribution of a halo at location $R_{\rm h}$ to the mean surface density $\Sigma$ of the cluster in radial bin $R_i$ is $\Sigma_i(M,R_{\rm h}) = \Sigma_{\rm misc}(R_i|M,R_{\rm h})$, where $\Sigma_{\rm misc}$ is a miscentered halo profile. Because of the Poisson-sampling assumption, the covariance matrix is generated by the mass profiles of individual halos, so that the correlated large scale structure contribution to the covariance matrix can be written as
\begin{equation}
	\label{eq:correlatedLSS_covariance}
    C_{ij}^{\rm cLSS} = \int (2\pi R_{\rm h} {\rm d}R_{\rm h}){\rm d}M\ \rho_{\rm h}(M,R_{\rm h}) 
    \Delta \Sigma_i(M,R_{\rm h}) \Delta \Sigma_j(M,R_{\rm h})\,.
\end{equation}
In practice, the above predicted covariance matrix is further rescaled by a constant factor calibrated on simulations. This is meant to account for additional variance not captured by linear bias and Poisson noise, due to filamentary structure and higher order statistics in the spatial distribution of the correlated halos (e.g. the non-zero three-point function).  A more detailed derivation of the above equation and its calibration is found in \citet{Gruen2015}.  

A very similar calculation can be made for characterizing the contribution due to halo ellipticity (to the covariance matrix -- for the effect on the mean signal, see \autoref{sec:triaxiality_and_projection_effects}).  If $\rho_{\rm ell}(q,\mu)$ is the distribution of the halo axis ratio $q$ and the line-of-sight orientation angle $\theta$ relative to the major axis such that $\cos \theta = \mu$, then one finds \citep{Gruen2015}
\begin{equation}
	\label{eq:ellipticity_covariance}
    C_{ij}^{\rm ell} = \int {\rm d}q{\rm d}\mu\  \rho_{\rm ell}(q,\mu) \Delta\Sigma_i\Delta\Sigma_j\,,
\end{equation}
where $\Delta\Sigma_i$ is the contribution to the bin $R_i$ under the assumption that the halo has an axis ratio $q$ and an orientation $\mu$.


\subsubsection{$M$--$c$ scatter, $M$-$\lambda$ scatter and miscentering}
\label{sec:mc_ml_miscentering_covariances}

Halos at a given mass have some intrinsic scatter in their $M$--$\lambda$ relation. A rough estimate of the intrinsic scatter in the mass--richness ($M$--$\lambda$) relation is $\sim$ 25 per cent \citep{RozoRykoff14_RM2, Farahi2018}, and it causes an increase in the variance of stacked measurements of \deltasig. This scatter causes an even larger increase in the variance, since it propagates into quantities that depend directly on the mass, including the $M-c$ relation. In addition, concentration \citep[e.g.][]{DiemerKravtsov15,Bhattacharya2013} and miscentering possess some intrinsic scatter from halo to halo themselves.

Scatter in the $M$--$\lambda$ relation causes variance on all scales, since the bias $b(M)$ directly depends on the mass. By comparison, scatter in the $M-c$ relation primarily affects small scales where the 1-halo term dominates. Similarly, some cluster centers are misidentified in our stacks, which creates additional covariance at small scales where the signal is substantially suppressed.

We modeled the combined contribution to the SAC from scatter in $M$--$\lambda$, scatter in concentration at fixed mass, and miscentering of individual clusters in our stacks by doing the following:
\begin{enumerate}
	\item For each cluster in our stack, assign a mass by inverting a fiducial $M$--$\lambda$ relation \citep{rmsva} and assuming 25 per cent scatter. This is not identical to 25 per cent scatter in the $M$--$\lambda$ relation, however this choice negligibly affects this component of the covariance matrix.
	\item For each cluster, assign a concentration (including scatter) based on \citet{DiemerKravtsov15}.
	\item For each cluster, make a draw from our centering prior described in \autoref{sec:miscentering_correction}. In other words, some fraction $\fmis$ of clusters in the stack are miscentered, and the distribution of the amount of miscentering is given by $p(\Rmis)$.
	\item Calculate \deltasig\ for each cluster and average these signals to generate a signal for the entire stack.
	\item Repeat this process many times, and use these independent realizations to estimate the corresponding covariance matrix between the various radial bins.
\end{enumerate}
Using \citet{Simet2017} as our fiducial $M$--$\lambda$ relation or using the \citet{Bhattacharya2013} mass--concentration relation had no impact on the final SAC matrix. We have also verified that using half as many realizations as our fiducial choice (1000) did not appreciably change the resulting covariance matrix. The same is true for changes in the richness scatter or miscentering model parameters within reasonable ranges.


\subsubsection{Semi-analytic covariance matrix}
\label{sec:semi_analytic_covariance_matrix}

\begin{figure}
	\includegraphics[width=\linewidth]{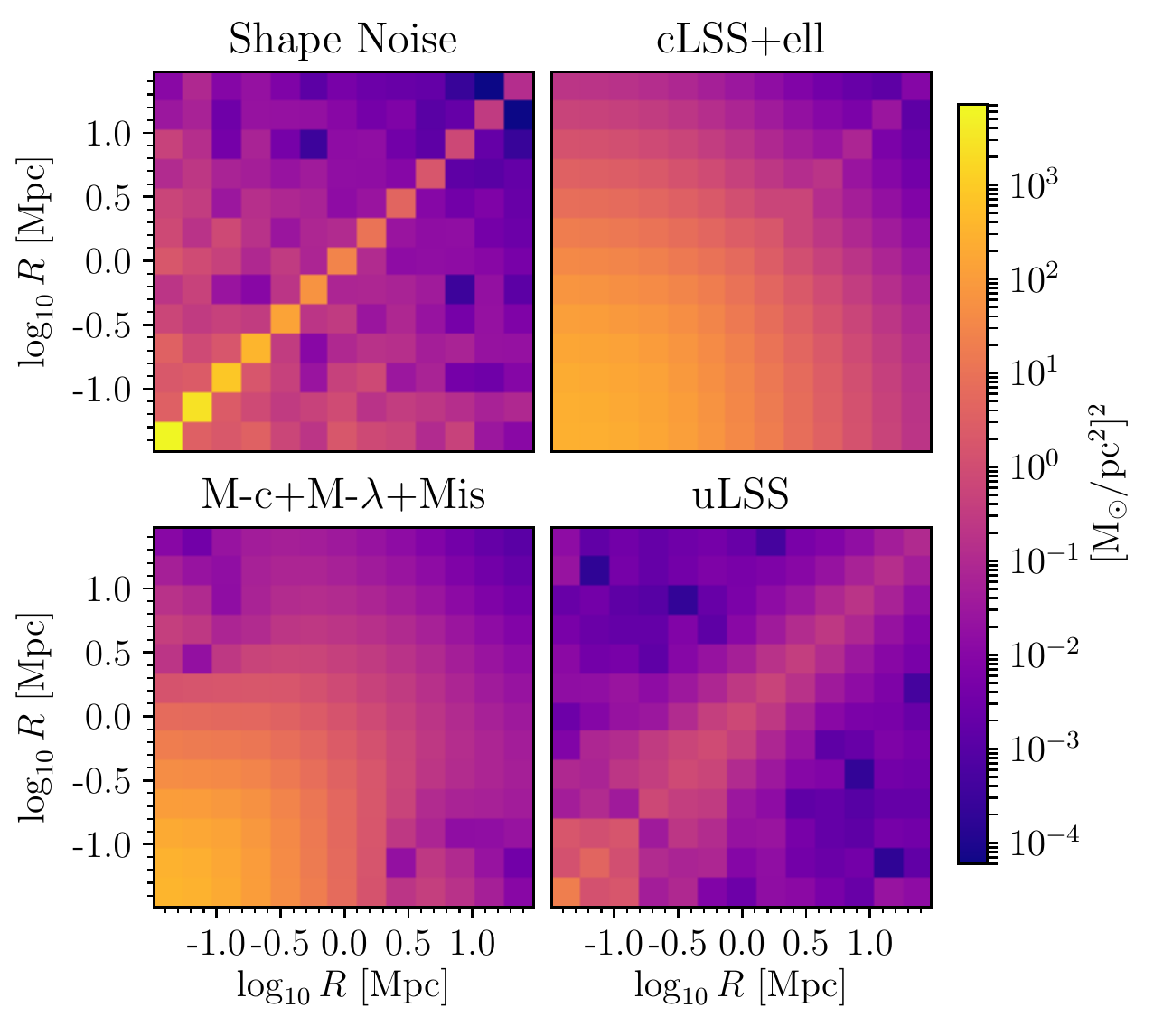}
	\caption{The four individual components to our Semi-Analytic Covariance (SAC) matrix. {\it Clockwise from top left}: Shape noise component from randomly rotating sources, correlated LSS and ellipticity component from integrating over configurations of the host cluster and its correlated halos, uncorrelated LSS from integrating over large scale structure, and finally scatter in the $M-\lambda$ relation, $M-c$ relation, and miscentering distribution. Dark colors correspond to low covariance and the colors are log scaled to show trends. Light colors are normalized to the total covariance in the SAC. See \autoref{sec:covariance_matrices} for details.}
    \label{fig:SAC_components}
\end{figure}

\begin{figure*}
	\includegraphics[width=0.8\linewidth]{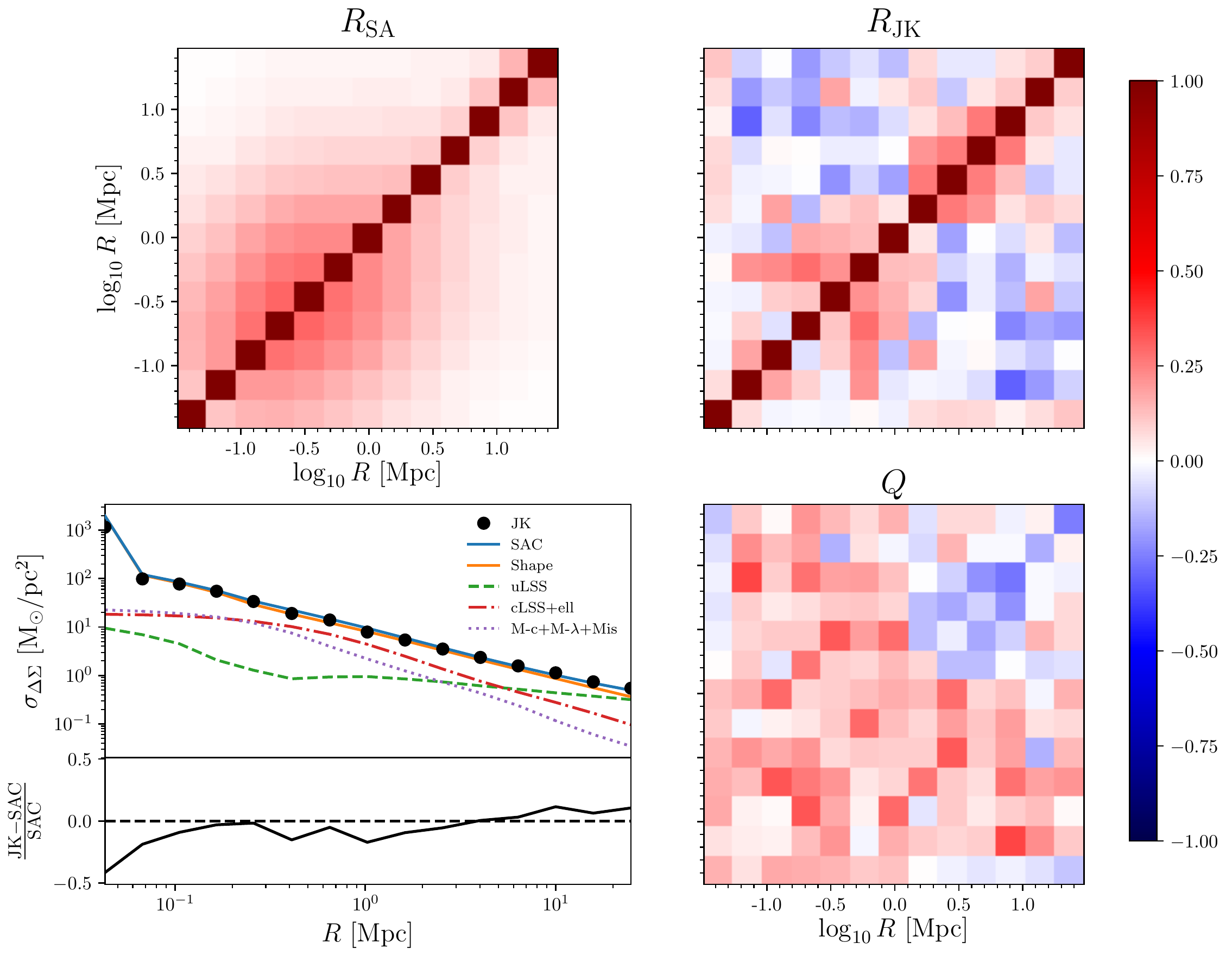}
	\caption{Comparison between the semi-analytic covariance matrix and the jackknife estimated covariance matrix. {\it Top left}: Correlation matrix of the SAC matrix. {\it Top-right:} Correlation matrix of the jackknife estimate.  {\it Bottom left:} Comparison of the errors in the SAC and jackknife estimate along with the contributions to the SAC error from each individual component. The line showing the SAC errors lies almost on top of the shape noise contribution, confirming that it is the dominant source of covariance. {\it Bottom right:} Residual matrix $Q$ (see \autoref{eq:Q_matrix_definition}) that represents the difference between the SAC and jackknife covariance matrices.  See \autoref{sec:covariance_matrices} for details.}
    \label{fig:SAC_jk_comparison_panels}
\end{figure*}

Following \citet{Gruen2015}, the full SAC matrix is obtained by adding each of the above contributions. The individual components described in the previous subsections are shown in \autoref{fig:SAC_components}.  \autoref{fig:SAC_jk_comparison_panels} demonstrates the differences between the SAC and jackknife covariance matrices. The top two panels show the correlation matrix $R$ of the SAC and $C^{\rm JK}$ respectively, where the correlation matrix is defined via
\begin{equation}
	\label{eq:correlation_matrix_definition}
    R_{ij} = \frac{C_{ij}}{\sqrt{C_{ii}C_{jj}}}\,.
\end{equation}
Similarly, to visualize the difference between $C^{SA}$ and $C^{JK}$ we define the residual matrix
\begin{equation}
	\label{eq:Q_matrix_definition}
    Q_{ij} = \frac{C_{ij}^{\rm SA}-C_{ij}^{\rm JK}}{\sqrt{C_{ii}^{\rm SA}C_{jj}^{\rm SA}}}.
\end{equation}
We show this residual matrix in the bottom right panel of \autoref{fig:SAC_jk_comparison_panels}. Finally, in the lower left panel we show the difference between the errors along the diagonal between the $C^{\rm JK}$ and the SAC, along with each of the contributions to the SAC; the lower panel shows the fractional difference between the diagonal entries. As expected, shape noise is the dominant contributor to the SAC matrix, with uncorrelated LSS becoming important at the largest scales. This explains why the choices we had to make in modeling the non-shape noise components did not significantly affect the resulting SAC matrix or the posteriors analysis.

Using the SACs in our analysis provides two major improvements: minimal bias from inverting the covariance matrix, and less overall noise in the off-diagonal elements which improves the mass measurement. In \citet{rmsva} we demonstrated that noise in the jackknife covariance matrix led to an increase of $\approx $30 per cent in the uncertainty of the mass of the stack. Using the SACs reduces the contribution of the covariance to the error budget by 10 per cent compared to the jackknife estimated covariance.

\section{Systematics}
\label{sec:systematics}

\subsection{Shear systematics}
\label{sec:shear_systematics}

The \metacal\ shear catalog and the associated calibration of the source redshift distributions \citep{Y1pz} passed a large number of tests performed by \citet{Y1shape} and \citet{Prat2017}. Here we briefly enumerate the constraints on the most relevant systematics, and refer the reader to the corresponding papers for a more detailed analysis.

We parametrize the various potential biases in the dataset as:
\begin{equation}
	\label{eq:shear_nbs_parametrization}
    g_i = (1\, +\, m_i)g_i^{\rm tr}\, + \, \alpha^{\rm PSF} e_i^{\rm PSF} \,+\, c_i\,,
\end{equation}
where $g_i^{\rm tr}$ is the true shear, while $g_i$ is the shear estimate, and $\alpha^{\rm PSF}$ relates to the contamination from the PSF ellipticity $e_i^{\rm PSF}$. 

In weak lensing surveys the three main sources of bias are commonly found to be model bias, noise bias, and selection bias (or representativeness bias). In order to account and correct for these sources of error, the \metacal\ algorithm performs a self-calibration on the actual data by shearing the galaxy images during the measurement, and using the thus calculated responses to correct the shear estimates. To quantify the effectiveness of this self-calibration, \cite{Y1shape} ran the \metacal\ pipeline on a set of simulated galaxy images using \textsc{GalSim} \citep{Rowe2014}.  The images were produced from high resolution galaxy images from the COSMOS sample, and processed to resemble the actual DES Y1 observations both in noise and PSF properties. Based on this test scenario \cite{Y1shape} found no significant multiplicative bias $m$ or additive bias $c$ present in the dataset.

\cite{Y1shape} further investigated the multiplicative biases due to blending of galaxy images, due to the potential leakage of stellar objects into the galaxy sample, and due to potential errors in the modeling of the PSF. They found blending as the only component with a net bias, with the other sources being consistent with zero, although contributing to the uncertainty on the value of $m$. The final multiplicative bias estimates were found to be $m = [1.2 \pm 1.3] \cdot 10^{-2}$ with a $1\sigma$ Gaussian error.  They found no evidence of a significant additive bias term.

\citet{Prat2017} tested for the presence of residual shear calibration biases in the DES Y1 galaxy-galaxy lensing analysis by splitting the source sample by various galaxy properties and parameters of the observational data. They showed that within the statistical uncertainty of the respective galaxy-galaxy lensing signals, and including the differences in redshift distributions induced by the splitting, no differential multiplicative biases between any of the splits are significantly detected.

In addition to the above calibrations during the construction of the shear catalog, we perform additional sanity checks relevant to stacked weak lensing measurements in the subsections below.


\subsubsection{Second order shear bias}
\label{sec:second_order_shear_bias}

Due to the larger tangential shear near massive clusters, this analysis is more strongly affected by non-linear shear response than previous DES Y1 lensing analyses (see the discussion in section~9 of \citealt{SheldonMETA}). This response biases cluster masses higher than they would be otherwise. To test this effect, we modify the measured $\Delta\Sigma$ profiles by adding the leading non-linear shear bias term, at third order in $\gamma_t=\Delta\Sigma\times\Sigma_{\rm crit}^{-1}$, as
\begin{equation}
	\label{eq:cubic_shear_test}
    \Delta\Sigma_{\rm obs}' = \frac{\Delta\Sigma_{\rm obs}\langle\Sigma_{\rm crit}^{-1}\rangle - \alpha_{\rm NL}\left(\Delta\Sigma_{\rm Model}\langle\Sigma_{\rm crit}^{-1}\rangle\right)^3}{\langle\Sigma_{\rm crit}^{-1}\rangle}\,,
\end{equation}
where $\Delta\Sigma_{\rm Model}$ is the optimized model discussed in \autoref{sec:modeling}. For the amplitude of non-linear shear bias we choose $\alpha_{\rm NL}=0.6$ \citep{SheldonMETA}. We model the profile of the highest richness stack at $z\in[0.2,0.35]$ where, for the source redshift distribution of DES Y1, this effect is strongest. The recovered mass changes by less than 1 per cent, demonstrating that our recovered mass--richness--redshift relation is robust to non-linear shear bias within our error budget.

The choice of $\alpha=0.6$ in our test is motivated by the image simulations used in \citet{SheldonMETA}. Other simulations find a range of values of similar magnitude. Since the effect is smaller than the overall shear uncertainty, yet its calibration is uncertain, we choose not to implement a correction in our final model.


\subsubsection{B-modes} 
\label{sec:bmodes}

Gravitational lensing due to localized mass distributions can only produce a net E-mode signal in the shear field, which corresponds to the tangential shear $\gamma_t$. This allows for a simple null test for the presence of systematics: any non-zero cross-shear (i.e. a non-zero B-mode) must be due to systematics. We compute the cross-shear by projecting the shears to the direction $45^\circ$ from the tangential direction. We estimate the stacked B-mode signal for all richness and redshift bins, and calculate the corresponding $\chi^2$ values using the jackknife estimate of the covariance matrix. 
We find $\chi^2 / 11 < 18/11$ for all richness bins with $\lambda > 20$, indicating that our measurement is consistent with no systematics at a $p > 0.1$ level.


\subsubsection{Random point test}
\label{sec:randoms}

In spite of not being detected by \cite{Y1shape} and \autoref{sec:bmodes}, additive shear systematics may be present in the data, which could manifest as net signals visible on all radial scales. In order to test for such potential systematics we measure the lensing signal around a set of random points chosen by the \redmapper\ algorithm \citep{Rykoff2016}. These points are selected via weighted random draws to mirror the distribution of DES Y1 \redmapper\ clusters both in angular distribution, as well as in redshift and richness.

As additive systematics would affect the lensing profiles of galaxy clusters and random points the same way, the systematic effect can be calibrated out by subtracting the profile of random points from the profile of clusters. 
While we find no significant net signal around random points, we nevertheless apply this calibration, and subtract the signal of $10^5$ random points from the $\widetilde{\Delta\Sigma}$ of each bin in richness and redshift. Thanks to the large number of random points used, this subtraction does not introduce significant noise to the measurement.

A motivation for subtracting the signal around random points from the measurement, regardless of the presence of systematics,  is presented by \cite{Singh2016}. They found that the random subtracted signal relates to the \emph{matter over-density field} around the lenses, while the un-subtracted lensing signal traces the \emph{matter density field}, which carries additional variance on large scales. Indeed, the precursor study of the present paper \citep{rmsva} found a similar trend. We note that when constructing our SAC matrix we always apply the random point subtraction described above to ensure that our covariance matrix properly accounts for the reduced covariance that this estimator enables. 


\subsection{Correction for cluster members in the shear catalog}
\label{sec:boost_factors}

Due to uncertainty in photometric redshift estimates, foreground galaxies can be included in the source catalog used in our lensing measurements. So long as the ensemble redshift distribution ${\rm d}n/{\rm d}z$ of the sources is properly estimated, this is accounted for in our analysis. In the projected vicinity of galaxy clusters there is however a systematic effect biasing the naive redshift estimates of galaxies: the presence of a large cluster member population and the associated large-scale matter overdensity localized at the cluster redshift. For rich clusters, these member galaxies could make up a significant fraction of all detected galaxies in a particular line of sight. Consequently, due to intrinsic imperfections in the selection, some of these galaxies leak into the source catalog used in the weak lensing measurement. Cluster member galaxies are randomly aligned \citep{Sifon2015}, meaning their contamination results in a measured lensing signal which is biased low due to the dilution of actual source galaxies within the catalog.

It is therefore important for weak lensing studies to characterize and correct this dilution when interpreting the measurements.\footnote{This  correction is also referred to as a \emph{boost factor} as the measured signal should be boosted to correct for the contamination} There have been several approaches in the literature to correct for the net effect of cluster member contamination. For instance, \cite{Sheldon04.1} estimated the correction factor from the transverse correlation of source galaxies around galaxy clusters, while \cite{Gruen2014} and \cite{rmsva} estimated the contamination rate based on the color or photometric redshift $p(z)$ information of galaxies in different radial separations from the cluster. One can also make simple color cuts \citep{Schrabback2018, Medezinski2010,Medezinski2018_colorcuts} or \photoz\ cuts \citep{Applegate2014} on the source population to mitigate the contamination, or estimate its effect based on the increased galaxy number density around the lenses \citep{Dietrich2017,Hoekstra2015,Simet2015}.

In this study we adopt the approach of our precursor analysis from the  Science Verification data release of DES \citep{rmsva}, in which we make use of the estimated $p(z)$ of the source galaxy sample to calculate the cluster contamination fraction $f_{\rm cl}$ along with a corresponding covariance matrix  ${\rm C_{\fcl}}$ estimated via spatial jackknifing, and use this quantity to recover the contamination corrected lensing profile:
\begin{equation}
	\label{eq:boost_correction}
	\widetilde{\Delta\Sigma}_{\rm corr}(R) = \frac{\widetilde{\Delta\Sigma}(R)}{1 - f_{\rm cl}(R)}\,.
\end{equation}
Using this $p(z)$ decomposition approach is further motivated by the complexity of the shear selection function in our analysis, which limits our ability to measure the correlation function of source galaxies. A detailed description of this method, along with validation on simulated DES-like mock observations is presented in an accompanying paper \citep{Varga2018}. The robustness of the method to the presence of intra-cluster light is confirmed in \citet[][their appendix A]{Gruen2017_PZmethods}.


\subsection{Photometric redshift systematics}
\label{sec:photometric_redshift_systematics}

\begin{figure}
	\includegraphics[width=\linewidth]{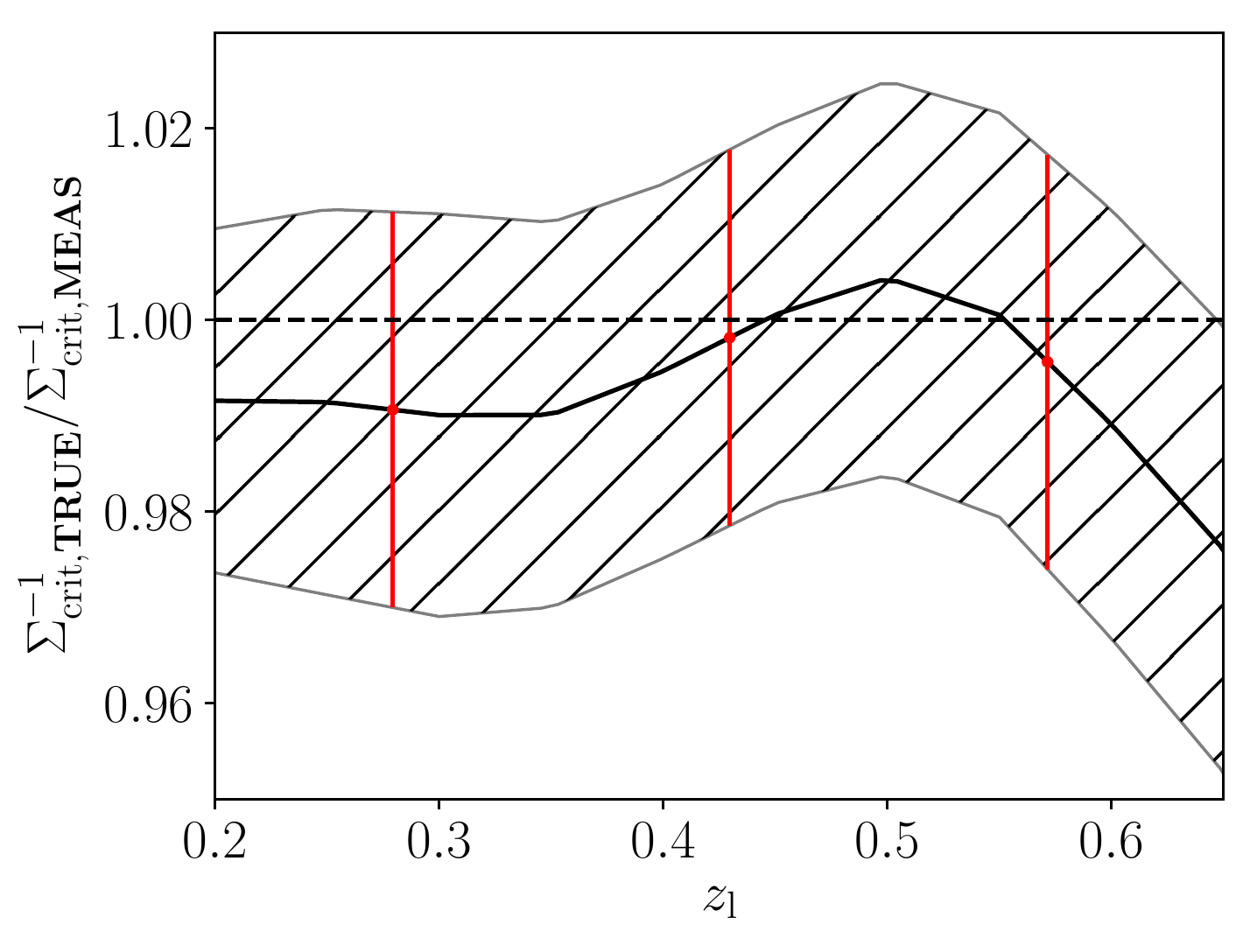}
	\caption{The photo-z correction factor to $\Sigma_{\rm crit}^{-1}$ as described in \autoref{sec:photometric_redshift_systematics}. The gray hatched region indicates the $1\sigma$ range of the correction factor. Red points with error bars show the correction factors applied to each redshift bin.}
    \label{fig:sci_correction}
\end{figure}

The redshift distribution of our selected source galaxies was estimated using BPZ \citep{Benitez2000} in the implementation of \citet{Y1pz}. In BPZ or similar photometric redshift estimation procedures, one assumes a variety of galaxy spectral energy distribution (SED) templates and priors for the relative abundance of galaxies as a function of luminosity and redshift. Any deviation from these assumptions in the DES source galaxy sample can cause biases in photometric redshift estimates which must be calibrated.

For the cosmology analyses of the lensing two-point functions \citep{Troxel2017,DESCosmicShearCosmology}, this calibration was performed in two independent ways, and with consistent results: by the redshift distributions of samples of galaxies with high-quality 30-band \photozs\ from COSMOS, matched to DES lensing source galaxies \citep{Y1pz}, and by the clustering of lensing source galaxies with \redmagic\ \citep{Rozo2016_redmagic} galaxies as a function of the redshift of the latter \citep{Davis2017,Gatti2017}.

For this work, we adapt the COSMOS calibration of \citet{Y1pz} to estimate the bias of our $\Delta\Sigma$ measurements, and the uncertainty in that bias. To this end, we select and weight galaxies from COSMOS in the same manner as for our measurements of the cluster $\Delta\Sigma$ profiles.

Following \citet[][their section 4.1]{Y1pz}, we randomly sample 200,000 galaxies in our data and match them to COSMOS galaxies according to their flux in each band and their intrinsic size. From this \emph{COSMOS resampling}, we select and weight galaxies as per \autoref{sec:practical_estimator} and \autoref{sec:data_vector}. From the COSMOS 30-band we calculate the weighted mean true $\Sigma_{\rm crit,TRUE}^{'-1}$. From noisy MOF $griz$ \bpz\ redshift distribution samples we get a mean $\Sigma_{\rm crit, MEAS}^{'-1}$ that relates the weighted mean tangential shear to the $\Delta\Sigma$ profile. As in the denominator of \autoref{eq:updated_deltasigma_estimate}, we use a weight $\omega\times\mathsf{R}$ for the means. Because the source selection, $\omega$ weight, and $\Sigma_{\rm crit}^{'}$ depends on lens redshift, we repeat this exercise for the range of cluster redshifts sampled by our catalog, $z_l=0.2\ldots0.65$. A bias in $\Sigma_{\rm crit}^{'-1}$ translates directly into a multiplicative bias in $\Delta\Sigma$.

We estimate four sources of uncertainty in the calibration of photometric redshift distributions \citep[see][their sections 4.2-4.5]{Y1pz}: 1) an uncertainty due to cosmic variance from the relative scatter of average $\Sigma_{\rm crit,TRUE}^{'-1}$ in the resampling of the 368 simulated COSMOS footprints, to which we add the (subdominant) statistical uncertainty due to the limited sample size from bootstrap resamplings in quadrature; 2) an uncertainty due to photometric zeropoint offsets from realizations of photometric zeropoint calibration offsets; 3) an uncertainty due to the morphology matching, which we estimate as half the difference between the estimated $\Sigma_{\rm crit,TRUE}^{'-1}$ of the sample with size+flux matching and that obtained without the size matching; and 4) a systematic uncertainty of the matching algorithm by a comparison between the fiducial $\Sigma_{\rm crit,TRUE}^{'-1}$ value and that of the aforementioned 368 resampled simulated COSMOS fields. Effects 1, 3, and 4 contribute to the systematic uncertainties with similar size, while effect 2 is smaller but not quite negligible. 

We define our model for the bias as 
\begin{equation}
	\label{eq:sigmacrit_ratio}
	\frac{\Sigma_{\rm crit, MEAS}^{'-1}}{\Sigma_{\rm crit, TRUE}^{'-1}} \equiv 1+\delta\, ,
\end{equation}
with the mean value given from the COSMOS analysis, and an uncertainty due to the four effects mentioned above.
This ratio depends on lens redshift through the selection/weighting of sources and the source redshift dependence of photo-$z$ bias. It is plotted across the entire lens redshift range in \autoref{fig:sci_correction}. The red points show the ratio at the mean redshifts of the bins used in our analysis. This multiplicative factor is fully degenerate with shear systematics (see \autoref{sec:shear_systematics}) and is assumed to be correlated across redshift bins. $\delta$ is incorporated in our analysis by a prior that varies between each stack. The variation between richness bins is small compared to the variation across cluster redshift bins.
\begin{equation}
	\label{eq:multiplicative_photoz_bias}
	\delta = \begin{cases}
		0.009 \pm 0.021 & {\rm for}\ z\in[0.2,0.35] \\
		0.002 \pm 0.020 & {\rm for}\ z\in[0.35,0.5] \\
		0.004 \pm 0.022 & {\rm for}\ z\in[0.5,0.65].
	\end{cases}
\end{equation}
This prior is combined with the prior on $m$ and included in the final likelihood as described in \autoref{sec:multiplicative_A}.

An additional concern is the effect of intra-cluster light leaking into source photometry used for redshift estimation. We test for this in \citet{Gruen2018_ICL} with the intra-cluster light measurements of \citet{Zhang2018_ICL}, finding negligible effects in the regimes relevant for this study.


\section{The stacked lensing signal}
\label{sec:modeling}

\subsection{Surface density model}
\label{sec:surface_density_model}

Our surface density model remains unchanged from \citet{rmsva}. The lensing signal is given by \autoref{eq:dsdef}. The quantities $\overline{\Sigma}(R)$ and $\overline{\Sigma}(<R)$ are given by
\begin{equation}
  \label{eq:Sigma}
  \overline{\Sigma}(R) = \int_{-\infty}^{+\infty}{\rm d}\chi\ \Delta\rho\left(\sqrt{R^2+\chi^2}\right)\,,
\end{equation}
where $R$ is the projected separation and $\chi$ the separation along the line of sight in comoving units and
\begin{equation}
  \label{eq:barSigma}
  \overline{\Sigma}(<R) = \frac{2}{R^2}\int_0^R{\rm d}R'\ R'\overline{\Sigma}(R')\,.
\end{equation}
If the shear signal is caused by halos of mass $M$, the average excess three dimensional matter density is given by
\begin{equation}
  \label{eq:density}
  \Delta\rho(r) = \rho(r)-\rho_{\rm m} = \rho_{\rm m}\xi_{\rm hm}(r\,|\, M)\,,
\end{equation}
where $\rho_{\rm m}=\Omega_{\rm m}\rho_{\rm crit}(1+z)^3$ is the mean matter density in physical units at the redshift of the sample, $\rho_{\rm crit}$ is the critical density at redshift zero, and $\xi_{\rm hm}(r\,|\,M)$ is the halo--matter correlation function at the halo redshift.

At small scales $\xi_{\rm hm}$ is dominated by the so-called ``1-halo'' term while at large scales it is dominated by the ``2-halo'' term. We use the \citet{zuetal14} update to the \citet{Hayashi08} model of $\xi_{\rm hm}$. This is
\begin{equation}
  \label{eq:hmcf}
  \xi_{\rm hm}(r\,|\, M) = \max\left\lbrace\xi_{\rm 1h}(r\,|\, M),\xi_{\rm 2h}(r\,|\, M)\right\rbrace\,.
\end{equation}
For the 1-halo term we use a \citet[][hereafter NFW]{Navarro96.1} density profile $\rho_{\rm NFW}(r\,|\, M)$
\begin{equation}
  \label{eq:1hcf}
  \xi_{\rm 1h}(r\,|\, M,c) = \frac{\rho_{\rm NFW}(r\,|\, M,c)}{\rho_m}-1\,,
\end{equation}
where 
\begin{equation}
	\label{eq:rho_nfw}
	\rho_{\rm NFW}(r\,|\, M,c) = \frac{\Omega_{\rm m}\rho_{\rm crit}\delta_c}{\left(r/r_s\right)\left(1+r/r_s\right)^2}\, .
\end{equation}
The concentration $c=r_{200m}/r_s$ is left as a free parameter, with a flat prior as per \autoref{tab:modeling_parameters}. This differs from the analysis in \citet{rmsva}, in which we forced the halo concentration to follow the concentration--mass relation of \citet{DiemerKravtsov15}. 

For the two-halo term $\xi_{\rm 2h}(r\,|\,M)$ we use the non-linear matter correlation function $\xi_{\rm nl}$ scaled by the halo bias $b(M)$ of \citet{Tinker2008} as
\begin{equation}
  \label{eq:2hcf}
  \xi_{\rm 2h}(r\,|\, M) = b(M)\xi_{\rm nl}(r)\,.
\end{equation}
$\xi_{\rm nl}$ is the 3D Fourier transform of the non-linear power spectrum $P_{\rm nl}$ \citep{Smith02Halofit,Takahashi12Halofit}, given by
\begin{equation}
  \label{eq:cf_transform}
  \xi_{\rm nl}(r) = \int_0^\infty \frac{{\rm d}k}{k}\ \frac{k^3P_{\rm nl}(k)}{2\pi^2}j_0(kr)\,,
\end{equation}
where $j_0(z)$ is the 0th spherical Bessel function of the first kind. The power spectrum is computed using the CLASS code\footnote{\url{http://class-code.net/}} \citep{Lesgourgues11CLASS1,Blas11CLASS2}. We repeated our analysis using the linear matter correlation function $\xi_{\rm lin}$ and found nearly identical results as discussed later in \autoref{sec:alternative_model}.


\subsection{Miscentering correction}
\label{sec:miscentering_correction}

We have thus far assumed that we can measure the stacked shear profile of clusters relative to the ``center'' of the halo as defined in an $N$-body simulation. Our simulations use the spherical overdensity algorithm \textsc{rockstar} as implemented in \citet{Behroozi2013}. If cluster centers are not properly identified, or are ``miscentered'', then the observed weak lensing signal in annuli around these clusters will be suppressed. As in \citet{rmsva}, we model the recovered weak lensing signal as a weighted sum of two independent contributions: a contribution $\Delta\Sigma_{\rm cen}$ from properly centered clusters, and a contribution $\Delta\Sigma_{\rm mis}$ from miscentered clusters,
\begin{equation}
	\label{eq:miscentering_model}
	\Delta\Sigma_{\rm model} = (1-\fmis)\Delta\Sigma_{\rm cen}+\fmis\Delta\Sigma_{\rm mis}\, .
\end{equation}
$\Delta\Sigma_{\rm cen}$ is given by \autoref{eq:dsdef}. When a cluster is miscentered by some radial offset $\Rmis$, the corresponding azimuthally averaged surface mass density is (e.g. \cite{Yang06, Johnston07})
\begin{equation}
	\label{eq:sigma_miscentered_single_cluster}
	\overline{\Sigma}_{\rm mis}(R\,|\, \Rmis) = \int_0^{2\pi}\frac{{\rm d}\theta}{2\pi}\ \overline{\Sigma}\left(\sqrt{R^2+\Rmis^2+2R\Rmis\cos\theta}\right)\,.
\end{equation}
Letting $p(\Rmis)$ be the distribution of radial offsets for miscentered clusters, the corresponding mean miscentered profile $\overline{\Sigma}_{\rm mis}$ is
\begin{equation}
	\label{eq:sigma_miscentered}
	\overline{\Sigma}_{\rm mis}(R) = \int {\rm d}\Rmis\ p(\Rmis) \overline{\Sigma}_{\rm mis}(R\,|\, \Rmis)\,.
\end{equation}
It is this quantity that we use to model the miscentered profile term in \autoref{eq:miscentering_model}.

\citet{Zhang2018, vonderLinden2018} measure the centering fraction and centering distribution of \redmapper\ clusters by comparing the reported centers to those derived from high resolution X-ray data (where available).  Here, we summarize their findings.   The miscentering distribution $p(\Rmis)$ has the form
\begin{align}
	\label{eq:miscentering_distribution}
	p(\Rmis) & = \frac{\Rmis}{(\tau R_\lambda)^2}\exp\left(-\frac{\Rmis}{\tau R_\lambda}\right)
\end{align}
where $R_\lambda$ is the cluster radius assigned by \redmapper, and $\tau=\Rmis/R_\lambda$. Note that this is a Gamma distribution, which is more heavily-tailed than the Rayleigh distribution used in \citet{Simet2017,rmsva}. This model choice is justified in \citet{Zhang2018, vonderLinden2018}. In the latter, the cluster sample is complete for $\lambda\in[25,35]$ and $z\in[0.08,0.12]$, and was selected to be representative of the SDSS \redmapper\ cluster population. We use a combination of the posteriors from those two works to set the priors of the miscentering parameters $\fmis$ and $\tau$, as detailed in \autoref{tab:modeling_parameters}. The prior uncertainties conservatively encompass the spread in best fitting values and the confidence intervals of both works. It corresponds to a miscentering fraction $\fmis=0.25 \pm 0.08$, that is, roughly $\approx 75$ per cent of the \redmapper\ clusters are correctly centered. Because the variation in $R_\lambda$ within each richness bin is mild, we ignore variations in $R_\lambda$ across the bin, and set $R_\lambda$ to the radius of a cluster whose richness is equal to the mean richness of the clusters in the bin. We have explicitly verified that if use the median rather than the mean cluster richness, the difference between our predicted profiles is insignificant relative to our statistical errors.  


\subsection{Multiplicative corrections}
\label{sec:multiplicative_corrections}

Multiplicative corrections to our model include boost factors, reduced shear, and shear+\photoz\ biases. These adjust our model according to
\begin{equation}
	\label{eq:multiplicative_corrections}
    \Delta\Sigma_{\rm full}(R) = \frac{\mathcal{A}_m\mathcal{G}(R)}{\calB(R)}\Delta\Sigma_{\rm model}\,.
\end{equation}
In this equation, $\mathcal{A}_m$ is the multiplicative correction due to shear and photometric redshift biases, $\mathcal{G}(R)$ is the multiplicative correction due to using reduced shear, and $\mathcal{B}(R)$ is the boost factor correction.

\subsubsection{Boost factor model}
\label{sec:boost_factor_model}

In section \autoref{sec:boost_factors}, we discussed how membership dilution biases the recovered weak lensing profile by a factor $1-\fcl$. This factor is known as a boost factor because to correct for it in the lensing profile, one would increase the measured signal. We decided to not apply this factor to our data, and instead dilute the theoretical profile. To marginalize over the statistical uncertainty in our boost factor measurements, we parameterize the boost factor $\calB \equiv (1-\fcl)^{-1}$ by constructing a model for the cluster-member contamination using a two component ($B_0$ and $R_s$) NFW profile:
\begin{equation}
  \label{eq:boost_model}
  \calB_{\rm model}(R) = 1+B_0\frac{1-F(x)}{x^2-1}\,,
\end{equation}
where $x=R/R_s$, and
\begin{align}
  \label{eq:boost_model2}
  F(x) = \left\{
  \begin{array}{lr}
    \frac{\tan^{-1}\sqrt{x^2-1}}{\sqrt{x^2-1}} & : x > 1\\
    1 & : x = 1\\
    \frac{\tanh^{-1}\sqrt{1-x^2}}{\sqrt{1-x^2}} & : x < 1
  \end{array}
  \right.\,.
\end{align}
We fit the boost factors measured in each bin along with the lensing profile of that bin. 
This introduces two additional parameters in our model, $B_0$ and $R_s$, for each richness and redshift bin.


\subsubsection{Reduced Shear}
\label{sec:reduced_shear}

We account for the fact that we measure the reduced shear $g$ rather than true shear $\gamma$, as seen in \autoref{eq:reduced_shear_definition}, by multiplying our lensing model by
\begin{equation}
  \label{eq:reduced_shear}
  \mathcal{G}(R) = \frac{1}{1-\kappa} = \frac{1}{1-\Sigma(R)\Sigma_{\rm crit}^{-1}}\,.
\end{equation}
Here, $\Sigma_{\rm crit}^{-1}$ is the same as that in \ref{sec:photometric_redshift_systematics} and $\Sigma(R)$ is 
\begin{equation}
	\label{eq:reduced_shear_sigma}
    \Sigma(R) = (1-\fmis)\Sigma_{\rm cen} + \fmis\Sigma_{\rm mis}\,,
\end{equation}
where $\Sigma_{\rm cen}$ is given by \autoref{eq:Sigma} and $\Sigma_{\rm mis}$ is given by \autoref{eq:sigma_miscentered}. This adjustment has a negligible effect on our results, and introduces no new free parameters to our analysis.


\subsubsection{Shear+photo-$z$ bias}
\label{sec:multiplicative_A}

The factor $\mathcal{A}_m = 1 + m + \delta$ combines the effects of shear ($m$, \autoref{sec:shear_systematics}) and \photoz\ ($\delta$, \autoref{sec:photometric_redshift_systematics}) systematic uncertainties. \citet{Y1shape} found a shear calibration of $m=0.012\pm0.013$. The \photoz\ bias comes from \citet{Y1pz} and varies between cluster stacks.

Since both $m$ and $\delta$ are assigned Gaussian priors, the width of the prior on $\mathcal{A}_m$ is obtained by adding the widths of the priors on $m$ and $\delta$ in quadrature.  We arrive at
\begin{equation}
	\label{eq:multiplicative_total_bias}
	\mathcal{A}_m = \begin{cases}
		1.021 \pm 0.025 & {\rm for}\ z\in[0.2,0.35] \\
		1.014 \pm 0.024 & {\rm for}\ z\in[0.35,0.5] \\
		1.016 \pm 0.025 & {\rm for}\ z\in[0.5,0.65].
	\end{cases}
\end{equation}
The typical systematic uncertainty of $\approx 2.5$ per cent represents a significant improvement over the typical systematic uncertainty of $\approx 3.8$ per cent we achieved in \citet{rmsva}.  This dramatic improvement in accuracy is primarily driven by the improved shear calibration achieved in the DES Y1 data with \metacal.

For the following data releases of DES, we anticipate that improvements in the treatment of blended objects can further reduce the multiplicative shear bias. This implies that uncertainties in the calibration of photometric redshift estimates will likely be our dominant measurement related systematic. Significant improvements on this will require either extended calibration data sets or a hierarchical treatment that uses survey data to inform redshift estimation consistently.


\subsection{Stacked mass corrections}
\label{sec:stacked_mass_correction}

We expect the masses we measure in \autoref{sec:complete_likelihood} to be biased with respect to the true mean mass of the stacks. This bias arises from two sources: our model presented above is not a true description of cluster lensing profiles, and effects due to triaxiality and projection. We account for both sources of bias by calculating a correction $\cal{C}$ applied to the expected mass of the stack $\mathcal{M}_{\rm true} = \mathcal{C}\langle M\rangle$, as detailed in the section below. This is applied after the lens modeling is complete, but before modeling the mass--observable relation from our stacked masses in \autoref{sec:mass_richness_relation}.

\subsubsection{Modeling systematics}
\label{sec:modeling_systematics}

The model presented above for $\Delta\Sigma$ is not perfect; our analytic model for the halo-mass correlation function in \autoref{eq:hmcf} does not match density profiles in simulations \citep{rmsva,Murata2018}, in particular in the transition between the 1-halo and 2-halo regimes. In lieu of a fully calibrated model, we correct for any bias imparted by our choice of model by using our likelihood analysis to estimate halo masses of synthetic data generated from $N$-body simulations. The halos are drawn from an $N$-body simulation of a flat $\Lambda$CDM cosmology run with \textsc{Gadget} \citep{Springel05}. The simulation uses 1400$^3$ particles in a box with $1050\ h^{-1}{\rm Mpc}$ on a side with periodic boundary conditions and for softening of $20\ h^{-1}{\rm kpc}$. The simulation was run with the cosmology $\Omega_m=0.318$, $h=0.6704$, $\Omega_b=0.049$, $\tau=0.08$, $n_s=0.962$, and $\sigma_8=0.835$. Halos of mass $10^{13}\ h^{-1}{\rm M}_\odot$ are resolved with 100 particles. We discard all information below 5 softening lengths, and verified that the choice of extrapolation scheme for describing the correlation function below this scale does not impact our results. Halos were defined using a spherical overdensity mass definition of 200 times the background density and were identified with the \textsc{ROCKSTAR} halo finder \citep{Behroozi2013}.

The simulation is used to construct the synthetic $\Delta\Sigma$ profiles of galaxy clusters at four different snapshots: $z\in[0,0.25,0.5,1]$. There were $\sim$420,000 halos at $z=1$ and $\sim$830,000 halos at $z=0$. We used snapshots instead of lightcones for two main reasons: we wanted to maximize the number of halos we had available to perform the calibration, and we found that the synthetic profiles to only weakly depend on redshift. Instead of splitting halos into mass subsets as in \citet{rmsva}, we assigned a richness to each halo by inverting the mass--richness relation of \citet{rmsva} and adding 25 per cent scatter. We then grouped our halos into richness subsets identical to how we grouped our clusters. For each of these halo subsets we measured the halo-matter correlation function with the \citet{Landy93} estimator as implemented in \textsc{Corrfunc}\footnote{\url{https://github.com/manodeep/Corrfunc}} \citep{Corrfunc}. We numerically integrate the halo-matter correlation function to obtain the $\Delta\Sigma$ profile as described in \autoref{sec:surface_density_model}. 

This $\Delta\Sigma$ profile contains none of the systematics that exist in the real data. To incorporate them, we modified this profile with the multiplicative corrections described in \autoref{sec:multiplicative_corrections} and miscentering corrections in \autoref{sec:miscentering_correction}. We took the central values of our priors in \autoref{tab:modeling_parameters} as well as values for $B_0$ and $R_s$ from modeling the boost factors independently and modified the simulated $\Delta\Sigma$ profile according to \autoref{eq:full_delta_sigma_model}. The observed mass $M_{\rm obs}$ for this simulated profile was obtained by using the same pipeline that we apply on the real data. When evaluating the likelihood in \autoref{eq:total_likelihood}, we used the semi-analytic covariance matrix corresponding to the nearest cluster subset in redshift.

\begin{figure}
	\includegraphics[width=\linewidth]{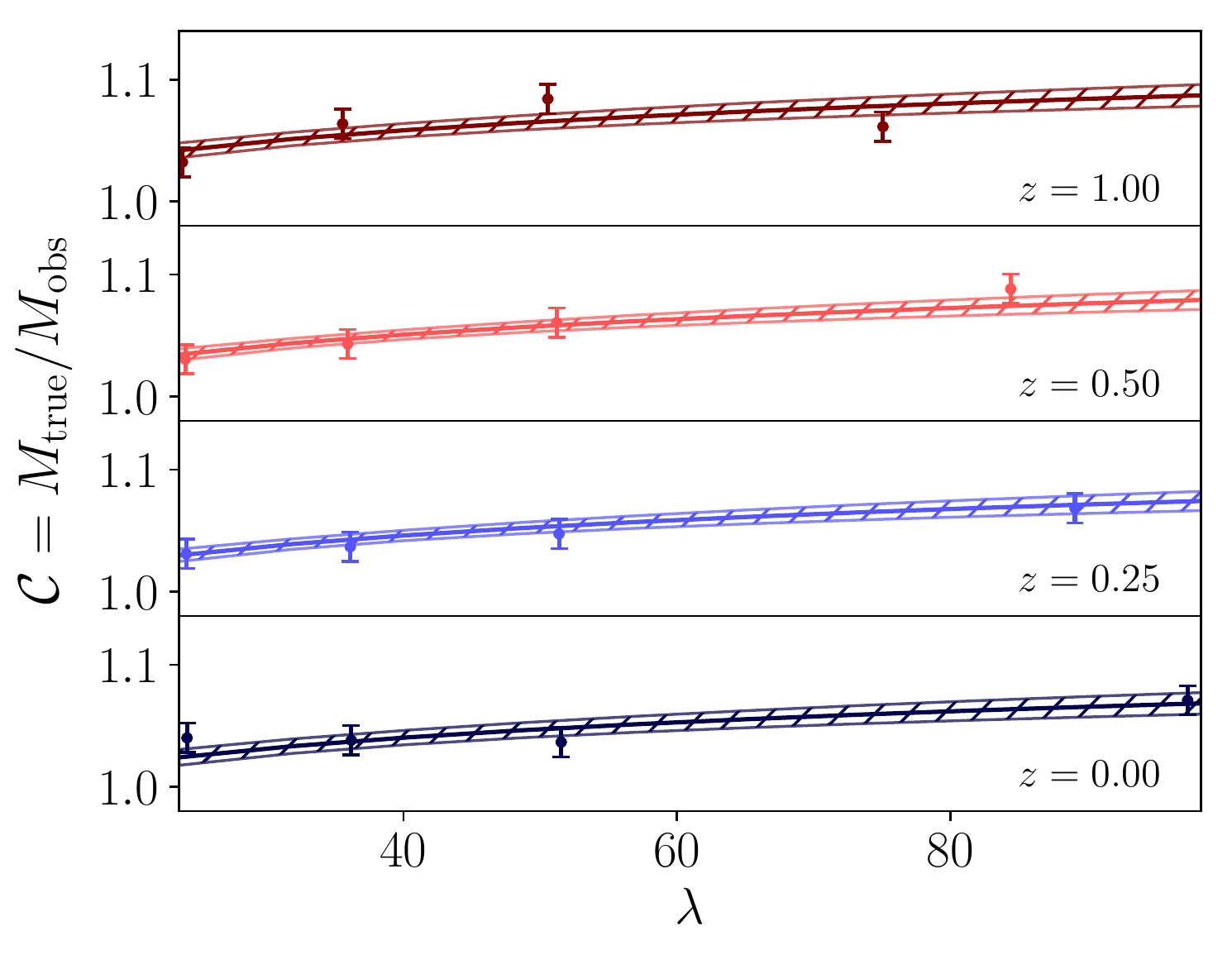}
	\caption{The mass calibration $\mathcal{C} = M_{\rm true}/M_{\rm obs}$ from adopting our model of the correlation function in \autoref{eq:hmcf} as a function of $\lambda$ and redshift. The solid line and hatched region are the best fit model and 1$\sigma$ uncertainty of the calibration. Error bars on the measured calibrations are the fitted intrinsic scatter $\sigma_{\mathcal{C}}$.}
	\label{fig:calibration}
\end{figure}

Denoting $M_{\rm true}$ as the mean mass of the halos in the simulated stack, the calibration for each simulated profile is seen in \autoref{fig:calibration}. The calibration $\mathcal{C}=M_{\rm true}/M_{\rm obs}$ was then modeled as a function of the mean richness of the simulated stack $\bar{\lambda}$ in the form
\begin{equation}
	\label{eq:calibration_model}
    \mathcal{C}(\bar{\lambda},z) = C_0\left(\frac{\bar{\lambda}}{\lambda_0}\right)^\alpha \left(\frac{1+z}{1+z_0}\right)^\beta\,.
\end{equation}
The free parameters in our fit are $C_0$, $\alpha$, and $\beta$ with pivot values at $z_0=0.5$ and $\lambda_0$=30, as well as the intrinsic scatter $\sigma_\mathcal{C}$ of the calibration. The mean model bias for our cluster stacks is $\approx 4$ per cent with $C_0=1.042 \pm 0.004$, $\alpha=0.03 \pm 0.006$, and $\beta=0.025\pm 0.012$ as well as $\sigma_\mathcal{C}=0.01$. We repeated this process while assuming different amounts of intrinsic scatter in the $M$--$\lambda$ relation from 10 per cent up to 45 per cent, as well as with no intrinsic scatter which is equivalent to the treatment of \citet{rmsva}. We found that the amount of model bias increased with scatter in the $M$--$\lambda$ relation. The model bias from \citet{rmsva} was recovered when no intrinsic scatter was present and using covariance matrices from that analysis.

We incorporated the dependence of the calibration on the intrinsic scatter in the $M$--$\lambda$ relation as follows. We took the calibration described above at 25 per cent scatter to be our fiducial model as estimated in \citet{RozoRykoff14_RM2}. In addition to the covariance between $C_0$, $\alpha$, and $\beta$, we add additional uncorrelated uncertainty to each of these terms equal to half of the difference between the mean values obtained for these parameters assuming 15 per cent and 45 per cent scatter. This increased the variance of all three parameters $C_0$, $\alpha$ and $\beta$ slightly. As discussed further in \ref{sec:future_improvements}, the calibration contributed $0.73$ per cent to the overall systematic uncertainty on the normalization of the $M$--$\lambda$ relation.

One effect which we have not explicitly accounted for is the impact of baryonic physics on the recovered weak lensing masses.  Baryonic cooling and feedback leads to mass redistribution in the central regions of galaxy clusters, with the impact of baryonic physics decreasing with increasing radius. Given that our fits allowed the concentration parameter of each cluster stack to float with no informative priors, we naively expect the impact of baryonic physics can be absorbed into the concentration of each cluster stack \citep{schalleretal15a}. This naive expectation is confirmed in the work of \citet{hensonetal17}, who found that while baryonic physics impact the recovered weak lensing masses, the relative bias between the recovered weak lensing mass and the true mass is roughly constant, independent of baryonic physics. Given that we measure masses at large radii ($R_{200m}$), and in light of the above results, we believe the impact of baryonic physics is likely to be less than $\sim 3\%$ or so. Future work in which we explicitly test our fitting routines using hydrodynamic simulations is clearly desirable.


\subsubsection{Triaxiality and projection effects}
\label{sec:triaxiality_and_projection_effects}

Photometric cluster selection preferentially selects halos that are oriented with their major axis along the line of sight. Similarly, cluster selection is affected by other objects along the line of sight, which increases both the observed cluster richness and the recovered lensing mass. These two effects have been studied closely elsewhere \citep{White2011,anguloetal12,nohcohn12,Dietrich2014}, and have competing effects on the recovered cluster masses. \citet{Dietrich2014} determined that triaxiality leads to an overestimation of the weak lensing mass and requires a correction factor of $0.96 \pm 0.02$, while \citet{Simet2017} argued projection effects require that the recovered masses be multiplied by a factor of $1.02 \pm 0.02$. Together, triaxiality and projection effects modify the recovered weak lensing masses by a multiplicative factor of $0.98 \pm 0.03$. Our treatment is identical to that of \citet{rmsva}, where additional details are provided. Although these two effects mildly depend on richness and redshift, we assume them to be constant in this analysis. We show the cumulative effect in \autoref{tab:error_budget}. For reference, we have estimated the number of galaxy clusters that have another \emph{cluster} within a $500\ {\rm kpc}$ radius along the line of sight.  The number of such cases with $\lambda \geq 20$ is about 30, or 0.4 per cent of our sample, and thus negligible.

These effects as well as the correction for model bias are applied to the masses after fitting the lensing and boost factor data as described in \autoref{sec:complete_likelihood}, but before modeling the $M$--$\lambda$ relation in \autoref{sec:mass_richness_relation}.


\subsection{The complete likelihood}
\label{sec:complete_likelihood}

The full model of the weak lensing profile is
\begin{equation}
	\label{eq:full_delta_sigma_model}
	\Delta\Sigma = \frac{\mathcal{A}_{m}\mathcal{G}(R)}{\mathcal{B}(r)}\left[(1-\fmis)\Delta\Sigma_{\rm cen}+\fmis\Delta\Sigma_{\rm mis}\right]\,.
\end{equation}
Written this way the model includes the multiplicative bias $\mathcal{A}_m$, the boost factor $\mathcal{B}(r)$, the reduced shear correction $\mathcal{G}(R)$, and miscentering effects $f_{mis}$ and $\Delta\Sigma_{\rm mis}$. Using the semi-analytic covariance matrices ${\rm C}_{\Delta\Sigma}$ described in \autoref{sec:semi_analytic_covariance_matrix}, the log-likelihood of the $k$th $\Delta\Sigma$ profile is
\begin{equation}
	\label{eq:deltasigma_likelihood}
    \ln \lkhd(\Delta\Sigma_k\,|\, M_k,c,\mathcal{A}_m,\Rmis,\fmis,B_0,R_s) \propto -\frac{1}{2}{\rm \bf D}_k^T{\rm C}^{-1}_{\Delta\Sigma}{\rm \bf D}_k
\end{equation}
where ${\rm \bf D} = (\widetilde{\Delta\Sigma} - \Delta\Sigma)$ and $\widetilde{\Delta\Sigma}$ is the measurement from \autoref{eq:updated_deltasigma_estimate}.

The boost factor covariance matrix ${\rm C_{\fcl}}$ is estimated from jackknifing. With this the corresponding log-likelihood of the measured $f_{{\rm cl},k}$ in cluster subset $k$ given the parameters in \autoref{eq:boost_model} is
\begin{equation}
  \label{eq:boost_likelihood}
  \ln \lkhd(f_{{\rm cl},k}\,|\, B_0, R_s) \propto -\frac{1}{2}{\rm \bf B}_k^T{\rm C}^{-1}_{\fcl}{\rm \bf B}_k
\end{equation}
where ${\rm \bf B}_k = (\calB-\calB_{\rm model})_k$. Each boost factor $\calB_k$ is fit in conjunction with the associated lensing profile for the $k$th subset, allowing us to account for any degeneracies between the parameters in the respective models.

The total log-likelihood for a single cluster subset is
\begin{equation}
	\label{eq:total_likelihood}
    \begin{split}
		\ln \lkhd_k =& \ln \lkhd(\Delta\Sigma_k\,|\, M_k,c,\mathcal{A}_m,\Rmis,\fmis,B_0,R_s) +\\
    		&\ln \lkhd(f_{{\rm cl},k}\,|\, B_0, R_s)\,.
    \end{split}
\end{equation}
Our goal is to constrain masses of independent cluster subsets $M_k$ and boost factor parameters. Constraints on the latter are informed by both their effect on the $\Delta\Sigma$ profile as well as independent measurements of $\fcl$. The weak lensing and boost factor profiles are fit simultaneously, but each cluster subset is fit independently of other subsets.


\subsection{Stacked cluster masses}
\label{sec:stacked_cluster_masses}

A complete list of the model parameters describing each cluster stack as well as their corresponding priors are summarized in \autoref{tab:modeling_parameters}. The likelihood is sampled using the package \textit{emcee}\footnote{\url{http://dan.iel.fm/emcee}} \citep{Foreman13}, which enables a parallelized exploration of the parameter space. We use 32 walkers with 10000 steps each, discarding the first 1000 steps of each walker as burn-in. We checked the convergence with independent runs of 5000 steps per walker that yielded identical results. After 14 steps the chains of single walkers become uncorrelated (with a correlation coefficient $|r|<0.1$). This is much shorter than the total length of the chain. As a result the number of independent draws between all walkers is $\approx 20500$. 

\begin{table}
	\setlength{\tabcolsep}{.4em}
	\caption{Parameters entering $\lkhd(\Delta\Sigma)$ (\autoref{eq:deltasigma_likelihood}) and $\lkhd(\calB)$(\autoref{eq:boost_likelihood}) Flat priors are specified with limits in square brackets, Gaussian priors with means $\pm$ standard deviations.}
	\begin{tabular}{lll}
		Parameter & Description & Prior \\ \hline
		$\log_{10}M_{\rm 200m}$ & Halo mass & $[11.0,18.0]$ \\
        $c_{\rm 200m}$ & Concentration & $[0,20]$\\
        $\tau$ & Dimensionless miscentering offset & $0.17\pm0.04$\\
		$f_{\rm mis}$ & Miscentered fraction & $0.25\pm0.08$\\
		$A_{m}$&Shape \& \photoz\ bias & \autoref{eq:multiplicative_total_bias}\\
		$B_0^{\rm cl}$ & Boost magnitude & $[0,\infty]$\\
		$R_s^{\rm cl}$ & Boost factor scale radius & $[0,\infty]$ \\
	\end{tabular}
    \label{tab:modeling_parameters}
\end{table}

The calibration correction described in \ref{sec:stacked_mass_correction} was applied to the recorded chains for each cluster subset. Specifically, for each point in the chain, we randomly sample the mass calibration factor $\mathcal{C}(\lambda,z)$ from its posteriors to adjust the mass. Further, we also apply the effect of triaxiality and projection effects (\autoref{sec:triaxiality_and_projection_effects}), both of which add 2 per cent to the uncertainty of each mass. In practice this is written as
\begin{equation}
	\label{eq:calibration_application}
    M = \mathcal{C}(\lambda,z)\mathcal{G}(0.96,0.02)\mathcal{G}(1.02,0.02)\times M_0\,,
\end{equation}
where $\mathcal{G}$ is a Gaussian and $M_0$ are the ``uncalibrated'' masses in the chains. In this fashion, our final posteriors are properly marginalized over the uncertainty in the calibration factor $\mathcal{C}$ as well as triaxiality and projection effects.

In order to characterize the contribution of both statistical and systematic uncertainties to our final results we perform our analysis three different times with three different sets of assumptions.  These three analyses we run are:
\begin{itemize}
	\item \verb|Full|: All parameters (concentration, Shear+\photoz, boost factors, miscentering) are allowed to vary within their priors. This is our fiducial analysis.
	\item \verb|FixedAm|: $\mathcal{A}_m$ is set to the center of its prior distribution but all other parameters are allowed to vary. This determines the contribution from the shape and \photoz\ uncertainties.
	\item \verb|OnlyMc|: Only mass and concentration are free. All other parameter priors are set to $\delta$-functions at their central values. This represents our statistical uncertainty on the mass.
\end{itemize}

\begin{figure*}
	\includegraphics[width=\linewidth]{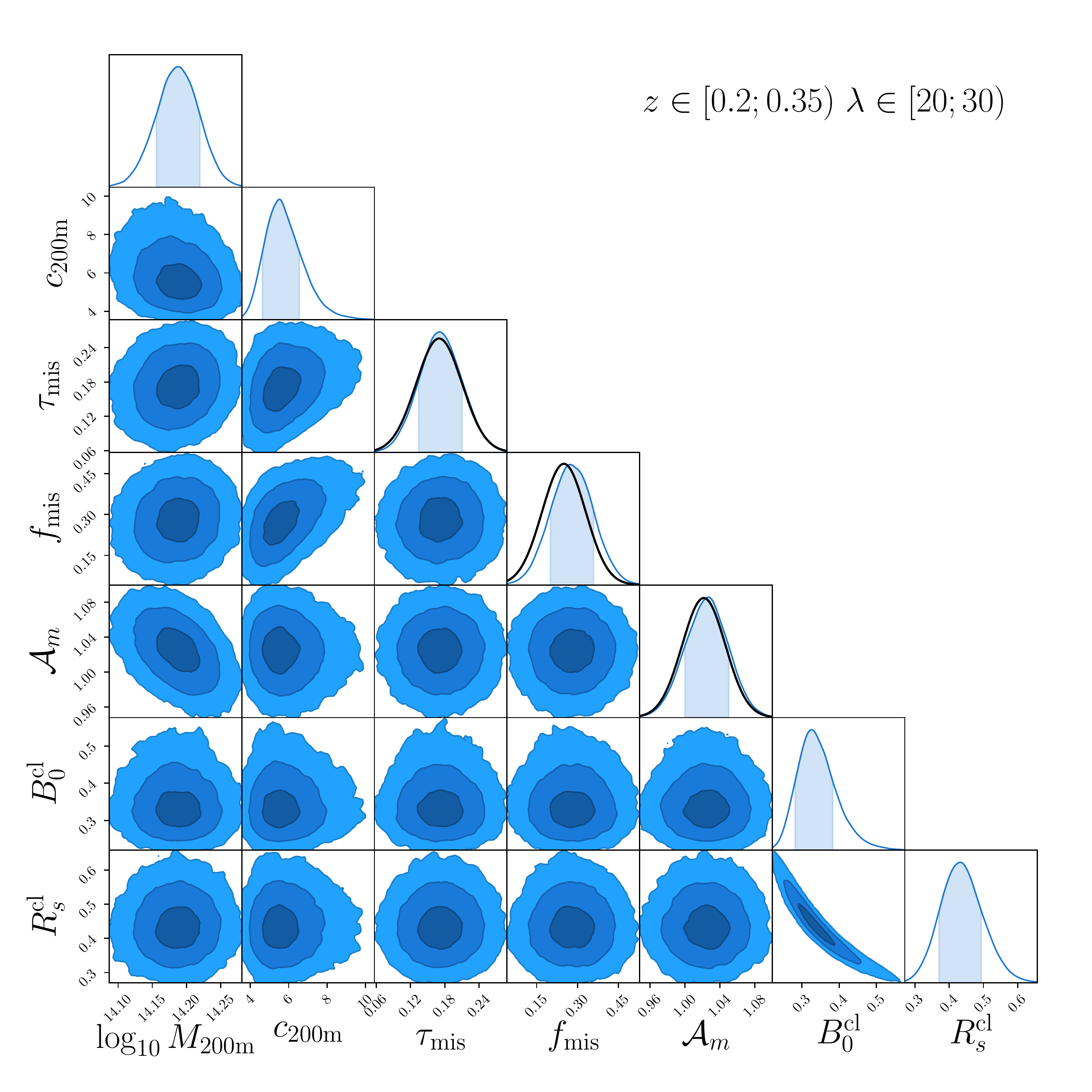}
	\caption{Posteriors for the parameters describing the lensing profile $\Delta\Sigma$ and the boost factor profile $\mathcal{B}$ for the bin $z \in [0.2,0.35)$, $\lambda \in [20,30)$. Contours show the $1\sigma$, $2\sigma$, and $3\sigma$ confidence areas. Black lines show the prior distributions. The mass presented here is uncalibrated, meaning it has not been corrected for modeling systematics, projection effects, or cluster triaxiality (see \autoref{sec:stacked_mass_correction}).}
    \label{fig:corner1}
\end{figure*}

\begin{figure}
	\includegraphics[width=\linewidth]{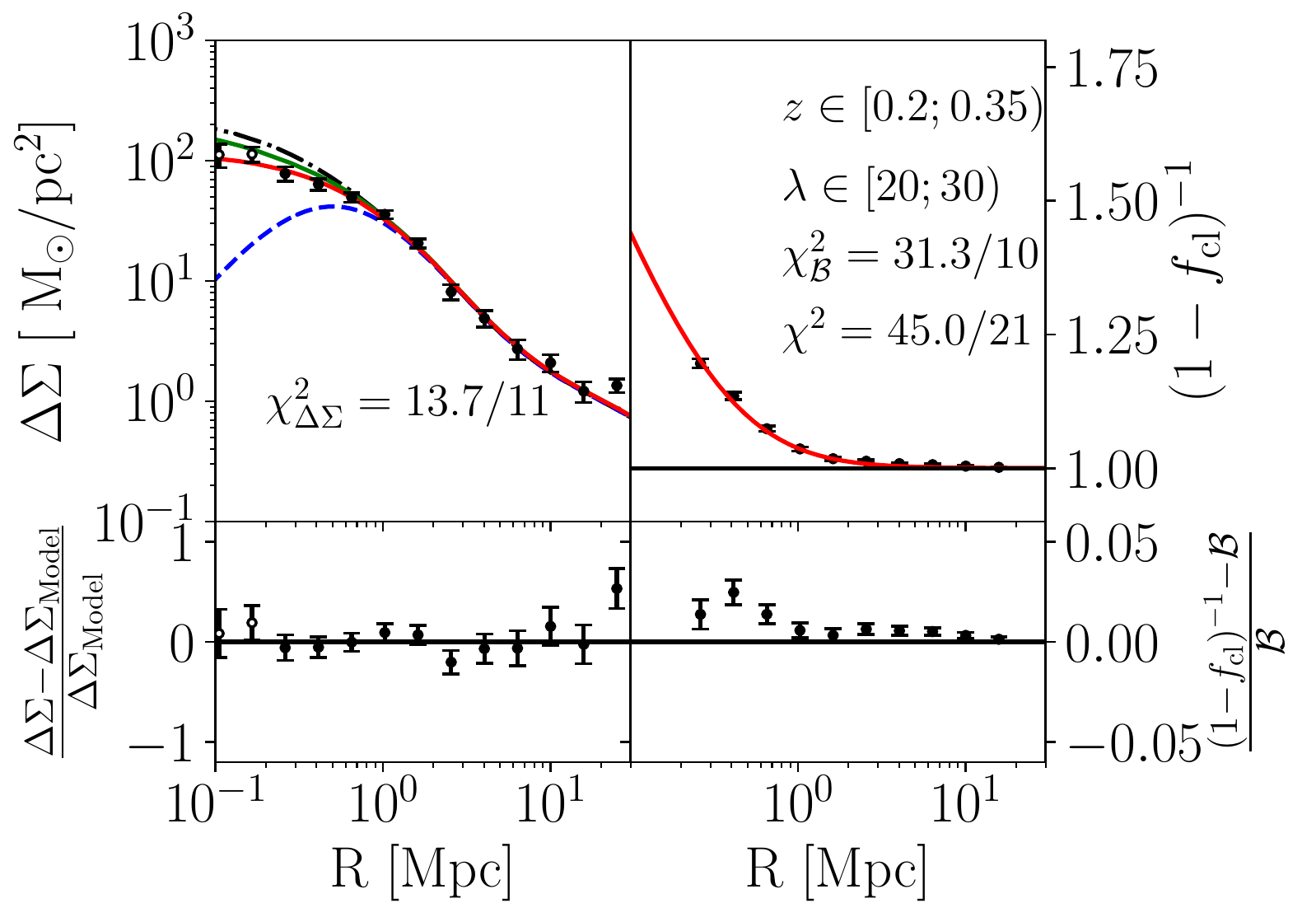}
	\caption{Fit with all components of the $\Delta\Sigma$ and $\mathcal{B}$ models for the cluster subset $z\in[0.2,0.35)$ and $\lambda\in[20,30)$. The top two panels show the best fit models in red compared to the data. Unfilled points are not included in the fit. {\it Top left}: the black dot-dashed line is $\Delta\Sigma_{\rm cen}$ while the blue dashed line is $\Delta\Sigma_{\rm mis}$. The weighted mean of these two yields the green solid line, and then applying the boost factor model, reduced shear, and multiplicative bias yields the final model in red. {\it Top right}: the red line is our NFW model for the boost factors. {\it Bottom}: the fractional difference between our data and models. The total $\chi^2$ is 45 with 21 degrees of freedom, which is acceptable despite the imperfect fit of our simple model to the boost factors. The boost factors are measured from the data with small uncertainty, which is why the small mismatch with respect to the best-fit model causes a relatively large $\chi^2$ but negligible effect on the recovered mass.}
    \label{fig:best_fit_example}
\end{figure}

\autoref{tab:posterior_masses} contains the results of the \verb|Full| analysis. Full posteriors from the cluster subset $z\in[0.2,0.35)$ and $\lambda\in[20,30)$ are shown in \autoref{fig:corner1}. The corresponding data and best-fit model are shown in \autoref{fig:best_fit_example}, where we also demonstrate the combined effects of miscentering, boost factors, reduced shear, and multiplicative bias. The best fit model for each richness and redshift bin is over-plotted on top of the weak lensing data in \autoref{fig:all_profiles}.

From the \verb|Full| analysis we can see the contribution of the various systematics to our final results. The boost factors amount to a correction of $\approx 2$ per cent to $\Delta\Sigma$ at $R=1\ {\rm Mpc}$. The posteriors on the miscentering parameters are equal to the priors, demonstrating that these parameters are only weakly constrained by the weak lensing data. In our earlier analysis \citep{rmsva} we found a weak correlation between $\fmis$ and $M$, which did not appear in this work. This was due to our use of the \citet{diemerkravtsov14} $M$--$c$ relation. We determined this by running one additional configuration in which the concentration was fixed by the \citet{diemerkravtsov14} $M$--$c$ relation, thus increasing the correlation between $\fmis$ and $M$. At present, the contribution of miscentering to the mass is sub-dominant to other sources of systematic uncertainties in our final error budget (cf.~\autoref{tab:error_budget}). The multiplicative bias $\mathcal{A}_m$ follows the prior and is degenerate with mass.

The \verb|OnlyMc| likelihood evaluation allows us to quantify the statistical and systematic uncertainties of the fiducial analysis. The difference in quadrature between the uncertainties in the \verb|Full| and \verb|OnlyMc| configurations represents the total systematic contribution to the error budget, while the \verb|OnlyMc| alone provides the statistical contribution. The central values for each cluster subset along with statistical and systematic contributions to the uncertainties are presented in \autoref{tab:posterior_masses}. 

\begin{table*}
  \caption{Calibrated masses for each richness-redshift stack. All masses are in units of $\log_{10}$M$_\odot$ using the $M_{\rm 200m}$ definition. Listed uncertainties are split into the symmetric 68 per cent confidence intervals of the systematic and statistical components, in that order. Adding the two in quadrature gives the total uncertainty on the mass.}
  \label{tab:posterior_masses}
    \begin{tabular}{llll}
		$\lambda$ & $z\in[0.2,0.35)$ & $z\in[0.35,0.5)$ & $z\in[0.5,0.65)$ \\ \hline
        $[20,30)$ & 14.191 $\pm$ 0.013 $\pm$ 0.032 & 14.162 $\pm$ 0.013 $\pm$ 0.033 & 14.083 $\pm$ 0.015 $\pm$ 0.048\\
		$[30,45)$ & 14.477 $\pm$ 0.014 $\pm$ 0.031 & 14.446 $\pm$ 0.014 $\pm$ 0.031 & 14.456 $\pm$ 0.015 $\pm$ 0.041\\
		$[45,60)$ & 14.608 $\pm$ 0.011 $\pm$ 0.044 & 14.643 $\pm$ 0.011 $\pm$ 0.044 & 14.648 $\pm$ 0.016 $\pm$ 0.056\\
		$[60,\infty)$ & 14.913 $\pm$ 0.014 $\pm$ 0.038 & 14.899 $\pm$ 0.015 $\pm$ 0.038 & 14.879 $\pm$ 0.023 $\pm$ 0.061\\
    \end{tabular}
\end{table*}


\section{The mass--richness--redshift relation}
\label{sec:mass_richness_relation}

The quantity we aim to constrain in this paper is the mean mass $\mathcal{M}(\lambda,z)$ of clusters of galaxies at a given observed richness $\lambda$ and redshift $z$, similar to what was done in \citet{Melchior15}. Note that this is different from constraining the mean (and possibly distribution) of richness at given mass, or the full distribution of mass at given richness, as done in e.g.~\citet{Simet2017,Murata2018}. In particular, we neither constrain nor require a model of the intrinsic scatter in richness, hence making this analysis largely independent from the choices in subsequent cluster cosmology studies based upon it.

We note that an assumed value of the intrinsic scatter is used in two places: the semi-analytic covariance matrices described in \autoref{sec:semi_analytic_covariance_matrix} and the calibration described in \autoref{sec:modeling_systematics}. For the covariance, this assumption had a negligible effect compared to the shape noise. While the overall calibration did depend on the amount of scatter, we took a conservative approach by treating the difference in calibration between assuming 15 per cent and 45 per cent scatter as a systematic uncertainty. In this way, our final results are not sensitive to the amount of assumed intrinsic scatter.

\subsection{Modeling the mass--richness relation}
\label{modeling_the_mass_richness_relation}
We fit a redshift-dependent power-law relation between cluster richness and cluster mass.  Specifically, we set
\begin{equation}
	\label{eq:mass_richness_redshift}
	\mathcal{M}(\lambda,z) \equiv \avg{M\,|\,\lambda,z} = M_0\left(\frac{\lambda}{\lambda_0}\right)^{F_\lambda}\left(\frac{1+z}{1+z_0}\right)^{G_z}\,,
\end{equation}
where $M_0$, $F_\lambda$, and $G_z$ are our model parameters.  We select pivot values $\lambda_0 = 40$ and $z_0 = 0.35$, which are very near the median values of the cluster sample. Note $\calM$ is the expectation value of the mass of a halo as a function of richness and redshift.

The expected mass of a given cluster subset $k$ represents a weighted mean of the masses of the individual clusters in that stack. We then have
\begin{equation}
	\label{eq:weighted_mass}
    \mathcal{M}_k = \frac{\sum_{j \in k}\hat{w}_j\mathcal{M}(\lambda_j,z_j)}{\sum_{j \in k } \hat{w}_j}\,.
\end{equation}
We take the weight $\hat{w}_j$ of the $j$th cluster to be the sum of the weights of all lens--source pairs $w_{j,i}$ around that cluster from $0.3$~Mpc and above and verified that the choice of radial range does not affect our recovered masses. Individual cluster weights $\hat{w}_j$ differ from unity. This is because 1) the lensing weight of each lens--source pair given by \autoref{eq:w_ideal} depends on the cluster redshift, and 2) in a given radial bin there are more sources associated with low redshift clusters since that bin subtends a larger angle on the sky compared to the same bin around a high redshift cluster. In other words, the mass in the bin is skewed toward the average mass of the lower redshift clusters in the bin.

The impact of the weak lensing weights on the stacked mass estimates can be characterized by the ratio
\begin{equation}
	\label{eq:weight_impact_factor}
	a=\frac{\mathcal{M}_0}{\mathcal{M}_{\hat{w}}}\,.
\end{equation}
We report the quantity $\log a$ in \autoref{tab:weight_impacts}. We chose to report $\log a$ rather than $a$ which has the computational advantage that one need only  add $\log a$ to the mass values in \autoref{tab:posterior_masses} to arrive at an estimate of the mean cluster masses of cluster in a bin with unit weighting (as opposed to mean weak lensing weighted masses). The corrections in \autoref{tab:weight_impacts} are used to correct the recovered cluster masses to unit-weighted masses in the DES Y1 analysis of cluster abundances (DES collaboration, in preparation).

\begin{table*}
  \caption{The logarithm of the mean mass correction factor $\log_{10}a$ from \autoref{eq:weight_impact_factor}. This represents a correction to the stacked cluster masses due to the fact that different clusters contribute to the measured mass in a different way than they contribute to $\Delta\Sigma$.}
  \label{tab:weight_impacts}
    \begin{tabular}{llll}
		$\lambda$ & $z\in[0.2,0.35)$ & $z\in[0.35,0.5)$ & $z\in[0.5,0.65)$ \\ \hline
        $[20,30)$ & \num{-1.372e-03} & \num{-8.744e-04} & \num{-4.501e-04} \\
		$[30,45)$ & \num{-2.979e-03} & \num{-3.278e-03} & \num{-6.660e-04} \\
		$[45,60)$ & \num{-8.258e-04} & \num{-7.856e-05} & \num{-1.903e-03} \\
		$[60,\infty)$ & \num{3.043e-03} & \num{-4.061e-03} & \num{6.264e-03} \\
    \end{tabular}
\end{table*}


\subsection{Mass covariance}
\label{sec:mass_covariance}

The purpose of our different chain configurations (\verb|Full|, \verb|FixedAm|, and \verb|OnlyMc|) is to allow us to estimate the contribution of each systematic to the final uncertainty on the mass calibration parameters $M_0$, $F_\lambda$, and $G_z$. In our analysis there are seven sources of systematic uncertainty: multiplicative shear bias, multiplicative \photoz\ bias, miscentering, boost factors, modeling systematics, triaxiality and projection.

We discuss how we combine all systematics to arrive at a full covariance matrix between our bins that respects the covariance in our systematic error budget. The reader will recall that the original chains we obtain from fitting the weak lensing profiles are processed via~\autoref{eq:calibration_application} to account for the effect of calibration, triaxiality, and projection effects. If we let $M_0$ denote our unprocessed chains, and $M$ denote the chains post-processing, in order to derive the statistical uncertainty only in our mass measurement we generate a new chain $\tilde M$ via
\begin{equation}
	\label{eq:mean_calibration}
    \tilde{M} = \bar{\mathcal{C}}\times 0.96 \times 1.02 \times M_0\,.
\end{equation}
The difference in the variance between chain $M$ in~\autoref{eq:calibration_application} and that of chain $\tilde M$ represents the uncertainty associated with calibration, triaxiality, and projection effects. We will use the $M$ without a $\sim$ to denote the chains post-processed as per~\autoref{eq:calibration_application}, and $\tilde M$ for chains post-processed as above.  

The \verb|OnlyMc| chain configuration contains only the statistical uncertainty in our analysis. For this reason, the covariance matrix describing the masses in this configuration is diagonal. We define the statistical uncertainty of the $i$th mass $\sigma^2_{i,{\rm stat}}$ 
\begin{equation}
	\label{eq:onlymc_covariance}
    \sigma^2_{i,{\rm stat}} = {\rm Var}\left(\tilde{M}^{\rm OnlyMc}_i\right)\,. 
\end{equation}
We also isolate the uncertainty associated with shear and \photoz\ systematics. To do so, we note the \verb|Full| chain configuration includes all sources of uncertainty. Consequently, the difference in the variance between this chain and the \verb|FixedAm| chain represents the uncertainty associated with shear and \photoz\ systematics.  Therefore, we define
\begin{equation}
	\label{eq:full_ampart}
    \sigma^2_{i,{\rm S+Pz}} = {\rm Var}(M^{\rm Full}_i) - {\rm Var}(M^{\rm FixedAm}_i)\,.
\end{equation}
Finally, the uncertainty associated with modeling systematics (calibration, triaxiality, and projection effects) takes the form
\begin{equation}
	\label{eq:full_calcov}
	\sigma^2_{i,{\rm mod}} = {\rm Var}(M^{\rm Full}_i) - {\rm Var}(\tilde{M}^{\rm Full}_i)\,.
\end{equation}
By construction, the full uncertainty $\sigma^2_{i,{\rm Tot}}$ satisfies
\begin{equation}
\Var(M^{\rm Full}_i ) = \sigma^2_{i,{\rm stat}} + \sigma^2_{i,{\rm S+Pz}} +\sigma^2_{i,{\rm mod}}\,.
\end{equation}

We now define three different covariance matrices.  First, $\mathsf{C}^{\rm stat}$ is diagonal, with $\mathsf{C}^{\rm stat}_{ii}=\sigma^2_{i,{\rm stat}}$.  When we fit the weak lensing masses using this covariance matrix, we recover the statistical uncertainty in our scaling relation parameters. Second, $\mathsf{C}^{\rm S+Pz}$ is defined via
\begin{eqnarray}
\mathsf{C}^{\rm S+Pz}_{ii} & = & \sigma^2_{i,{\rm stat}} + \sigma^2_{i,{\rm S+Pz}} \\
\mathsf{C}^{\rm S+Pz}_{ij} & = & \left[ \sigma^2_{i,{\rm S+Pz}}\sigma^2_{j,{\rm S+Pz}} \right]^{1/2} .
\end{eqnarray}
Note the shear and \photoz\ components of the uncertainty are perfectly correlated across all bins.  Fitting the weak lensing mass with this covariance matrix, and subtracting the statistical uncertainty in quadrature, enables us to calculate the uncertainty in our scaling relation parameters associated with shear and \photoz\ systematics. Third, $\mathsf{C}^{\rm Full}$ is defined via
\begin{eqnarray}
\mathsf{C}^{\rm Full}_{ii} & = & \sigma^2_{i,{\rm stat}} + \sigma^2_{i,{\rm S+Pz}} + \sigma^2_{i,{\rm mod}} \\
\mathsf{C}^{\rm Full}_{ij} & = & \left[ \left( \sigma^2_{i,{\rm S+Pz}}+\sigma^2_{i,{\rm mod}} \right)
	\left( \sigma^2_{j,{\rm S+Pz}}+\sigma^2_{j,{\rm mod}} \right) \right]^{1/2} \,.
\end{eqnarray}
Just like the shear and \photoz\ calibration uncertainties, modeling systematics are assumed to be perfectly correlated across all bins. The posteriors for the scaling relation parameters derived with this covariance matrix represent our full error budget, and the difference in quadrature between these errors and those obtained using the covariance matrix $\mathsf{C}^{\rm S+Pz}$ give us the error budget associated with modeling systematics.


\subsection{Likelihood for the mass--observable relation}
\label{sec:likelihood_for_the_MOR}

\begin{figure}
    \includegraphics[width=\linewidth]{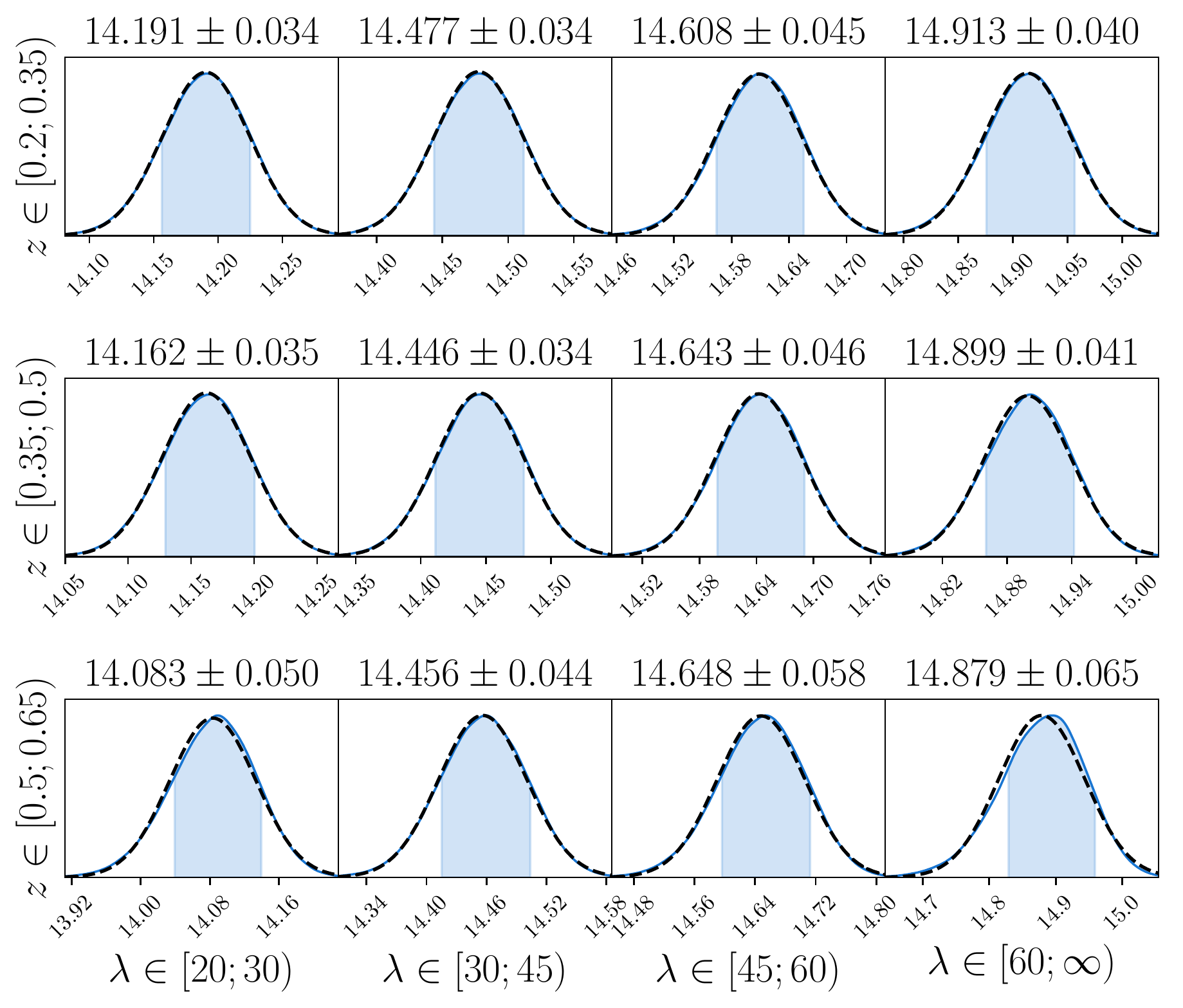}
	\caption{The calibrated posteriors of the masses for each cluster stack. Uncertainties appear above each panel, and are highlighted by the blue shaded regions. Gaussian approximations to these posteriors appear as black dashed lines.}
	\label{fig:mass_posteriors}
\end{figure}

We model the likelihood of the recovered weak lensing masses as Gaussian in the log. This is illustrated in \autoref{fig:mass_posteriors}, which shows the posterior for each of the 12 cluster bins with $\lambda \ge 20$, along with the corresponding Gaussian approximation. We rely on the $\lambda \ge 20$ cluster bins exclusively as it is only these systems for which we are confident we can unambiguously map halos to clusters and vice-versa. The full likelihood function is
\begin{equation}
  \label{eq:mass_richness_likelihood}
  \ln \lkhd(\Mobs\,|\,M_0,F_\lambda,G_z) \propto - \frac{1}{2} (\Delta \log M)^T\ \mathsf{C}_M^{-1}\ (\Delta \log M)\,,
\end{equation}
where $\mathsf{C}_M$ is the covariance between the mass bins for a particular configuration (see \autoref{sec:mass_covariance}). In the above equation $\Delta \log M$ is the difference between the measured mass of each cluster subset and our model for the expected mass given as per \autoref{eq:mass_richness_redshift}. Thus, for the $k$th bin
\begin{equation}
	\label{eq:delta_logmass}
	\Delta \log M_k = \log M_k - \log \mathcal{M}_k\,.
\end{equation}

\begin{figure}
  \includegraphics[width=\linewidth]{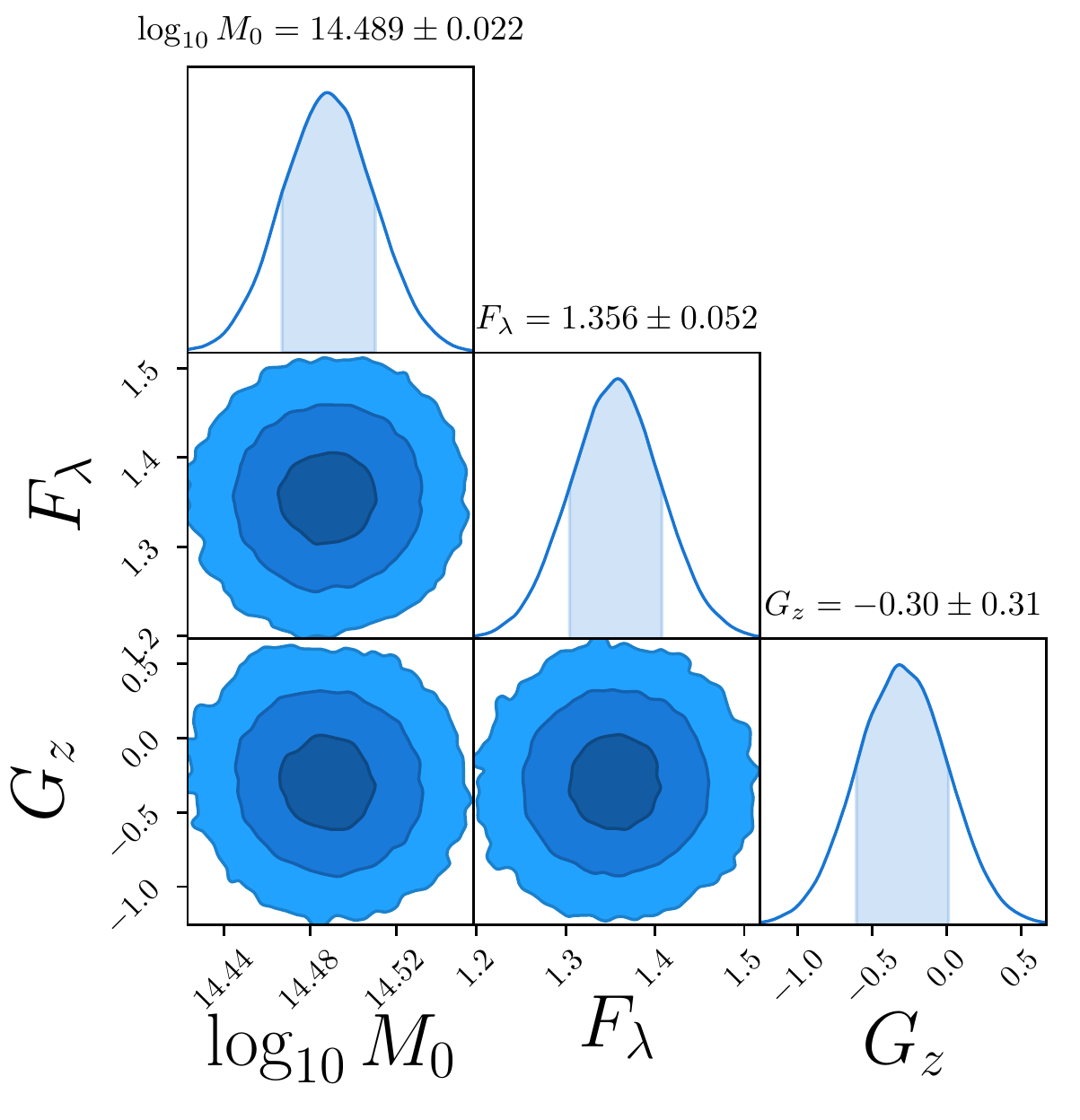}
  \caption{Parameters of the $M$--$\lambda$--$z$ relation. Contours show the $1\sigma$, $2\sigma$ and $3\sigma$ confidence areas from the \texttt{Full} run.}
  \label{fig:corner2}
\end{figure}

We sample the posterior of the MOR parameters using \textit{emcee} with 48 walkers taking 10000 steps each, discarding the first 1000 steps of each walker as burn-in. \autoref{tab:mass_richness_parameters} summarizes the posteriors of our model parameters, while \autoref{fig:corner2} shows the corresponding confidence contours. All parameters in the $M$--$\lambda$--$z$ relation have flat priors.

We explicitly enforce correlated uncertainties of shear, \photoz, modeling systematics, and triaxiality and projection effects. Miscentering and boost factors are considered independent across cluster subsets. These independent uncorrelated systematics will tend to average out across bins. 

In order to distinguish between the systematic and statistical contribution to the error budget on the $M$--$\lambda$--$z$ relation parameters, we repeat the analysis using the statistical errors from the \verb|OnlyMc| run. That is, we calculate \autoref{eq:mass_richness_likelihood} using only the uncertainties measured from the \verb|OnlyMc| run, or the $\mathsf{C}^{\rm stat}_M$ covariance matrix. The central values of the measured masses from the \verb|OnlyMc| run are nearly identical to the \verb|Full| run, as are the parameters in the $M$--$\lambda$--$z$ relation. The difference in quadrature between the two uncertainties represents the systematic contribution while the excess uncertainty from the \verb|OnlyMc| run is the statistical contribution. These uncertainties are reported in \autoref{tab:mass_richness_parameters}.

Our results imply that galaxy clusters of richness $\lambda = 40$ at redshift $z=0.35$ have a mean mass of $\log_{10} \calM = 14.489\ \pm 0.011\ {\rm (stat)}\ \pm 0.019\ {\rm (sys)}$. The richness scaling is slightly steeper than linear at $F_\lambda = 1.356 \pm 0.051\ {\rm (stat)}\ \pm 0.008\ {\rm (sys)}$, while the mass shows a weak redshift dependence of $G_z = -0.30 \pm 0.30\ {\rm (stat)}\ \pm 0.06\ {\rm (sys)}$ consistent with no evolution. This amounts to a 5.0 per cent calibration (2.4 per cent statistical, 4.3 per cent systematic), of the normalization of the $M$--$\lambda$--$z$ relation.

In \citet{rmsva}, we found that the dominant systematic uncertainty stemmed from shear and \photoz\ systematics, as was the case in \citet{Simet2017}. By repeating our analysis with the \verb|FixedAm| run, which includes all systematics {\it except} $\mathcal{A}_m$, we are able to quantify the contribution from these sources. We found that the posterior distributions from the $M$--$\lambda$--$z$ relation are significantly reduced, and that shear and \photoz\ systematics alone account for 48 per cent of the systematic uncertainty. This means that the remaining 52 per cent of the systematic uncertainty is due to modeling systematics, projection effects, and cluster triaxiality.

\begin{table}
\setlength{\tabcolsep}{.6em}
	\caption{Parameters of the $M$--$\lambda$--$z$ relation from \autoref{eq:mass_richness_likelihood} with their posteriors. The mass is defined as $M_{200{\rm m}}$ in units of $\msun$. The pivot richness and pivot redshift correspond to the median values of the cluster sample. Uncertainties are the 68 per cent confidence intervals and are split into statistical (first) and systematic (second).}
	\label{tab:mass_richness_parameters}
    \begin{tabular}{llll}
    	Parameter & Description & Posterior \\ \hline
        $\log_{10}M_0$ & Mass pivot & $14.489 \pm 0.011 \pm 0.019$\\
		$F_\lambda$ & Richness scaling & $1.356 \pm 0.051 \pm 0.008$\\
		$G_z$ & Redshift scaling & $-0.30 \pm 0.30 \pm 0.06$\\
    \end{tabular}
\end{table}

\begin{figure}
  \includegraphics[width=\linewidth]{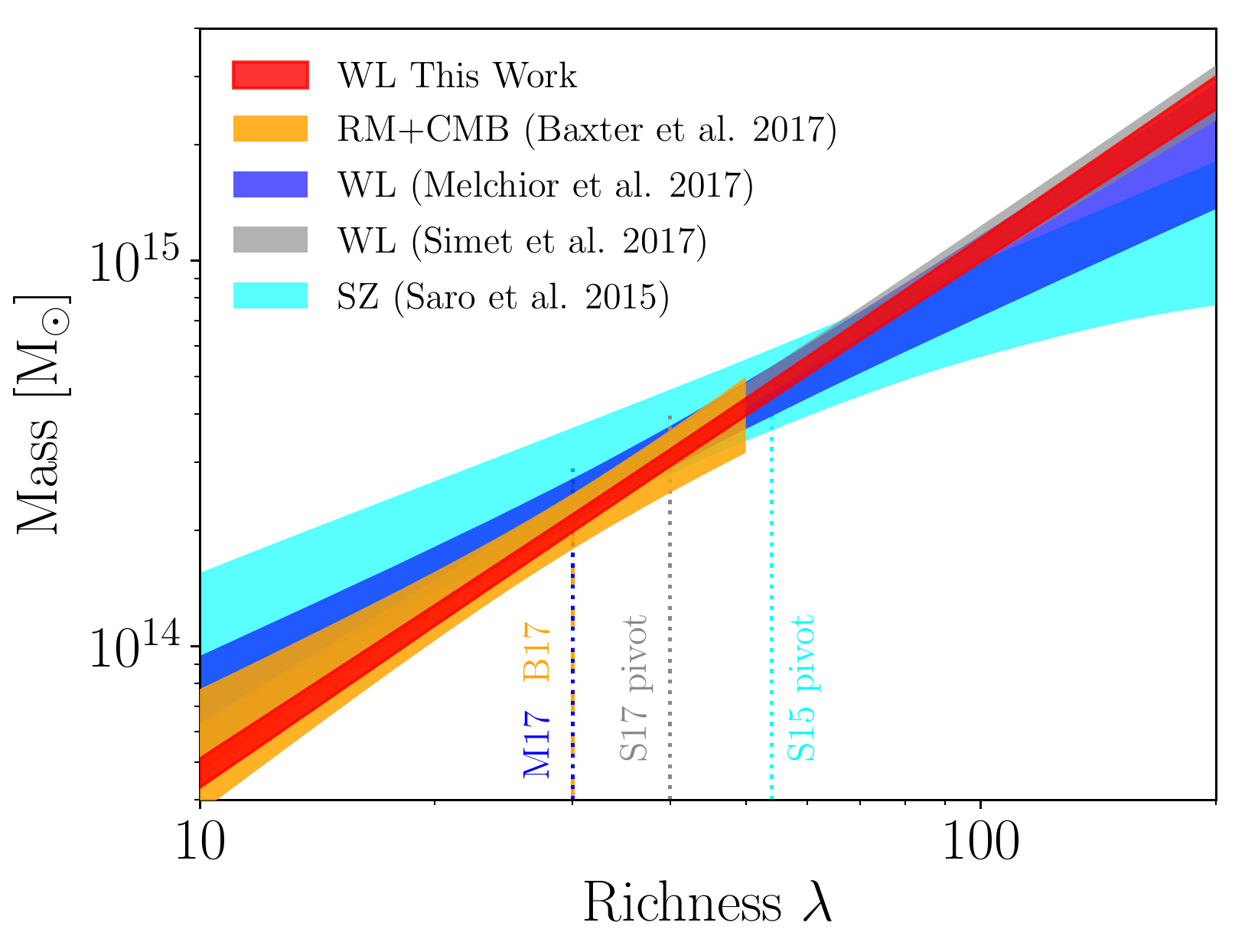}
  \caption{Best fit model for $M$--$\lambda$ relation evaluated at the pivot redshift of our model, $z_0=0.35$, compared to other measurements. Our pivot richness is at $\lambda_0=40$. The previous DES result is in blue, from \citet{rmsva}, while the relation measured in this analysis is in red. The analysis by \citet{Baxter2018} in orange used the same clusters as this work and found a consistent scaling relation over the richness range it probed.}
  \label{fig:mass_richness}
\end{figure}

\subsubsection{Alternative model using $\xi_{\rm lin}$}
\label{sec:alternative_model}

\citet{Hayashi08} used a similar model to ours, but with the \emph{linear} matter correlation function for their 2-halo term. This causes very different behavior near the 1-halo to 2-halo transition region, which can affect the fitting procedure, as discussed in \cite{rmsva}. We repeated our entire analysis, including recomputing the calibration, using $\xi_{\rm lin}$ in place of $\xi_{\rm nl}$. The masses of the stacks changed by less than 1 per cent, as did the normalization of the $M$--$\lambda$ relation $\log_{10}M_0$. This means that our approach of calibrating the masses is largely robust to our choice of model.

\subsubsection{Additional tests}
\label{sec:additional_tests}

We performed additional tests to verify our results. To ensure against possible small-scale systematic effects, we repeated our analysis with a more conservative radial cut of 500 kpc rather than 200 kpc. The resulting $M$--$\lambda$--$z$ relation changed only in the mass scale, with $M_0$ changing by 0.2$\sigma$.

We also tested against possible differences in modeling systematics between large and small scales. By dividing each $\Delta\Sigma$ profile at 2 Mpc into large and small scale samples we could fit these regimes independently. While the constraining power was greatly diminished, the recovered masses were consistent with each other and the fiducial value within errors. No trend was observed in the differences between the recovered masses in any of these tests compared to the fiducial masses in \autoref{tab:posterior_masses}.

Lastly, we tested an extension of \autoref{eq:mass_richness_redshift} where $F_\lambda(z) = F_{\lambda,0} + zF_{\lambda,1}$ and found $F_{\lambda,1}$ consistent with 0 at the 1.2$\sigma$ level. Therefore, if any redshift evolution exists in the richness scaling, we are unable to resolve the behavior at present.


\section{Comparison to Results in the Literature}
\label{sec:comparisons}

We compare our calibration of the $M$--$\lambda$ relation to previous results from the literature. The specific richness--mass relations we consider are summarized in \autoref{tab:scaling_relations}, and we describe below the origin of each of these.
\begin{itemize}
\item \citet{rmsva} was the precursor to this analysis. In that work, we calibrated the mass--richness relation of \redmapper\ clusters in the DES Science Verification data. A detailed description of the changes between that analysis and this one appears in the next section. 
\item \citet{Baxter2018} used the lensing of the Cosmic Microwave Background as measured by the South Pole Telescope to measure the mass--richness relation of DES Y1 \redmapper\ clusters. Their analysis focused on 7066 clusters with richness $20 \leq \lambda \leq 40$. The upper limit was set to avoid potential biases in the recovered masses from contamination by thermal Sunyuaev-Zel'dovich emission by the clusters. 
\item \citet{Simet2017} measured the mass--richness relation of \redmapper\ clusters found in the Sloan Digital Sky Survey (SDSS). While their analysis is similar in spirit to ours, there are numerous methodological differences, including modeling choices (Simet et al. only fit the 1-halo term in the lensing profile), different radial scales used in the fit, a different shape catalog, and different photometric redshift catalogs.
\item \citet{Murata2018} measured the richness--mass relation of SDSS \redmapper\ clusters assuming a {\it Planck} cosmology. We compute the mean mass at $\lambda=40$ as well as the local slope at this point in the scaling relation. As demonstrated in \citet{Murata2018}, their work and \citet{Simet2017} are consistent with each other, despite the fact that they used different models for $\Delta\Sigma$, different radial scales and slightly different richness bins. Of special note is the fact that while \citet{Simet2017} modeled only the 1-halo term using an NFW profile (along with a calibration step to correct for any biases introduced by this choice), \citet{Murata2018} used an emulator approach to simultaneously model the 1-halo and 2-halo terms of the lensing profile. The authors constrained the richness--mass relation using both lensing and cluster abundance data, and the use of the emulator effectively fixed the concentration--mass relation. These differences add significant information relative to a lensing-only analysis. Finally, the posteriors we had available did not include the effects of \photoz\ or shear uncertainty in the error budget.  Together, these difference result in error bars that are tighter than our own.
\item \citet{Baxter2016} analyzed the cluster clustering of SDSS \redmapper\ clusters. By measuring the angular correlation function of clusters they were able to constrain the amplitude of the mass scaling relation to 18 per cent, in which their dominant systematic was uncertainty in the bias--mass relation.
\item \citet{farahi16} measured masses using stacked pairwise velocity dispersion measurements of SDSS \redmapper\ clusters. Their measurements serve as a good cross check against other analyses of SDSS clusters, but found that they are ultimately less precise due to large uncertainties in velocity bias.
\item \citet{Saro2015} measured the mass--richness relation of galaxy clusters by assuming a {\it Planck} cosmology to determine the observable--mass relation of clusters from the South Pole Telescope \citep{Bleem2015}. They then matched these SPT clusters to \redmapper\ clusters from the DES Science Verification data, and use the overlap sample to determine the richness--mass relation. We invert the relation using the method of \citet{Evrard2014} in order to show the comparison in \autoref{fig:mass_richness}.
\item \citet{Mantz2016} compared the scaling relation measured from the Weighting the Giants mass estimates for individual \redmapper\ clusters in SDSS from \citet{Applegate2014} to that of the \citet{Simet2017} analysis. They found the two scaling relations in good agreement, which is also the case when compared to our measurement.
\item \citet{GeachPeacock2017} constrained the mass--richness relation of \redmapper\ clusters found in SDSS using convergence profiles measured from {\it Planck} data. They constrain the normalization of the scaling relation at the $\sim 11.5\%$ level, but are unable to reach similar precision for the scaling index. Mass calibrations from this type of measurement are expected to improve in the future as both optical cluster catalogs expand and CMB lensing maps improve.
\item \citet{vanUitert2016} focused on significantly lower richness clusters found with a different algorithm. For these reasons, a direct comparison is not possible, however they were able to constrain the mass--richness scaling relation at low redshifts for their cluster finder at the ~5\% level using lensing and cluster-satellite correlations.
\end{itemize}

\autoref{tab:scaling_relations} summarizes these scaling relations. Critically, the richness definition $\lambda$ is sensitive to the details of image processing, source detection, choice of magnitudes, etc, and can therefore vary systematically from one survey to the next. We explicitly correct for this impact cross-matching DES Y1 clusters to DES SV and SDSS \redmapper\ clusters, and measuring the richness offset.  We find
\begin{eqnarray}
	\label{eq:richness_conversion}
	\lambda_{\rm DES\ SV} & = & (1.08 \pm 0.16) \lambda_{\rm DES\ Y1} \\
	\lambda_{\rm SDSS} & = & (0.93 \pm 0.14) \lambda_{\rm DES\ Y1}\,.
\end{eqnarray}
In these equations, the error is the standard deviation in the richness ratio, not the error on the mean. We apply these corrections to the SDSS and DES SV scaling relations before comparing to our result. So, for instance, if the scaling relation for data set $X$ takes the form 
\begin{equation}
	\avg{M|\lambda_X} = A\lambda_X^\alpha
\end{equation}
and the ratio $\lambda_X/\lambda_{\rm DES\ Y1}=r$, then the scaling relation for Y1 richnesses is
\begin{equation}
\avg{M|\lambda_{\rm DES\ Y1}} = A r^\alpha \lambda_{\rm DES\ Y1}^\alpha .
\end{equation}
Finally, all scaling relations that do not explicitly incorporate redshift evolution are transported from their quoted pivot redshift to our chosen pivot redshift $z=0.35$ using our best fit redshift evolution.

\begin{figure*}
  \includegraphics[width=\linewidth]{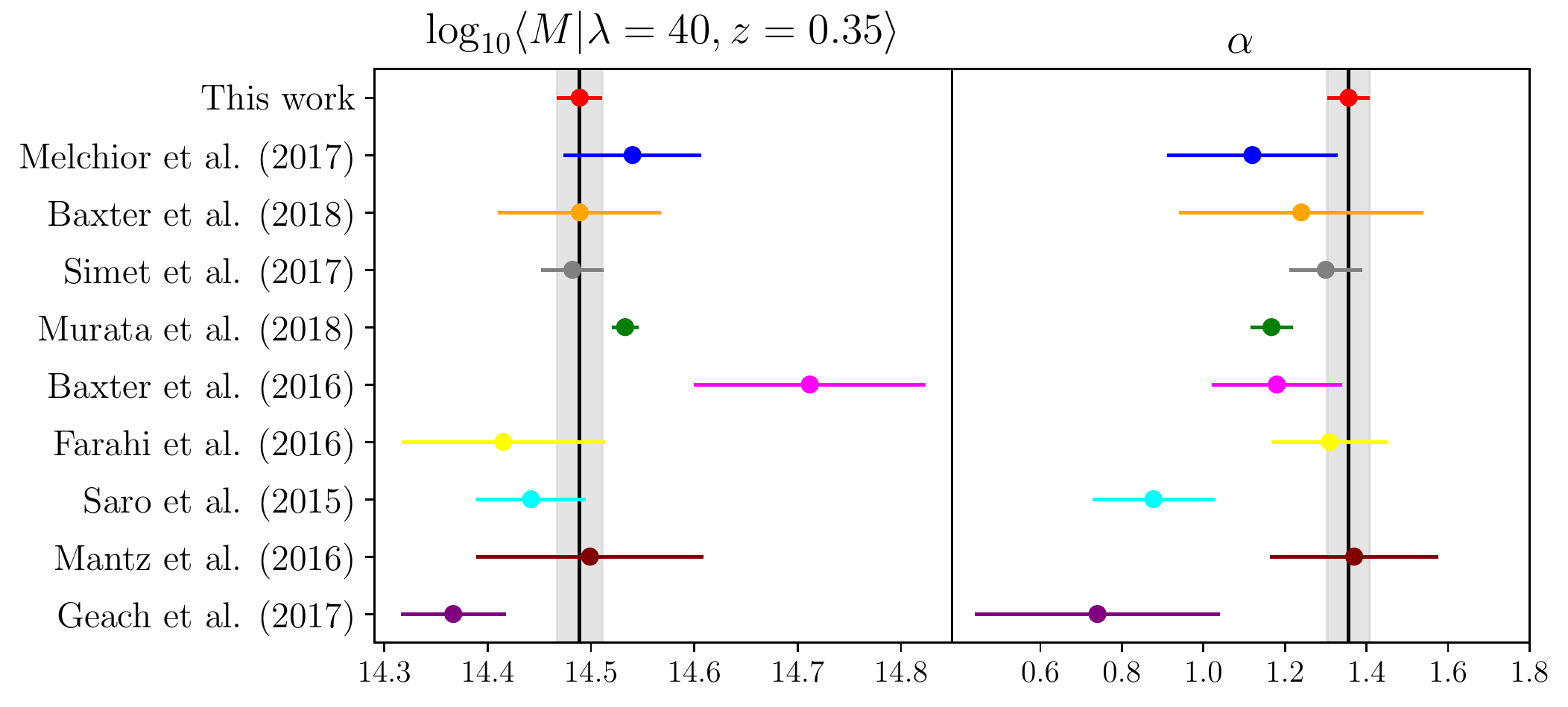}
  \caption{Comparison of the predicted mass at $\lambda=40$ and $z=0.35$ as well as the richness scaling relation between this work ({\it gray bands}) and other results from the literature.}
  \label{fig:comparison_summary}
\end{figure*}

\autoref{fig:comparison_summary} and \autoref{tab:scaling_relations} show the mass at $\lambda=40$ and $z=0.35$ as well as the richness scaling index for each of the scaling relations described above.

\begin{table*}
\caption{\redmapper\ scaling relation comparisons from the literature. Of note, the \citet{Simet2017} results have changed slightly (Simet, private communication). We evaluate $\log_{10}\langle M|\lambda=40, z=0.35 \rangle$ of the other scaling relations in order to compare them to our result, applying the richness correction given by \autoref{eq:richness_conversion}. When necessary, we use the method presented in \citet{Evrard2014} to convert from richness--mass relations to mass--richness relations. All masses are $M_{\rm 200m}$.}
\begin{tabular}{llll}
Authors & Description & $\log\langle M|\lambda=40,z=0.35 \rangle\ [{\rm M}_\odot]$ & Richness scaling index $F_\lambda$ \\ \hline
This work & weak lensing calibration using DES Y1 & $14.489 \pm 0.022$ & $1.356 \pm 0.052$\\
\citet{rmsva} & weak lensing calibration using DES SV & $14.540 \pm 0.067$ & $1.12 \pm 0.21$ \\
\citet{Baxter2018} & CMB lensing calibration using DES Y1 & $14.49 \pm 0.31$ & $1.24 \pm 0.30$ \\
\citet{Baxter2016} & cluster clustering using SDSS & $14.7 \pm 0.1$ & $1.18 \pm 0.16$ \\
\citet{Simet2017} & weak lensing calibration using SDSS & $14.48 \pm 0.03$ & $1.30 \pm 0.09$\\
\citet{Murata2018} & weak lensing calibration using SDSS & $14.533 \pm 0.013$ & $1.167 \pm 0.052$ \\
\citet{farahi16} & pairwise velocity dispersion using SDSS & $14.42 \pm 0.10$ & $1.31 \pm 0.14$ \\
\citet{Saro2015} & SZE mass calibration using SPT and DES SV & $14.44 \pm 0.05$ & $0.91 \pm 0.18$ \\
\citet{Mantz2016} & weak lensing of individual WtG clusters & $14.42 \pm 0.11$ & $1.36 \pm 0.21$ \\
\citet{GeachPeacock2017} & CMB lensing calibration using SDSS & $14.37 \pm 0.05$ & $0.74 \pm 0.3$\\
\end{tabular}
\label{tab:scaling_relations}
\end{table*}

\section{Systematic Improvements From DES SV to DES Year 1, and from Year 1 to Year 5}
\label{sec:future_improvements}

Our current analysis has multiple significant improvements relative to \citet{rmsva}, the precursor to this work.  Specifically:
\begin{itemize}
\item Shear calibration related errors on mass decreased from 4 per cent to 1.7 per cent, based primarily on the data driven correction of shear biases with \textsc{metacalibration}. This implies that the shear calibration uncertainty is no longer the dominant source of systematic error in our weak lensing analysis.  
\item The largest contribution to the systematic uncertainty is now \photoz\ errors. In the purely COSMOS-based calibration applied in this work, we find only a minimal improvement between SV and Y1.
\item The $\approx15$ per cent increase in uncertainty due to using noisy jackknife estimates of the covariance matrix \citep{Dodelson2013} of the weak lensing profiles was entirely removed through the use of a semi-analytic covariance matrix. 
\item Uncertainty from modeling systematics decreased from 2 per cent to 0.73 per cent. In \citet{rmsva}, the model calibration corrections were computed by stacking halos in mass bins. By contrast, our current analysis  assigns richness according to a fiducial richness--mass relation, allowing us to stack in richness bins and to therefore accurately compute the correction for a richness bin. Relative to our SV analysis, the amplitude of this correction increased, while its uncertainty was reduced, from $1.00 \pm 0.02$ to $1.042 \pm 0.004$. The increase in the correction is primarily due to the richness binning of the halos as well as the fact that by using the Y1 covariance matrix, the impact of small differences on large scales between our analytical model and numerical simulations is amplified, leading to larger correction factors. More importantly, however, the uncertainty on this correction was greatly reduced. 
This is due to the semi-analytic covariance matrix as well as allowing for intrinsic scatter of the calibration factors. While we use the semi-analytic covariance matrix for calibration on the simulated profiles, the matrix does not adequately describe the uncertainty in any systematic differences between the model and real data. Additionally, we now fit for the associated systematic uncertainty from the dispersion in the calibration data. Both of these factors result in a decrease in the systematic error budget.
\item The mass--concentration relation of galaxy clusters is allowed to float in this analysis, while it was held fixed in \citet{rmsva}.  The fixed mass--concentration relation in our SV analysis was necessitated by the relatively low S/N of the weak lensing measurements. By contrast, our current analysis enables us to marginalize over concentration, which in turn should make our results significantly more robust to the impact of baryonic physics in the cores of galaxy clusters (see section \ref{sec:modeling_systematics}).
\end{itemize}

All in all, the reduced statistical and systematic uncertainty in our analysis has reduced the error in the amplitude of the mass--richness relation from 11.2 per cent to 5.0 per cent.  Unlike our analysis in \citet{rmsva}, our current constraints are close to systematics limited.  Without improved systematics between now and the end of the survey, the improved statistics of the Year 5 data will only decrease our total error budget from 5.0 per cent to $\approx 4$ per cent. Evidently, further reducing systematic uncertainties in future weak lensing mass calibration analysis is imperative.

Photometric redshift errors currently dominate the systematic error budget.  Significant improvements in the weak lensing mass calibration of galaxy clusters will require new algorithms for calibrating photometric redshifts. A joint constraint from high-fidelity \photozs\ of matched reference samples and clustering redshifts \citep{Gatti2017,Davis2017}, as done in \citet{Y1pz}, is not feasible for higher redshift sources due to the limited redshift range of available spectroscopic or \redmagic{} catalogs. Alternatively, source selection criteria that take into consideration photometric redshift uncertainty could lead to a desirable trade off between statistical and systematic uncertainty. With some combination of these approaches, reducing the photometric redshift uncertainty by a factor of two for a year 5 analysis seems plausible.

Related to this, in our current analysis we have assumed that all systematics are perfectly correlated across all redshift and richness bins. This is likely {\it too} conservative. In particular, photometric redshift systematics are unlikely to be perfectly correlated across all redshift bins: the selection of the source population, and their relative signal contribution as a function of source redshift, differ as a function of lens redshift.  Adequately characterizing the covariance in the systematic uncertainties associated with photometric redshift errors seems like a relatively simple way to significantly decrease our systematic error budget. For instance, if one were to assume that the photometric redshift systematics were entirely uncorrelated, the associated systematic would be reduced by a factor of $1/\sqrt{3}$, rendering photometric redshift errors sub-dominant. This is clearly unrealistic, but it does illustrate that characterizing the covariance in the systematics may lead to significant reductions in the total error budget.

Following photometric redshift uncertainties, three different effects come in at the $\approx 2$ per cent level: shear systematics, triaxiality effects, and projection effects. Of these, shear systematics are the least problematic. We fully expect shear calibration uncertainties will continue to decrease over the coming years, and they will no longer be a major source of error for cluster mass calibration. By contrast, the current systematic error estimates for triaxiality and projection effects clearly demonstrate that there is a significant need for a detailed study of these on weak lensing mass profiles, such as in the recent work of \citet{Osato2017}.  

Additional, but less urgent, upgrades to our analysis are also possible. For instance, following \citet{Murata2018}, an emulator based approach to modeling the halo-matter correlation function or $\Delta\Sigma$ directly can potentially greatly reduce the modeling calibration and its contribution to the uncertainty. Centering errors will also continue to decrease as the availability of multi-wavelength data continues to increase.

Finally, systematics that we have thus far ignored need to be better addressed. For instance, intrinsic alignment by cluster member galaxies even if its effect is very small \citep{Sifon2015}, which impact membership dilution estimates. Likewise, a study of the impact on baryonic physics on our weak lensing calibration methodology is necessary. While we expect these sources of error to be subdominant in our present study, quantifying the systematic error associated with these effects will be increasingly important in the future.


\section{Summary and Conclusions}
\label{sec:summary}

We measured the stacked weak lensing signal of \redmapper\ clusters in the DES Y1 data. The clusters were divided into 21 subsets of richness and redshift. The mean mass of each cluster stack was estimated for those subsets with $\lambda \geq 20$ and $0.2 \leq z \leq 0.65$. Our model incorporated:
\begin{itemize}[label=\textbullet, leftmargin=*]
	\item Shear measurement systematics (\autoref{sec:shear_systematics}),
    \item Source photometric redshift uncertainties (\autoref{sec:photometric_redshift_systematics}),
    \item Source sample dilution by cluster members (\autoref{sec:boost_factors}, \autoref{sec:boost_factor_model}),
    \item Cluster miscentering (\autoref{sec:miscentering_correction}),
    \item Model calibration systematics (\autoref{sec:stacked_mass_correction}),
    \item Triaxiality \& projection effects (\autoref{sec:triaxiality_and_projection_effects}).
\end{itemize}
The mean masses of the cluster subsets were used to determine the mean cluster mass as a function of richness and redshift according to \autoref{eq:mass_richness_redshift}. We emphasize that the full analysis was performed blindly: the paper underwent internal review by the DES collaboration prior to unblinding, and no changes to the analysis were made post-unblinding. 

We summarize our constraints on the scaling relation as follows:
for clusters at our pivot richness of $\lambda_0=40$ and pivot redshift of $z_0=0.35$, the mean cluster mass is
\begin{equation}
	M_0 = [3.081 \pm 0.075 \pm 0.133] \cdot 10^{14}\ \msun.
\end{equation}
The slope $F_\lambda$ for the mass--richness relation is
\begin{equation}
	F_\lambda = 1.356 \pm 0.051 \pm 0.008,
\end{equation}
and the slope $G_z$ governing the redshift evolution of the mass--richness relation is
\begin{equation}
	G_z = -0.30 \pm 0.30 \pm 0.06,
\end{equation}
where the first and second set of errors correspond to statistical and systematic errors, respectively.  The full scaling relation is given by \autoref{eq:mass_richness_redshift}.
This scaling relation is in excellent agreement with, while being significantly more precise and accurate than, previous results from the literature: \citet{Mantz2016, Saro2015,Simet2017,Baxter2018,Murata2018,rmsva}. 

The 5.0 per cent constraint on the amplitude of the mass--richness relation is systematics dominated, with our systematic error alone reaching 4.3 per cent. We stress the systematic uncertainty in the shear and photometric redshift catalogs have been extensively tested and validated, so we are confident our systematic error budget is robust. Halo triaxiality and line of sight projections are now an important contributor to the total systematic error, and represent a critical path for minimizing the overall error budget for future analyses

Mass calibration remains the limiting factor for the ability of testing cosmological models with cluster counts.  Nevertheless, this work represents a significant step forward: we were able to reduce the systematic error budget from 6.1 per cent in DES Science Verification to 4.3 per cent in DES Year 1. While we will need to achieve similar level of improvements for future analyses including DES Year 5 and LSST Year 1 to significantly improve upon our results, we are confident that we will be able to rise to the challenge: the story of weak lensing mass calibration is one of ever decreasing systematic errors, a trend that to this day shows no signs of abating. 

\begin{table*}
	\caption{Systematic error budget on the amplitude of the mass--richness relation as measured with the DES Y1 data compared to the DES SV result of \citet{rmsva}. The shear (\autoref{sec:shear_systematics}), \photoz\ (\autoref{sec:photometric_redshift_systematics}), modeling systematics (\autoref{sec:modeling_systematics}), triaxiality, and projection effects (\autoref{sec:triaxiality_and_projection_effects}) are conservatively taken to be perfectly correlated between cluster stacks. Membership dilution (\autoref{sec:boost_factors}) and miscentering (\autoref{sec:miscentering_correction}) are assumed to be independent. Statistical and systematic errors are added in quadrature to arrive at the total error.}
	\label{tab:error_budget}
	\setlength{\tabcolsep}{1.5em}
	\begin{tabular}{lll}
		Source of systematic & SV Amplitude uncertainty & Y1 Amplitude Uncertainty \\
		\hline
		Shear measurement                  &  4\%        & 1.7\% \\
		Photometric redshifts              &  3\%        & 2.6\% \\ 
		Modeling systematics               &  2\%        & 0.73\% \\
		Cluster triaxiality                &  2\%        & 2.0\% \\
		Line-of-sight projections          &  2\%        & 2.0\% \\
		Membership dilution + miscentering & $\leq 1\%$  & 0.78\% \\
		\hline
		{\bf Total Systematics} & 6.1\% & 4.3\% \\
		{\bf Total Statistical} & 9.4\% & 2.4\% \\
		{\bf Total} & 11.2\% & 5.0\% \\
	\end{tabular}
\end{table*}

\section*{Acknowledgments}

This paper has gone through internal review by the DES collaboration. TM and ER are supported by DOE grant DE-SC0015975. TNV and SS are supported by the SFB-Transregio 33 `The Dark Universe' by the Deutsche Forschungs\-gemeinschaft (DFG) and the DFG cluster of excellence `Origin and Structure of the Universe'. ER acknowledges additional support by the Sloan Foundation, grant FG-2016-6443. Support for DG was provided by NASA through Einstein Postdoctoral Fellowship grant number PF5-160138 awarded by the Chandra X-ray Center, which is operated by the Smithsonian Astrophysical Observatory for NASA under contract NAS8-03060.
TJ, DH, and SE are supported by DOE grant DE-SC0010107.
KR and SB acknowledge support from the UK Science and Technology Facilities Council via Research Training Grants ST/P000252/1 and ST/N504452/1, respectively. CV is supported by the Mexican National Council for Science and Technology via grant number 411117.
AvdL is supported by DoE grant DE-SC0018053.

Part of our computations have been carried out on the computing facilities of the Computational Center for Particle and Astrophysics (C2PAP). A portion of this research was carried out at the Jet Propulsion Laboratory, California Institute of Technology, under a contract with the National Aeronautics and Space Administration.

Funding for the DES Projects has been provided by the U.S. Department of Energy, the U.S. National Science Foundation, the Ministry of Science and Education of Spain, 
the Science and Technology Facilities Council of the United Kingdom, the Higher Education Funding Council for England, the National Center for Supercomputing 
Applications at the University of Illinois at Urbana-Champaign, the Kavli Institute of Cosmological Physics at the University of Chicago, 
the Center for Cosmology and Astro-Particle Physics at the Ohio State University,
the Mitchell Institute for Fundamental Physics and Astronomy at Texas A\&M University, Financiadora de Estudos e Projetos, 
Funda{\c c}{\~a}o Carlos Chagas Filho de Amparo {\`a} Pesquisa do Estado do Rio de Janeiro, Conselho Nacional de Desenvolvimento Cient{\'i}fico e Tecnol{\'o}gico and 
the Minist{\'e}rio da Ci{\^e}ncia, Tecnologia e Inova{\c c}{\~a}o, the Deutsche Forschungsgemeinschaft and the Collaborating Institutions in the Dark Energy Survey. 

The Collaborating Institutions are Argonne National Laboratory, the University of California at Santa Cruz, the University of Cambridge, Centro de Investigaciones Energ{\'e}ticas, 
Medioambientales y Tecnol{\'o}gicas-Madrid, the University of Chicago, University College London, the DES-Brazil Consortium, the University of Edinburgh, 
the Eidgen{\"o}ssische Technische Hochschule (ETH) Z{\"u}rich, 
Fermi National Accelerator Laboratory, the University of Illinois at Urbana-Champaign, the Institut de Ci{\`e}ncies de l'Espai (IEEC/CSIC), 
the Institut de F{\'i}sica d'Altes Energies, Lawrence Berkeley National Laboratory, the Ludwig-Maximilians Universit{\"a}t M{\"u}nchen and the associated Excellence Cluster Universe, 
the University of Michigan, the National Optical Astronomy Observatory, the University of Nottingham, The Ohio State University, the University of Pennsylvania, the University of Portsmouth, 
SLAC National Accelerator Laboratory, Stanford University, the University of Sussex, Texas A\&M University, and the OzDES Membership Consortium.

Based in part on observations at Cerro Tololo Inter-American Observatory, National Optical Astronomy Observatory, which is operated by the Association of 
Universities for Research in Astronomy (AURA) under a cooperative agreement with the National Science Foundation.

The DES data management system is supported by the National Science Foundation under Grant Numbers AST-1138766 and AST-1536171.
The DES participants from Spanish institutions are partially supported by MINECO under grants AYA2015-71825, ESP2015-66861, FPA2015-68048, SEV-2016-0588, SEV-2016-0597, and MDM-2015-0509, 
some of which include ERDF funds from the European Union. IFAE is partially funded by the CERCA program of the Generalitat de Catalunya.
Research leading to these results has received funding from the European Research
Council under the European Union's Seventh Framework Program (FP7/2007-2013) including ERC grant agreements 240672, 291329, and 306478.
We  acknowledge support from the Australian Research Council Centre of Excellence for All-sky Astrophysics (CAASTRO), through project number CE110001020, and the Brazilian Instituto Nacional de Ci\^encia
e Tecnologia (INCT) e-Universe (CNPq grant 465376/2014-2).

This manuscript has been authored by Fermi Research Alliance, LLC under Contract No. DE-AC02-07CH11359 with the U.S. Department of Energy, Office of Science, Office of High Energy Physics. The United States Government retains and the publisher, by accepting the article for publication, acknowledges that the United States Government retains a non-exclusive, paid-up, irrevocable, world-wide license to publish or reproduce the published form of this manuscript, or allow others to do so, for United States Government purposes.

This work used simulations and computations performed using computational resources at SLAC and at NERSC.

\bibliographystyle{mn2e_adsurl}
\bibliography{astroref}

\begin{thebibliography}{}
  \providecommand{\doi}[1]{\href{http://dx.doi.org/#1}{doi:#1}}
  \providecommand{\eprint}[1]{\href{http://arxiv.org/abs/#1}{arXiv:#1}}

\bibitem[\protect\citeauthoryear{Annis et~al.,}{Annis et~al.}{2014}]{S82}
\href{http://stacks.iop.org/0004-637X/794/i=2/a=120}{Annis J.  et~al., 2014,
  The Astrophysical Journal, 794, 120}

\bibitem[\protect\citeauthoryear{{Angulo} et~al.,}{{Angulo}
  et~al.}{2012}]{anguloetal12}
\href{http://adsabs.harvard.edu/abs/2012MNRAS.426.2046A}{{Angulo} R.~E.,
  et~al., 2012, \mnras, 426, 2046}

\bibitem[\protect\citeauthoryear{{Applegate} et~al.,}{{Applegate}
  et~al.}{2014a}]{applegateetal14}
\href{http://adsabs.harvard.edu/abs/2014MNRAS.439...48A}{{Applegate} D.~E.
  et~al., 2014a, \mnras, 439, 48}

\bibitem[\protect\citeauthoryear{{Applegate} et~al.,}{{Applegate}
  et~al.}{2014b}]{Applegate2014}
\href{http://adsabs.harvard.edu/abs/2014MNRAS.439...48A}{{Applegate} D.~E.
  et~al., 2014b, \mnras, 439, 48}

\bibitem[\protect\citeauthoryear{{Baxter} et~al.,}{{Baxter}
  et~al.}{2016}]{Baxter2016}
\href{http://adsabs.harvard.edu/abs/2016arXiv160400048B}{{Baxter} E.~J.
  et~al., 2016, arXiv:1604.00048}

\bibitem[\protect\citeauthoryear{{Baxter} et~al.,}{{Baxter}
  et~al.}{2018}]{Baxter2018}
\href{http://adsabs.harvard.edu/abs/2018MNRAS.tmp..309B}{{Baxter} E.~J.
  et~al., 2018, \mnras}

\bibitem[\protect\citeauthoryear{{Behroozi}, {Wechsler} \& {Wu}}{{Behroozi}
  et~al.}{2013}]{Behroozi2013}
\href{http://adsabs.harvard.edu/abs/2013ApJ...762..109B}{{Behroozi} P.~S.,
  {Wechsler} R.~H.,    {Wu} H.-Y.,  2013, \apj, 762, 109}

\bibitem[\protect\citeauthoryear{{Ben{\'{\i}}tez}}{{Ben{\'{\i}}tez}}{2000}]{Benitez2000}
\href{http://adsabs.harvard.edu/abs/2000ApJ...536..571B}{{Ben{\'{\i}}tez} N.,
  2000, \apj, 536, 571}

\bibitem[\protect\citeauthoryear{{Bhattacharya} et~al.,}{{Bhattacharya}
  et~al.}{2013}]{Bhattacharya2013}
\href{http://adsabs.harvard.edu/abs/2013ApJ...766...32B}{{Bhattacharya} S.
  et~al., 2013, \apj, 766, 32}

\bibitem[\protect\citeauthoryear{{Blas}, {Lesgourgues} \& {Tram}}{{Blas}
  et~al.}{2011}]{Blas11CLASS2}
\href{http://adsabs.harvard.edu/abs/2011JCAP...07..034B}{{Blas} D.,
  {Lesgourgues} J.,    {Tram} T.,  2011, \jcap, 7, 034}

\bibitem[\protect\citeauthoryear{{Bleem} et~al.,}{{Bleem}
  et~al.}{2015}]{Bleem2015}
\href{http://adsabs.harvard.edu/abs/2015ApJS..216...27B}{{Bleem} L.~E.  et~al.,
  2015, \apjs, 216, 27}

\bibitem[\protect\citeauthoryear{{Carlstrom} et~al.,}{{Carlstrom}
  et~al.}{2011}]{Carlstrom11.1}
\href{http://adsabs.harvard.edu/abs/2011PASP..123..568C}{{Carlstrom} J.~E.
  et~al., 2011, \pasp, 123, 568}

\bibitem[\protect\citeauthoryear{{Chang} et~al.,}{{Chang}
  et~al.}{2018}]{Chang2017}
\href{http://adsabs.harvard.edu/abs/2018MNRAS.475.3165C}{{Chang} C.  et~al.,
  2018, \mnras, 475, 3165}

\bibitem[\protect\citeauthoryear{{Coe} et~al.,}{{Coe} et~al.}{2006}]{coe06}
\href{http://adsabs.harvard.edu/abs/2006AJ....132..926C}{{Coe} D.  et~al.,
  2006, \aj, 132, 926}

\bibitem[\protect\citeauthoryear{{DES Collaboration} et~al.,}{{DES
  Collaboration} et~al.}{2017}]{desy1kp}
\href{http://adsabs.harvard.edu/abs/2017arXiv170801530D}{{DES Collaboration}
  et~al., 2017, arXiv:1708.01530}

\bibitem[\protect\citeauthoryear{{Dark Energy Survey Collaboration}}{{Dark
  Energy Survey Collaboration}}{2016}]{DESCosmicShearCosmology}
\href{http://adsabs.harvard.edu/abs/2016PhRvD..94b2001A}{{Dark Energy Survey
  Collaboration} 2016, \prd, 94, 022001}

\bibitem[\protect\citeauthoryear{{Davis} et~al.,}{{Davis}
  et~al.}{2017}]{Davis2017}
\href{http://adsabs.harvard.edu/abs/2017arXiv170708256D}{{Davis} C.  et~al.,
  2017, arXiv:1707.08256}

\bibitem[\protect\citeauthoryear{{DeRose} et~al.,}{{DeRose}
  et~al.}{2018}]{DeRose2018}
{DeRose} J.  et~al., 2018, in prep.

\bibitem[\protect\citeauthoryear{{Diemer} \& {Kravtsov}}{{Diemer} \&
  {Kravtsov}}{2014}]{diemerkravtsov14}
\href{http://adsabs.harvard.edu/abs/2014ApJ...789....1D}{{Diemer} B.,
  {Kravtsov} A.~V.,  2014, \apj, 789, 1}

\bibitem[\protect\citeauthoryear{{Diemer} \& {Kravtsov}}{{Diemer} \&
  {Kravtsov}}{2015}]{DiemerKravtsov15}
\href{http://adsabs.harvard.edu/abs/2015ApJ...799..108D}{{Diemer} B.,
  {Kravtsov} A.~V.,  2015, \apj, 799, 108}

\bibitem[\protect\citeauthoryear{{Dietrich} et~al.,}{{Dietrich}
  et~al.}{2014}]{Dietrich2014}
\href{http://adsabs.harvard.edu/abs/2014MNRAS.443.1713D}{{Dietrich} J.~P.
  et~al., 2014, \mnras, 443, 1713}

\bibitem[\protect\citeauthoryear{{Dietrich} et~al.,}{{Dietrich}
  et~al.}{2017}]{Dietrich2017}
\href{http://adsabs.harvard.edu/abs/2017arXiv171105344D}{{Dietrich} J.~P.
  et~al., 2017, arXiv:1711.05344}

\bibitem[\protect\citeauthoryear{{Dodelson} \& {Schneider}}{{Dodelson} \&
  {Schneider}}{2013}]{Dodelson2013}
\href{http://adsabs.harvard.edu/abs/2013PhRvD..88f3537D}{{Dodelson} S.,
  {Schneider} M.~D.,  2013, 88, 063537}

\bibitem[\protect\citeauthoryear{{Dodelson} et~al.,}{{Dodelson}
  et~al.}{2016}]{CosmicVisions16}
\href{http://adsabs.harvard.edu/abs/2016arXiv160407626D}{{Dodelson} S.  et~al.,
  2016, arXiv:1604.07626}

\bibitem[\protect\citeauthoryear{{Drlica-Wagner} et~al.,}{{Drlica-Wagner}
  et~al.}{2018}]{Y1gold}
\href{http://adsabs.harvard.edu/abs/2018ApJS..235...33D}{{Drlica-Wagner} A.
  et~al., 2018, \apjs, 235, 33}

\bibitem[\protect\citeauthoryear{{Efron}}{{Efron}}{1982}]{Efron82.1}
\href{http://adsabs.harvard.edu/abs/1982jbor.book.....E}{{Efron} B.,  1982,
  {The Jackknife, the Bootstrap and other resampling plans}.
 CBMS-NSF Regional Conference Series in Applied Mathematics (SIAM)}

\bibitem[\protect\citeauthoryear{{Evrard} et~al.,}{{Evrard}
  et~al.}{2014}]{Evrard2014}
\href{http://adsabs.harvard.edu/abs/2014MNRAS.441.3562E}{{Evrard} A.~E.
  et~al., 2014, \mnras, 441, 3562}

\bibitem[\protect\citeauthoryear{{Farahi} et~al.,}{{Farahi}
  et~al.}{2018}]{Farahi2018}
{Farahi} A.,  et~al., 2018, in prep.

\bibitem[\protect\citeauthoryear{{Farahi} et~al.,}{{Farahi}
  et~al.}{2016}]{farahi16}
\href{http://adsabs.harvard.edu/abs/2016MNRAS.460.3900F}{{Farahi} A.  et~al.,
  2016, \mnras, 460, 3900}

\bibitem[\protect\citeauthoryear{{Flaugher} et~al.,}{{Flaugher}
  et~al.}{2015}]{Flaugher2015}
\href{http://adsabs.harvard.edu/abs/2015AJ....150..150F}{{Flaugher} B.  et~al.,
  2015, \aj, 150, 150}

\bibitem[\protect\citeauthoryear{{Foreman-Mackey} et~al.,}{{Foreman-Mackey}
  et~al.}{2013}]{Foreman13}
\href{http://adsabs.harvard.edu/abs/2013PASP..125..306F}{{Foreman-Mackey} D.
  et~al., 2013, \pasp, 125, 306}

\bibitem[\protect\citeauthoryear{{Friedrich} et~al.,}{{Friedrich}
  et~al.}{2016}]{Friedrich2016}
\href{http://adsabs.harvard.edu/abs/2016MNRAS.456.2662F}{{Friedrich} O.
  et~al., 2016, \mnras, 456, 2662}

\bibitem[\protect\citeauthoryear{{Friedrich} et~al.,}{{Friedrich}
  et~al.}{2017}]{Friedrich2017b}
\href{http://adsabs.harvard.edu/abs/2017arXiv171005162F}{{Friedrich} O.
  et~al., 2017, arXiv:1710.05162}

\bibitem[\protect\citeauthoryear{{Gatti} et~al.,}{{Gatti}
  et~al.}{2017}]{Gatti2017}
\href{http://adsabs.harvard.edu/abs/2017arXiv170900992G}{{Gatti} M.  et~al.,
  2017, arXiv:1709.00992}

\bibitem[\protect\citeauthoryear{{Geach} \& {Peacock}}{{Geach} \&
  {Peacock}}{2017}]{GeachPeacock2017}
\href{http://adsabs.harvard.edu/abs/2017NatAs...1..795G}{{Geach} J.~E.,
  {Peacock} J.~A.,  2017, Nature Astronomy, 1, 795}

\bibitem[\protect\citeauthoryear{{Gruen} et~al.,}{{Gruen}
  et~al.}{2018}]{Gruen2018_ICL}
{Gruen} D.,  et~al., 2018, in prep.

\bibitem[\protect\citeauthoryear{{Gruen} \& {Brimioulle}}{{Gruen} \&
  {Brimioulle}}{2017}]{Gruen2017_PZmethods}
\href{http://adsabs.harvard.edu/abs/2017MNRAS.468..769G}{{Gruen} D.,
  {Brimioulle} F.,  2017, \mnras, 468, 769}

\bibitem[\protect\citeauthoryear{{Gruen} et~al.,}{{Gruen}
  et~al.}{2014}]{Gruen2014}
\href{http://adsabs.harvard.edu/abs/2014MNRAS.442.1507G}{{Gruen} D.  et~al.,
  2014, \mnras, 442, 1507}

\bibitem[\protect\citeauthoryear{{Gruen} et~al.,}{{Gruen}
  et~al.}{2015}]{Gruen2015}
\href{http://adsabs.harvard.edu/abs/2015MNRAS.449.4264G}{{Gruen} D.  et~al.,
  2015, \mnras, 449, 4264}

\bibitem[\protect\citeauthoryear{{Gruen} et~al.,}{{Gruen}
  et~al.}{2018}]{Gruen2018_splits}
\href{http://adsabs.harvard.edu/abs/2018PhRvD..98b3507G}{{Gruen} D.  et~al.,
  2018, \prd, 98, 023507}

\bibitem[\protect\citeauthoryear{{Hayashi} \& {White}}{{Hayashi} \&
  {White}}{2008}]{Hayashi08}
\href{http://adsabs.harvard.edu/abs/2008MNRAS.388....2H}{{Hayashi} E.,  {White}
  S.~D.~M.,  2008, \mnras, 388, 2}

\bibitem[\protect\citeauthoryear{{Henson} et~al.,}{{Henson}
  et~al.}{2017}]{hensonetal17}
\href{http://adsabs.harvard.edu/abs/2017MNRAS.465.3361H}{{Henson} M.~A.
  et~al., 2017, \mnras, 465, 3361}

\bibitem[\protect\citeauthoryear{{Hoekstra}}{{Hoekstra}}{2003}]{Hoekstra03}
\href{http://adsabs.harvard.edu/abs/2003MNRAS.339.1155H}{{Hoekstra} H.,  2003,
  \mnras, 339, 1155}

\bibitem[\protect\citeauthoryear{{Hoekstra} et~al.,}{{Hoekstra}
  et~al.}{2015}]{Hoekstra2015}
\href{http://adsabs.harvard.edu/abs/2015MNRAS.449..685H}{{Hoekstra} H.  et~al.,
  2015, \mnras, 449, 685}

\bibitem[\protect\citeauthoryear{{Hoyle} et~al.,}{{Hoyle} et~al.}{2017}]{Y1pz}
\href{http://adsabs.harvard.edu/abs/2017arXiv170801532H}{{Hoyle} B.  et~al.,
  2017, arXiv:1708.01532}

\bibitem[\protect\citeauthoryear{{Huff} \& {Mandelbaum}}{{Huff} \&
  {Mandelbaum}}{2017}]{HuffMETA}
\href{http://adsabs.harvard.edu/abs/2017arXiv170202600H}{{Huff} E.,
  {Mandelbaum} R.,  2017, arXiv:1702.02600}

\bibitem[\protect\citeauthoryear{{Jarvis} et~al.,}{{Jarvis}
  et~al.}{2016}]{Jarvis2016}
\href{http://adsabs.harvard.edu/abs/2016MNRAS.460.2245J}{{Jarvis} M.  et~al.,
  2016, \mnras, 460, 2245}

\bibitem[\protect\citeauthoryear{{Johnston} et~al.,}{{Johnston}
  et~al.}{2007}]{Johnston07}
\href{http://adsabs.harvard.edu/abs/2007arXiv0709.1159J}{{Johnston} D.~E.
  et~al., 2007, arXiv:0709.1159}

\bibitem[\protect\citeauthoryear{{Landy} \& {Szalay}}{{Landy} \&
  {Szalay}}{1993}]{Landy93}
\href{http://adsabs.harvard.edu/abs/1993ApJ...412...64L}{{Landy} S.~D.,
  {Szalay} A.~S.,  1993, \apj, 412, 64}

\bibitem[\protect\citeauthoryear{{Lesgourgues}}{{Lesgourgues}}{2011}]{Lesgourgues11CLASS1}
\href{http://adsabs.harvard.edu/abs/2011arXiv1104.2932L}{{Lesgourgues} J.,
  2011, arXiv:1104.2932}

\bibitem[\protect\citeauthoryear{{Mantz} et~al.,}{{Mantz}
  et~al.}{2015a}]{mantzetal15}
\href{http://adsabs.harvard.edu/abs/2015MNRAS.446.2205M}{{Mantz} A.~B.  et~al.,
  2015a, \mnras, 446, 2205}

\bibitem[\protect\citeauthoryear{{Mantz} et~al.,}{{Mantz}
  et~al.}{2015b}]{Mantz2015}
\href{http://adsabs.harvard.edu/abs/2015MNRAS.446.2205M}{{Mantz} A.~B.  et~al.,
  2015b, \mnras, 446, 2205}

\bibitem[\protect\citeauthoryear{{Mantz} et~al.,}{{Mantz}
  et~al.}{2016}]{Mantz2016}
\href{http://adsabs.harvard.edu/abs/2016MNRAS.463.3582M}{{Mantz} A.~B.  et~al.,
  2016, \mnras, 463, 3582}

\bibitem[\protect\citeauthoryear{{Medezinski} et~al.,}{{Medezinski}
  et~al.}{2010}]{Medezinski2010}
\href{http://adsabs.harvard.edu/abs/2010MNRAS.405..257M}{{Medezinski} E.
  et~al., 2010, \mnras, 405, 257}

\bibitem[\protect\citeauthoryear{{Medezinski} et~al.,}{{Medezinski}
  et~al.}{2018a}]{Medezinski2018_colorcuts}
\href{http://adsabs.harvard.edu/abs/2018PASJ...70...30M}{{Medezinski} E.
  et~al., 2018a, Publ. Aston. Soc. Jpn., 70, 30}

\bibitem[\protect\citeauthoryear{{Medezinski} et~al.,}{{Medezinski}
  et~al.}{2018b}]{Medezinski2018_PlanckHSC}
\href{http://adsabs.harvard.edu/abs/2018PASJ...70S..28M}{{Medezinski} E.
  et~al., 2018b, Publ. Aston. Soc. Jpn., 70, S28}

\bibitem[\protect\citeauthoryear{{Melchior} et~al.,}{{Melchior}
  et~al.}{2015}]{Melchior15}
\href{http://adsabs.harvard.edu/abs/2015MNRAS.449.2219M}{{Melchior} P.  et~al.,
  2015, \mnras, 449, 2219}

\bibitem[\protect\citeauthoryear{{Melchior} et~al.,}{{Melchior}
  et~al.}{2017}]{rmsva}
\href{http://adsabs.harvard.edu/abs/2017MNRAS.469.4899M}{{Melchior} P.  et~al.,
  2017, \mnras, 469, 4899}

\bibitem[\protect\citeauthoryear{{Miyatake} et~al.,}{{Miyatake}
  et~al.}{2018}]{Miyatake2018_ACTPolHSC}
\href{http://adsabs.harvard.edu/abs/2018arXiv180405873M}{{Miyatake} H.  et~al.,
  2018, arXiv:1804.05873}

\bibitem[\protect\citeauthoryear{{Murata} et~al.,}{{Murata}
  et~al.}{2018}]{Murata2018}
\href{http://adsabs.harvard.edu/abs/2018ApJ...854..120M}{{Murata} R.  et~al.,
  2018, \apj, 854, 120}

\bibitem[\protect\citeauthoryear{{Navarro}, {Frenk} \& {White}}{{Navarro}
  et~al.}{1996}]{Navarro96.1}
\href{http://adsabs.harvard.edu/abs/1996ApJ...462..563N}{{Navarro} J.~F.,
  {Frenk} C.~S.,    {White} S.~D.~M.,  1996, \apj, 462, 563}

\bibitem[\protect\citeauthoryear{{Noh} \& {Cohn}}{{Noh} \&
  {Cohn}}{2012}]{nohcohn12}
\href{http://adsabs.harvard.edu/abs/2012MNRAS.426.1829N}{{Noh} Y.,  {Cohn}
  J.~D.,  2012, \mnras, 426, 1829}

\bibitem[\protect\citeauthoryear{{Okabe} \& {Smith}}{{Okabe} \&
  {Smith}}{2016}]{okabesmith16}
\href{http://adsabs.harvard.edu/abs/2016MNRAS.461.3794O}{{Okabe} N.,  {Smith}
  G.~P.,  2016, \mnras, 461, 3794}

\bibitem[\protect\citeauthoryear{{Osato} et~al.,}{{Osato}
  et~al.}{2017}]{Osato2017}
\href{http://adsabs.harvard.edu/abs/2017arXiv171200094O}{{Osato} K.  et~al.,
  2017, arXiv:1712.00094}

\bibitem[\protect\citeauthoryear{{Planck Collaboration} et~al.,}{{Planck
  Collaboration} et~al.}{2016}]{planck_clusters_15}
\href{http://adsabs.harvard.edu/abs/2016A%26A...594A..24P}{{Planck
  Collaboration} et~al., 2016, \aap, 594, A24}

\bibitem[\protect\citeauthoryear{{Prat} et~al.,}{{Prat}
  et~al.}{2017}]{Prat2017}
\href{http://adsabs.harvard.edu/abs/2017arXiv170801537P}{{Prat} J.  et~al.,
  2017, arXiv:1708.01537}

\bibitem[\protect\citeauthoryear{{Rowe} et~al.,}{{Rowe}
  et~al.}{2015}]{Rowe2014}
\href{http://adsabs.harvard.edu/abs/2015A%26C....10..121R}{{Rowe} B.~T.~P.
  et~al., 2015, Astronomy and Computing, 10, 121}

\bibitem[\protect\citeauthoryear{{Rozo} \& {Rykoff}}{{Rozo} \&
  {Rykoff}}{2014}]{RozoRykoff14_RM2}
\href{http://adsabs.harvard.edu/abs/2014ApJ...783...80R}{{Rozo} E.,  {Rykoff}
  E.~S.,  2014, \apj, 783, 80}

\bibitem[\protect\citeauthoryear{{Rozo} et~al.,}{{Rozo}
  et~al.}{2010}]{rozoetal10}
\href{http://adsabs.harvard.edu/abs/2010ApJ...708..645R}{{Rozo} E.  et~al.,
  2010, \apj, 708, 645}

\bibitem[\protect\citeauthoryear{{Rozo} et~al.,}{{Rozo}
  et~al.}{2016}]{Rozo2016_redmagic}
\href{http://adsabs.harvard.edu/abs/2016MNRAS.461.1431R}{{Rozo} E.  et~al.,
  2016, \mnras, 461, 1431}

\bibitem[\protect\citeauthoryear{{Rykoff} et~al.,}{{Rykoff}
  et~al.}{2012}]{Rykoff12}
\href{http://adsabs.harvard.edu/abs/2012ApJ...746..178R}{{Rykoff} E.~S.
  et~al., 2012, \apj, 746, 178}

\bibitem[\protect\citeauthoryear{{Rykoff} et~al.,}{{Rykoff}
  et~al.}{2014}]{Rykoff2014_RM1}
\href{http://adsabs.harvard.edu/abs/2014ApJ...785..104R}{{Rykoff} E.~S.
  et~al., 2014, \apj, 785, 104}

\bibitem[\protect\citeauthoryear{{Rykoff} et~al.,}{{Rykoff}
  et~al.}{2016}]{Rykoff2016}
\href{http://adsabs.harvard.edu/abs/2016ApJS..224....1R}{{Rykoff} E.~S.
  et~al., 2016, \apjs, 224, 1}

\bibitem[\protect\citeauthoryear{{Saro} et~al.,}{{Saro}
  et~al.}{2015}]{Saro2015}
\href{http://adsabs.harvard.edu/abs/2015MNRAS.454.2305S}{{Saro} A.  et~al.,
  2015, \mnras, 454, 2305}

\bibitem[\protect\citeauthoryear{{Schaller} et~al.,}{{Schaller}
  et~al.}{2015}]{schalleretal15a}
\href{http://adsabs.harvard.edu/abs/2015MNRAS.451.1247S}{{Schaller} M.  et~al.,
  2015, \mnras, 451, 1247}

\bibitem[\protect\citeauthoryear{{Schneider} et~al.,}{{Schneider}
  et~al.}{1998}]{Schneider1998}
\href{http://adsabs.harvard.edu/abs/1998MNRAS.296..873S}{{Schneider} P.
  et~al., 1998, \mnras, 296, 873}

\bibitem[\protect\citeauthoryear{{Schrabback} et~al.,}{{Schrabback}
  et~al.}{2018}]{Schrabback2018}
\href{http://adsabs.harvard.edu/abs/2018A%26A...610A..85S}{{Schrabback} T.
  et~al., 2018, \aap, 610, A85}

\bibitem[\protect\citeauthoryear{{Sheldon}}{{Sheldon}}{2015}]{ngmix2015}
\href{http://adsabs.harvard.edu/abs/2015ascl.soft08008S}{{Sheldon} E.,  2015,
  {NGMIX: Gaussian mixture models for 2D images}, Astrophysics Source Code
  Library, \eprint {ascl} {1508.008}}

\bibitem[\protect\citeauthoryear{{Sheldon} \& {Huff}}{{Sheldon} \&
  {Huff}}{2017}]{SheldonMETA}
\href{http://adsabs.harvard.edu/abs/2017ApJ...841...24S}{{Sheldon} E.~S.,
  {Huff} E.~M.,  2017, \apj, 841, 24}

\bibitem[\protect\citeauthoryear{{Sheldon} et~al.,}{{Sheldon}
  et~al.}{2004}]{Sheldon04.1}
\href{http://adsabs.harvard.edu/abs/2004AJ....127.2544S}{{Sheldon} E.~S.
  et~al., 2004, \aj, 127, 2544}

\bibitem[\protect\citeauthoryear{{Sif{\'o}n} et~al.,}{{Sif{\'o}n}
  et~al.}{2015}]{Sifon2015}
\href{http://adsabs.harvard.edu/abs/2015A%26A...575A..48S}{{Sif{\'o}n} C.
  et~al., 2015, \aap, 575, A48}

\bibitem[\protect\citeauthoryear{{Simet} et~al.,}{{Simet}
  et~al.}{2015}]{Simet2015}
\href{http://adsabs.harvard.edu/abs/2015arXiv150201024S}{{Simet} M.  et~al.,
  2015, arXiv:1502.01024}

\bibitem[\protect\citeauthoryear{{Simet} et~al.,}{{Simet}
  et~al.}{2017}]{Simet2017}
\href{http://adsabs.harvard.edu/abs/2017MNRAS.466.3103S}{{Simet} M.  et~al.,
  2017, \mnras, 466, 3103}

\bibitem[\protect\citeauthoryear{{Singh} et~al.,}{{Singh}
  et~al.}{2016}]{Singh2016}
\href{http://adsabs.harvard.edu/abs/2016arXiv161100752S}{{Singh} S.  et~al.,
  2016, arXiv:1611.00752}

\bibitem[\protect\citeauthoryear{{Sinha} \& {Garrison}}{{Sinha} \&
  {Garrison}}{2017}]{Corrfunc}
\href{http://adsabs.harvard.edu/abs/2017ascl.soft03003S}{{Sinha} M.,
  {Garrison} L.,  2017, {Corrfunc: Blazing fast correlation functions on the
  CPU}, Astrophysics Source Code Library, \eprint {ascl} {1703.003}}

\bibitem[\protect\citeauthoryear{{Smith} et~al.,}{{Smith}
  et~al.}{2003}]{Smith02Halofit}
\href{http://adsabs.harvard.edu/abs/2003MNRAS.341.1311S}{{Smith} R.~E.  et~al.,
  2003, \mnras, 341, 1311}

\bibitem[\protect\citeauthoryear{{Springel}}{{Springel}}{2005}]{Springel05}
\href{http://adsabs.harvard.edu/abs/2005MNRAS.364.1105S}{{Springel} V.,  2005,
  \mnras, 364, 1105}

\bibitem[\protect\citeauthoryear{{Takahashi} et~al.,}{{Takahashi}
  et~al.}{2012}]{Takahashi12Halofit}
\href{http://adsabs.harvard.edu/abs/2012ApJ...761..152T}{{Takahashi} R.
  et~al., 2012, \apj, 761, 152}

\bibitem[\protect\citeauthoryear{{Tinker} et~al.,}{{Tinker}
  et~al.}{2008}]{Tinker2008}
\href{http://adsabs.harvard.edu/abs/2008ApJ...688..709T}{{Tinker} J.  et~al.,
  2008, 688, 709}

\bibitem[\protect\citeauthoryear{{Troxel} et~al.,}{{Troxel}
  et~al.}{2017}]{Troxel2017}
\href{http://adsabs.harvard.edu/abs/2017arXiv170801538T}{{Troxel} M.~A.
  et~al., 2017, arXiv:1708.01538}

\bibitem[\protect\citeauthoryear{{Umetsu} et~al.,}{{Umetsu}
  et~al.}{2011}]{Umetsu2011}
\href{http://adsabs.harvard.edu/abs/2011ApJ...738...41U}{{Umetsu} K.  et~al.,
  2011, \apj, 738, 41}

\bibitem[\protect\citeauthoryear{{Varga} et~al.,}{{Varga}
  et~al.}{2018}]{Varga2018}
{Varga} T.~N.  et~al., 2018, in prep.

\bibitem[\protect\citeauthoryear{{Wechsler}, {DeRose} \& {Busha}}{{Wechsler}
  et~al.}{2018}]{Wechsler2018}
{Wechsler} R.,  {DeRose} J.,    {Busha} 2018, in prep.

\bibitem[\protect\citeauthoryear{{White} et~al.,}{{White}
  et~al.}{2011}]{White2011}
\href{http://adsabs.harvard.edu/abs/2011ApJ...728..126W}{{White} M.,  et~al.,
  2011, \apj, 728, 126}

\bibitem[\protect\citeauthoryear{{Yang} et~al.,}{{Yang} et~al.}{2006}]{Yang06}
\href{http://adsabs.harvard.edu/abs/2006MNRAS.373.1159Y}{{Yang} X.  et~al.,
  2006, \mnras, 373, 1159}

\bibitem[\protect\citeauthoryear{{Zhang} et~al.,}{{Zhang}
  et~al.}{2018b}]{Zhang2018_ICL}
{Zhang} Y.,  et~al., 2018b, in prep.

\bibitem[\protect\citeauthoryear{{Zhang} et~al.,}{{Zhang}
  et~al.}{2018a}]{Zhang2018}
{Zhang} Y.,  et~al., 2018a, in prep.

\bibitem[\protect\citeauthoryear{{Zuntz} et~al.,}{{Zuntz}
  et~al.}{2013}]{Zuntz13}
\href{http://adsabs.harvard.edu/abs/2013MNRAS.434.1604Z}{{Zuntz} J.  et~al.,
  2013, \mnras, 434, 1604}

\bibitem[\protect\citeauthoryear{{Zuntz} et~al.,}{{Zuntz}
  et~al.}{2017}]{Y1shape}
\href{http://adsabs.harvard.edu/abs/2017arXiv170801533Z}{{Zuntz} J.  et~al.,
  2017, arXiv:1708.01533}

\bibitem[\protect\citeauthoryear{{Zu} et~al.,}{{Zu} et~al.}{2014}]{zuetal14}
\href{http://adsabs.harvard.edu/abs/2014MNRAS.439.1628Z}{{Zu} Y.  et~al., 2014,
  \mnras, 439, 1628}

\bibitem[\protect\citeauthoryear{{van Uitert} et~al.,}{{van Uitert}
  et~al.}{2016}]{vanUitert2016}
\href{http://adsabs.harvard.edu/abs/2016A%26A...586A..43V}{{van Uitert} E.
  et~al., 2016, \aap, 586, A43}

\bibitem[\protect\citeauthoryear{{von der Linden} et~al.,}{{von der Linden}
  et~al.}{2018}]{vonderLinden2018}
{von der Linden} A.,  et~al., 2018, in prep.

\bibitem[\protect\citeauthoryear{{von der Linden} et~al.,}{{von der Linden}
  et~al.}{2014a}]{WtGI}
\href{http://adsabs.harvard.edu/abs/2014MNRAS.439....2V}{{von der Linden} A.
  et~al., 2014a, \mnras, 439, 2}

\bibitem[\protect\citeauthoryear{{von der Linden} et~al.,}{{von der Linden}
  et~al.}{2014b}]{vonderLinden14}
\href{http://adsabs.harvard.edu/abs/2014MNRAS.443.1973V}{{von der Linden} A.
  et~al., 2014b, \mnras, 443, 1973}

\end{thebibliography}

\section*{Affiliations}
$^{1}$ Department of Physics, University of Arizona, Tucson, AZ 85721, USA\\
$^{2}$ Max Planck Institute for Extraterrestrial Physics, Giessenbachstrasse, 85748 Garching, Germany\\
$^{3}$ Universit\"ats-Sternwarte, Fakult\"at f\"ur Physik, Ludwig-Maximilians Universit\"at M\"unchen, Scheinerstr. 1, 81679 M\"unchen, Germany\\
$^{4}$ Kavli Institute for Particle Astrophysics \& Cosmology, P. O. Box 2450, Stanford University, Stanford, CA 94305, USA\\
$^{5}$ SLAC National Accelerator Laboratory, Menlo Park, CA 94025, USA\\
$^{6}$ Department of Physics and Astronomy, University of Pennsylvania, Philadelphia, PA 19104, USA\\
$^{7}$ Department of Astrophysical Sciences, Princeton University, Peyton Hall, Princeton, NJ 08544, USA\\
$^{8}$ Department of Physics, Stanford University, 382 Via Pueblo Mall, Stanford, CA 94305, USA\\
$^{9}$ Excellence Cluster Universe, Boltzmannstr.\ 2, 85748 Garching, Germany\\
$^{10}$ Faculty of Physics, Ludwig-Maximilians-Universit\"at, Scheinerstr. 1, 81679 Munich, Germany\\
$^{11}$ Brookhaven National Laboratory, Bldg 510, Upton, NY 11973, USA\\
$^{12}$ Fermi National Accelerator Laboratory, P. O. Box 500, Batavia, IL 60510, USA\\
$^{13}$ Department of Physics and Astronomy, Stony Brook University, Stony Brook, NY 11794, USA\\
$^{14}$ Santa Cruz Institute for Particle Physics, Santa Cruz, CA 95064, USA\\
$^{15}$ Department of Physics and Astronomy, Pevensey Building, University of Sussex, Brighton, BN1 9QH, UK\\
$^{16}$ Department of Physics, University of Michigan, Ann Arbor, MI 48109, USA\\
$^{17}$ Department of Physics \& Astronomy, University College London, Gower Street, London, WC1E 6BT, UK\\
$^{18}$ Department of Physics, ETH Zurich, Wolfgang-Pauli-Strasse 16, CH-8093 Zurich, Switzerland\\
$^{19}$ Department of Physics, Carnegie Mellon University, Pittsburgh, Pennsylvania 15312, USA\\
$^{20}$ Center for Cosmology and Astro-Particle Physics, The Ohio State University, Columbus, OH 43210, USA\\
$^{21}$ Department of Physics, The Ohio State University, Columbus, OH 43210, USA\\
$^{22}$ Jet Propulsion Laboratory, California Institute of Technology, 4800 Oak Grove Dr., Pasadena, CA 91109, USA\\
$^{23}$ University of California, Riverside, 900 University Avenue, Riverside, CA 92521, USA\\
$^{24}$ Brandeis University, Physics Department, 415 South Street, Waltham MA 02453\\
$^{25}$ Institute for Astronomy, University of Edinburgh, Edinburgh EH9 3HJ, UK\\
$^{26}$ Cerro Tololo Inter-American Observatory, National Optical Astronomy Observatory, Casilla 603, La Serena, Chile\\
$^{27}$ Department of Physics and Electronics, Rhodes University, PO Box 94, Grahamstown, 6140, South Africa\\
$^{28}$ Institute of Cosmology \& Gravitation, University of Portsmouth, Portsmouth, PO1 3FX, UK\\
$^{29}$ Jodrell Bank Center for Astrophysics, School of Physics and Astronomy, University of Manchester, Oxford Road, Manchester, M13 9PL, UK\\
$^{30}$ Laborat\'orio Interinstitucional de e-Astronomia - LIneA, Rua Gal. Jos\'e Cristino 77, Rio de Janeiro, RJ - 20921-400, Brazil\\
$^{31}$ Observat\'orio Nacional, Rua Gal. Jos\'e Cristino 77, Rio de Janeiro, RJ - 20921-400, Brazil\\
$^{32}$ Department of Astronomy, University of Illinois at Urbana-Champaign, 1002 W. Green Street, Urbana, IL 61801, USA\\
$^{33}$ National Center for Supercomputing Applications, 1205 West Clark St., Urbana, IL 61801, USA\\
$^{34}$ Institut de F\'{\i}sica d'Altes Energies (IFAE), The Barcelona Institute of Science and Technology, Campus UAB, 08193 Bellaterra (Barcelona) Spain\\
$^{35}$ Institut d'Estudis Espacials de Catalunya (IEEC), 08193 Barcelona, Spain\\
$^{36}$ Institute of Space Sciences (ICE, CSIC),  Campus UAB, Carrer de Can Magrans, s/n,  08193 Barcelona, Spain\\
$^{37}$ Centro de Investigaciones Energ\'eticas, Medioambientales y Tecnol\'ogicas (CIEMAT), Madrid, Spain\\
$^{38}$ Department of Astronomy, University of Michigan, Ann Arbor, MI 48109, USA\\
$^{39}$ Kavli Institute for Cosmological Physics, University of Chicago, Chicago, IL 60637, USA\\
$^{40}$ Instituto de Fisica Teorica UAM/CSIC, Universidad Autonoma de Madrid, 28049 Madrid, Spain\\
$^{41}$ Institute of Astronomy, University of Cambridge, Madingley Road, Cambridge CB3 0HA, UK\\
$^{42}$ Kavli Institute for Cosmology, University of Cambridge, Madingley Road, Cambridge CB3 0HA, UK\\
$^{43}$ Harvard-Smithsonian Center for Astrophysics, Cambridge, MA 02138, USA\\
$^{44}$ Department of Astronomy/Steward Observatory, 933 North Cherry Avenue, Tucson, AZ 85721-0065, USA\\
$^{45}$ Australian Astronomical Observatory, North Ryde, NSW 2113, Australia\\
$^{46}$ Departamento de F\'isica Matem\'atica, Instituto de F\'isica, Universidade de S\~ao Paulo, CP 66318, S\~ao Paulo, SP, 05314-970, Brazil\\
$^{47}$ George P. and Cynthia Woods Mitchell Institute for Fundamental Physics and Astronomy, and Department of Physics and Astronomy, Texas A\&M University, College Station, TX 77843,  USA\\
$^{48}$ Instituci\'o Catalana de Recerca i Estudis Avan\c{c}ats, E-08010 Barcelona, Spain\\
$^{49}$ School of Physics and Astronomy, University of Southampton,  Southampton, SO17 1BJ, UK\\
$^{50}$ Instituto de F\'isica Gleb Wataghin, Universidade Estadual de Campinas, 13083-859, Campinas, SP, Brazil\\
$^{51}$ Computer Science and Mathematics Division, Oak Ridge National Laboratory, Oak Ridge, TN 37831\\
$^{52}$ Argonne National Laboratory, 9700 South Cass Avenue, Lemont, IL 60439, USA

\appendix

\section{The redMaPPer v6.4.17 cluster catalog}
\label{app:rm}

The full \redmapper{} DES Y1A1 catalogs will be available at {\tt
 http://risa.stanford.edu/redmapper/} in FITS format.  The catalogs will also
be available from the online journal in machine-readable formats.  We note that
this is of the same format as \citep{Rykoff2016}, and we point the reader to
that paper for further details.  The cluster catalog is described in
\autoref{tab:y1a1cat}, and the associated members in
\autoref{tab:y1a1mem}.  The catalog is the ``full'' catalog, with all
clusters with $\lambda>20$, and the volume-limited subset is flagged with the
{\tt VLIM} flag.  The map of the maximum redshift of the volume-limited
catalog is described in \autoref{tab:zmaskcat}, and the random points are described
in \autoref{tab:randcat}.

\begin{table*}
\caption{\redmapper{} Y1A1 Cluster Catalog Format}
\begin{tabular}{lll}
Name & Format & Description \\ \hline
ID & INT(4) & \redmapper{} Cluster Identification Number\\
VLIM & INT(2) & One if in cosmology catalog, 0 otherwise\\
NAME & CHAR(20) & \redmapper{} Cluster Name\\
RA & FLOAT(8) & Right ascension in decimal degrees (J2000)\\
DEC & FLOAT(8) & Declination in decimal degrees (J2000)\\
Z\_LAMBDA & FLOAT(4) & Cluster \photoz $z_\lambda$\\
Z\_LAMBDA\_ERR & FLOAT(4) & Gaussian error estimate for $z_\lambda$\\
LAMBDA & FLOAT(4) & Richness estimate $\lambda$\\
LAMBDA\_ERR & FLOAT(4) & Gaussian error estimate for $z_\lambda$\\
S & FLOAT(4) & Richness scale factor\\
Z\_SPEC & FLOAT(4) & Spectroscopic redshift for most likely center (-1.0 if not available)\\
COADD\_OBJECTS\_ID & INT(8) & DES {\tt COADD\_OBJECTS\_ID} identification number \\
MAG\_CM\_G & FLOAT(4) & $g$ MAG\_CM magnitude for most likely central
galaxy (SLR corrected)\\
MAGERR\_CM\_G & FLOAT(4) & error on $g$ MAG\_CM magnitude\\
MAG\_CM\_R & FLOAT(4) & $r$ MAG\_CM magnitude for most likely central
galaxy (SLR corrected)\\
MAGERR\_CM\_R & FLOAT(4) & error on $g$ MAG\_CM magnitude\\
MAG\_CM\_I & FLOAT(4) & $i$ MAG\_CM magnitude for most likely central
galaxy (SLR corrected)\\
MAGERR\_CM\_I & FLOAT(4) & error on $g$ MAG\_CM magnitude\\
MAG\_CM\_Z & FLOAT(4) & $z$ MAG\_CM magnitude for most likely central
galaxy (SLR corrected)\\
MAGERR\_CM\_Z & FLOAT(4) & error on $g$ MAG\_CM magnitude\\
ZLUM & FLOAT(4) & Total membership-weighted $z$-band luminosity (units of
$L_*$)\\
P\_CEN[5] & 5$\times$FLOAT(4) & Centering probability $P_{\mathrm{cen}}$ for 5 most likely centrals\\
RA\_CEN[5] & 5$\times$FLOAT(8) & R.A. for 5 most likely centrals\\
DEC\_CEN[5] & 5$\times$FLOAT(8) & Decl. for 5 most likely centrals\\
ID\_CEN[5] & 5$\times$INT(8) & DES {\tt COADD\_OBJECTS\_ID} identification number for
5 most likely centrals\\
PZBINS[21] & 21$\times$FLOAT(4) & Redshift points at which $P(z)$ is evaluated\\
PZ[21] & 21$\times$FLOAT(4) & $P(z)$ evaluated at redshift points given by
PZBINS\\
\end{tabular}
\label{tab:y1a1cat}
\end{table*}

\begin{table*}
\caption{\redmapper{} DES Y1A1 Member Catalog Format}
\begin{tabular}{lll}
Name & Format & Description \\ \hline
ID & INT(4) & \redmapper{} Cluster Identification Number\\
RA & FLOAT(8) & Right ascension in decimal degrees (J2000)\\
DEC & FLOAT(8) & Declination in decimal degrees (J2000)\\
R & FLOAT(4) & Distance from cluster center ($h^{-1}\,\mathrm{Mpc}$)\\
P & FLOAT(4) & Membership probability\\
P\_FREE & FLOAT(4) & Probability that member is not a member of a higher ranked
cluster\\
THETA\_L & FLOAT(4) & Luminosity ($z$-band) weight\\
THETA\_R & FLOAT(4) & Radial weight\\
MAG\_CM\_G & FLOAT(4) & $g$ MAG\_CM magnitude (SLR corrected)\\
MAGERR\_CM\_G & FLOAT(4) & error on $g$ MAG\_CM magnitude\\
MAG\_CM\_R & FLOAT(4) & $r$ MAG\_CM magnitude (SLR corrected)\\
MAGERR\_CM\_R & FLOAT(4) & error on $r$ MAG\_CM magnitude\\
MAG\_CM\_I & FLOAT(4) & $i$ MAG\_CM magnitude (SLR corrected)\\
MAGERR\_CM\_I & FLOAT(4) & error on $i$ MAG\_CM magnitude\\
MAG\_CM\_Z & FLOAT(4) & $z$ MAG\_CM magnitude (SLR corrected)\\
MAGERR\_CM\_Z & FLOAT(4) & error on $z$ MAG\_CM magnitude\\
Z\_SPEC & FLOAT(4) & Spectroscopic redshift (-1.0 if not available)\\
COADD\_OBJECTS\_ID & INT(8) & DES {\tt COADD\_OBJECTS\_ID} identification number\\
\end{tabular}
\label{tab:y1a1mem}
\end{table*}

\begin{table*}
\caption{\redmapper{} $z_{\mathrm{max}}$ Map Format}
\begin{tabular}{lll}
Name & Format & Description \\ \hline
HPIX & INT(8) & {\tt HEALPIX} ring-ordered pixel number ({\tt NSIDE=4096})\\
ZMAX & FLOAT(4) & Maximum redshift of a cluster centered in this pixel\\
FRACGOOD & FLOAT(4) & Fraction of pixel area that is not masked\\
\end{tabular}
\label{tab:zmaskcat}
\end{table*}

\begin{table*}
\caption{\redmapper{} Random Points Catalog Format}
\begin{tabular}{lll}
Name & Format & Description \\ \hline
RA & FLOAT(8) & Right ascension in decimal degrees (J2000)\\
DEC & FLOAT(8) & Declination in decimal degrees (J2000)\\
Z & FLOAT(4) & Redshift of random point\\
LAMBDA & FLOAT(4) & Richness of random point\\
WEIGHT & FLOAT(4) & Weight of random point\\
\end{tabular}
\label{tab:randcat}
\end{table*}

\section{Parameter posteriors}
\label{app:posteriors}

When fitting the weak lensing profiles some parameters are not constrained by a prior and are also not shared between cluster bins. These are the halo concentration $c$, the boost factor amplitude $B_0$, and the boost factor scale radius $R_s$. \autoref{tab:appendix_parameter_posteriors} shows the posteriors for these three parameters for each cluster bin. As seen in \autoref{fig:corner1} $B_0$ and $R_s$ are highly degenerate.
\begin{table*}
	\caption{Lensing profile parameters not constrained by priors or shared between cluster bins. Uncertainties are the 68 per cent confidence intervals. Note that in the highest redshift and richness bin, the boost factor profile model scale radius had a bimodal distribution, and is not well constrained. This did not affect the mass estimate at all.}
    \begin{tabular}{llll}
    $\lambda$ & $z\in[0.2,0.35)$ & $z\in[0.35,0.5)$ & $z\in[0.5,0.65)$ \\ \hline
    Concentration $c$ & & & \\ \hline
    $[20,30)$ & 5.81 $\pm$ 1.03 & 5.68 $\pm$ 1.14 & 4.76 $\pm$ 1.62\\
	$[30,45)$ & 4.53 $\pm$ 0.74 & 6.24 $\pm$ 1.08 & 3.61 $\pm$ 0.72\\
	$[45,60)$ & 4.38 $\pm$ 0.96 & 5.41 $\pm$ 1.17 & 4.76 $\pm$ 1.21\\
	$[60,\infty)$ & 4.65 $\pm$ 0.82 & 3.19 $\pm$ 0.56 & 3.73 $\pm$ 1.02\\
    Boost factor amplitude $B_0$ & & & \\ \hline
    $[20,30)$ & 0.34 $\pm$ 0.05 & 0.05 $\pm$ 0.01 & 0.13 $\pm$ 0.05\\
	$[30,45)$ & 0.37 $\pm$ 0.06 & 0.14 $\pm$ 0.04 & 0.13 $\pm$ 0.08\\
	$[45,60)$ & 0.27 $\pm$ 0.05 & 0.05 $\pm$ 0.02 & 0.09 $\pm$ 0.06\\
	$[60,\infty)$ & 0.23 $\pm$ 0.03 & 0.21 $\pm$ 0.17 & 0.03 $\pm$ 0.04\\
    Boost factor scale radius $R_s\ [{\rm Mpc}]$ & & & \\ \hline
    $[20,30)$ & 0.44 $\pm$ 0.06 & 0.89 $\pm$ 0.24 & 0.38 $\pm$ 0.11\\
	$[30,45)$ & 0.50 $\pm$ 0.07 & 0.44 $\pm$ 0.10 & 0.44 $\pm$ 0.18\\
	$[45,60)$ & 0.80 $\pm$ 0.15 & 1.72 $\pm$ 0.95 & 0.85 $\pm$ 0.37\\
	$[60,\infty)$ & 1.37 $\pm$ 0.21 & 0.51 $\pm$ 0.23 & 35.94 $\pm$ 29.69\\
    \end{tabular}
	\label{tab:appendix_parameter_posteriors}
\end{table*}


\label{lastpage}
\end{document}